\documentclass[prd,aps,showpacs,preprintnumbers,amsmath,amssymb,superscriptaddress,twocolumn,floatfix]{revtex4}
\newcommand{\Tev}   {\ensuremath{\mathrm{Te\kern -0.1em V}}}
\newcommand{\Gev}   {\ensuremath{\mathrm{Ge\kern -0.1em V}}}
\newcommand{\Kev}   {\ensuremath{\mathrm{Ke\kern -0.1em V}}}
\newcommand{\Gevc}  {\ensuremath{\mathrm{Ge\kern -0.1em V/c}}}
\newcommand{\Gevcc} {\ensuremath{\mathrm{Ge\kern -0.1em V/c^2}}}
\newcommand{\Mev}   {\ensuremath{\mathrm{Me\kern -0.1em V}}}
\newcommand{\Mevc}  {\ensuremath{\mathrm{Me\kern -0.1em V/c}}}
\newcommand{\Mevcc} {\ensuremath{\mathrm{Me\kern -0.1em V/c^2}}}
\newcommand{\Tesla} {\ensuremath{\mathrm{T}}}
\newcommand{\ipb}   {\ensuremath{\mathrm{pb^{-1}}}}
\newcommand{\pb}    {\ensuremath{\mathrm{pb}}}
\newcommand{\met}   {\hbox{$E$\kern-0.5em\lower-0.1ex\hbox{/}}_T}
\newcommand{\m}     {\ensuremath{\mathrm{m}}}
\newcommand{\cm}    {\ensuremath{\mathrm{cm}}}

\newcommand{\microns} {\ensuremath{\mathrm{\mu m}}}

\newcommand{\hz}     {\ensuremath{\mathrm{Hz}}}

\newcommand{\mhz}     {\ensuremath{\mathrm{MHz}}}

\newcommand{\COT}   {COT}
\newcommand{\SVX}   {SVX}
\newcommand{\CEM}   {CEM}
\newcommand{\CES}   {CES}
\newcommand{\EM}    {EM}
\newcommand{\HAD}   {HAD}
\newcommand{\CMU}   {CMU}
\newcommand{\CMX}   {CMX}
\newcommand{\CMP}   {CMP}
\newcommand{\CMUP}  {CMUP}
\newcommand{\QCD}   {QCD}
\newcommand{\ISL}   {ISL}
\newcommand{\XFT}   {XFT}
\newcommand{\LUNO}  {L1}
\newcommand{\LDUE}  {L2}
\newcommand{\LTRE}  {L3}
\newcommand{\CDF}   {CDF}
\newcommand{\CDFII} {CDF II}

\newcommand{\PYTHIA}    {\tt{PYTHIA}}
\newcommand{\PYTHIAv}   {\tt{PYTHIA v6.2}}
\newcommand{\HERWIG}    {\tt{HERWIG}}

\newcommand{\HERWIGv}   {\tt{HERWIG v6.4}}
\newcommand{\QQv}       {\tt{QQ v9.1}}
\newcommand{\ALPGEN}    {\tt{ALPGEN}}
\newcommand{\CTEQ}      {\tt{CTEQ}}
\newcommand{\CTEQcL}    {\tt{CTEQ5L}}
\newcommand{\GEANTt}    {\tt{GEANT3}}
\newcommand{\GARFIELD}  {\tt{GARFIELD}}
\newcommand{\GFLASH}    {\tt{GFLASH}}
\newcommand{\PDF}       {\tt{PDF}}
\newcommand{\NLOCTEQsM} {\tt{NLO CTEQ6M}}
\newcommand{\MRST}      {\tt{MRST}}
\newcommand{\ISR}       {\tt{ISR}}
\newcommand{\FSR}       {\tt{FSR}}

\newcommand{\JP}    {jet probability}
\newcommand{\etal}  {{\it{et al.}}}  

\def\Journal#1#2#3#4{{#1} {\bf #2}, #3 (#4)}
 
\def\PRD{Phys. Rev. D}
\def\NIMA{Nucl. Instrum. Methods A}
\def\PRL{Phys. Rev. Lett.}
\def\PLB{Phys. Lett. B}
\def\EPJ{Eur. Phys. J}
\def\IEEETNS{IEEE Trans. Nucl. Sci.}
\def\CPCD{Comput. Phys. Commun.}

\usepackage{graphicx}
\usepackage{dcolumn}
\usepackage{bm}
\usepackage{setspace}

\begin{document}
\title{\boldmath Measurement of the $t\bar t$ Production Cross Section in $p\bar p$ collisions at $\sqrt{s}$ = 1.96 TeV using Lepton+Jets Events with Jet Probability b-tagging}

\affiliation{Institute of Physics, Academia Sinica, Taipei, Taiwan 11529, Republic of China} 
\affiliation{Argonne National Laboratory, Argonne, Illinois 60439} 
\affiliation{Institut de Fisica d'Altes Energies, Universitat Autonoma de Barcelona, E-08193, Bellaterra (Barcelona), Spain} 
\affiliation{Baylor University, Waco, Texas  76798} 
\affiliation{Istituto Nazionale di Fisica Nucleare, University of Bologna, I-40127 Bologna, Italy} 
\affiliation{Brandeis University, Waltham, Massachusetts 02254} 
\affiliation{University of California, Davis, Davis, California  95616} 
\affiliation{University of California, Los Angeles, Los Angeles, California  90024} 
\affiliation{University of California, San Diego, La Jolla, California  92093} 
\affiliation{University of California, Santa Barbara, Santa Barbara, California 93106} 
\affiliation{Instituto de Fisica de Cantabria, CSIC-University of Cantabria, 39005 Santander, Spain} 
\affiliation{Carnegie Mellon University, Pittsburgh, PA  15213} 
\affiliation{Enrico Fermi Institute, University of Chicago, Chicago, Illinois 60637} 
\affiliation{Joint Institute for Nuclear Research, RU-141980 Dubna, Russia} 
\affiliation{Duke University, Durham, North Carolina  27708} 
\affiliation{Fermi National Accelerator Laboratory, Batavia, Illinois 60510} 
\affiliation{University of Florida, Gainesville, Florida  32611} 
\affiliation{Laboratori Nazionali di Frascati, Istituto Nazionale di Fisica Nucleare, I-00044 Frascati, Italy} 
\affiliation{University of Geneva, CH-1211 Geneva 4, Switzerland} 
\affiliation{Glasgow University, Glasgow G12 8QQ, United Kingdom} 
\affiliation{Harvard University, Cambridge, Massachusetts 02138} 
\affiliation{Division of High Energy Physics, Department of Physics, University of Helsinki and Helsinki Institute of Physics, FIN-00014, Helsinki, Finland} 
\affiliation{University of Illinois, Urbana, Illinois 61801} 
\affiliation{The Johns Hopkins University, Baltimore, Maryland 21218} 
\affiliation{Institut f\"{u}r Experimentelle Kernphysik, Universit\"{a}t Karlsruhe, 76128 Karlsruhe, Germany} 
\affiliation{High Energy Accelerator Research Organization (KEK), Tsukuba, Ibaraki 305, Japan} 
\affiliation{Center for High Energy Physics: Kyungpook National University, Taegu 702-701, Korea; Seoul National University, Seoul 151-742, Korea; and SungKyunKwan University, Suwon 440-746, Korea} 
\affiliation{Ernest Orlando Lawrence Berkeley National Laboratory, Berkeley, California 94720} 
\affiliation{University of Liverpool, Liverpool L69 7ZE, United Kingdom} 
\affiliation{University College London, London WC1E 6BT, United Kingdom} 
\affiliation{Centro de Investigaciones Energeticas Medioambientales y Tecnologicas, E-28040 Madrid, Spain} 
\affiliation{Massachusetts Institute of Technology, Cambridge, Massachusetts  02139} 
\affiliation{Institute of Particle Physics: McGill University, Montr\'{e}al, Canada H3A~2T8; and University of Toronto, Toronto, Canada M5S~1A7} 
\affiliation{University of Michigan, Ann Arbor, Michigan 48109} 
\affiliation{Michigan State University, East Lansing, Michigan  48824} 
\affiliation{Institution for Theoretical and Experimental Physics, ITEP, Moscow 117259, Russia} 
\affiliation{University of New Mexico, Albuquerque, New Mexico 87131} 
\affiliation{Northwestern University, Evanston, Illinois  60208} 
\affiliation{The Ohio State University, Columbus, Ohio  43210} 
\affiliation{Okayama University, Okayama 700-8530, Japan} 
\affiliation{Osaka City University, Osaka 588, Japan} 
\affiliation{University of Oxford, Oxford OX1 3RH, United Kingdom} 
\affiliation{University of Padova, Istituto Nazionale di Fisica Nucleare, Sezione di Padova-Trento, I-35131 Padova, Italy} 
\affiliation{LPNHE, Universite Pierre et Marie Curie/IN2P3-CNRS, UMR7585, Paris, F-75252 France} 
\affiliation{University of Pennsylvania, Philadelphia, Pennsylvania 19104} 
\affiliation{Istituto Nazionale di Fisica Nucleare Pisa, Universities of Pisa, Siena and Scuola Normale Superiore, I-56127 Pisa, Italy} 
\affiliation{University of Pittsburgh, Pittsburgh, Pennsylvania 15260} 
\affiliation{Purdue University, West Lafayette, Indiana 47907} 
\affiliation{University of Rochester, Rochester, New York 14627} 
\affiliation{The Rockefeller University, New York, New York 10021} 
\affiliation{Istituto Nazionale di Fisica Nucleare, Sezione di Roma 1, University of Rome ``La Sapienza," I-00185 Roma, Italy} 
\affiliation{Rutgers University, Piscataway, New Jersey 08855} 
\affiliation{Texas A\&M University, College Station, Texas 77843} 
\affiliation{Istituto Nazionale di Fisica Nucleare, University of Trieste/\ Udine, Italy} 
\affiliation{University of Tsukuba, Tsukuba, Ibaraki 305, Japan} 
\affiliation{Tufts University, Medford, Massachusetts 02155} 
\affiliation{Waseda University, Tokyo 169, Japan} 
\affiliation{Wayne State University, Detroit, Michigan  48201} 
\affiliation{University of Wisconsin, Madison, Wisconsin 53706} 
\affiliation{Yale University, New Haven, Connecticut 06520} 
\author{A.~Abulencia}
\affiliation{University of Illinois, Urbana, Illinois 61801}
\author{D.~Acosta}
\affiliation{University of Florida, Gainesville, Florida  32611}
\author{J.~Adelman}
\affiliation{Enrico Fermi Institute, University of Chicago, Chicago, Illinois 60637}
\author{T.~Affolder}
\affiliation{University of California, Santa Barbara, Santa Barbara, California 93106}
\author{T.~Akimoto}
\affiliation{University of Tsukuba, Tsukuba, Ibaraki 305, Japan}
\author{M.G.~Albrow}
\affiliation{Fermi National Accelerator Laboratory, Batavia, Illinois 60510}
\author{D.~Ambrose}
\affiliation{Fermi National Accelerator Laboratory, Batavia, Illinois 60510}
\author{S.~Amerio}
\affiliation{University of Padova, Istituto Nazionale di Fisica Nucleare, Sezione di Padova-Trento, I-35131 Padova, Italy}
\author{D.~Amidei}
\affiliation{University of Michigan, Ann Arbor, Michigan 48109}
\author{A.~Anastassov}
\affiliation{Rutgers University, Piscataway, New Jersey 08855}
% ...
\collaboration{CDF Collaboration}
\noaffiliation

\date{\today}

\begin{abstract}
We present a measurement of the $t\bar t$ production cross section using events
with one charged lepton and jets from $p\bar p$ collisions at a center-of-mass 
energy of 1.96~$\Tev$. A $b$-tagging algorithm based on the probability of 
displaced tracks coming from the event interaction vertex is 
applied to identify $b$ quarks from top decay. 
Using 318~$\ipb$ of data collected with the {\CDFII}  
detector, we measure the $t\bar t$ production cross section in events with at 
least one restrictive (tight) $b$-tagged jet and obtain
$8.9^{+1.0}_{-1.0}$(stat.)$^{+1.1}_{-1.0}$(syst.)~\pb. 
The cross section value a\-ssu\-mes a top quark mass of $m_{t}=178~\Gevcc$ in the acceptance 
corrections. The dependence of the cross section on $m_t$ is presented in the 
paper.
This result is consistent with other \CDF\ measurements of the $t\bar t$ cross section
using different samples and analysis techniques, and has similar systematic uncertainties. 
We have also performed consistency checks by using
the $b$-tagging probability function to vary the signal to background
ratio and also using events that have at least two $b$-tagged jets.
\end{abstract}

\maketitle

\section{\label{sec:introd}$\mathbf {Introduction}$}
The top quark is the most massive fundamental particle observed so far, 
and the study of its properties is interesting for several reasons ranging
from its possible special role in electroweak symmetry breaking to its 
sensitivity to physics beyond the standard model (SM).
In particular, the measurement of the top quark pair production cross
section $\sigma_{t\bar{t}}$ is of interest as a test of {\QCD}
predictions.  Recent {\QCD}
calculations done with perturbation theory to next-to-leading order
predict $\sigma_{t\bar{t}}$ with an uncertainty of less than
15\%~\cite{bib:ref1, bib:ref11}, which motivate measurements of 
comparable precision.

Top quark pairs in the SM are produced via either
quark-antiquark annihilation or gluon-gluon fusion in hadron colliders. At the 
Fermilab Tevatron collider, with a center-of-mass energy of 1.96~$\Tev$ in 
$p\bar p$ collisions, about 85\% of the total top pair production comes from
quark-antiquark annihilation. At this center-of-mass energy, the calculated 
cross section, for the combined Tevatron Run I top mass of 
178~$\Gevcc$~\cite{bib:runimass}, is 6.1$_{-0.8}^{+0.6}~\pb$~\cite{bib:ref1} and
decreases by approximately 0.2~$\pb$ for each increase of 1$~\Gevcc$ in the
value of the top mass over the range $170 ~\Gevcc < m_{t} < 190
~\Gevcc$. 
The standard model top quark decays
to a $W$ boson and a $b$ quark almost 100\% of the time, resulting in a
final state from $t\bar{t}$ production of two $W$ bosons and two $b$
jets from $b$ quark fragmentation. When one $W$ decays
leptonically and the other $W$ decays to quarks, the $t\bar{t}$ event 
typically contains a high momentum charged
lepton, an undetected neutrino and four
high transverse momentum jets, two of which originate from $b$
quarks. The undetected neutrino results in an imbalance of
the transverse energy of the event, labeled
as ``missing $E_T$'' ($\met$). This decay mode is called  ``lepton+jets''.

In this paper, we report a measurement of the cross section for pair production
of top quarks in the lepton+jets channel in 318~pb$^{-1}$ of $p\bar{p}$
collision data at $\sqrt{s}$ = 1.96~$\Tev$.
The data were recorded between March 2002 and August 2004, during Run II of 
the Tevatron, by the {\CDFII} detector, a general purpose detector
which combines charged particle tra\-ckers, sampling calorimeters, and
muon detectors. Processes in which a $W$ boson is produced in association with 
several jets with large transverse momentum can be misidentified as 
$t\bar{t}$, since they have the same signature. In order to separate the
$t\bar{t}$ events from this background, we develop a method to tag
$b$-jets based on tracking information from the silicon detector. 
The main event selection requires at least one tight (more restrictive) $b$
tag in the event. As a cross check, we also measure the cross section u\-sing
events with a loose (less restrictive) $b$ tag and events which have at
least two tight or at least two loose $b$ tags. 
Background contributions from heavy flavor production processes,
such as $Wb\bar{b}$, $Wc\bar{c}$ or $Wc$, misidentified $W$ bosons,
electroweak processes, single top production, and mistagged jets
are estimated u\-sing a
combination of Monte Carlo calculations and independent measurements
in control data samples. An excess over background in the number of events 
that contain a lepton, missing energy and three or more jets with at
least one $b$-tag is assumed to be a signal of $t\bar{t}$ production and is 
used to measure the production cross section $\sigma_{t\bar{t}}$.

Previous measurements~\cite{bib:ref2} at $\sqrt{s}=1.8$~TeV gave a
production cross section consistent with the standard
model prediction. 
Recent {\CDF} measurements at $\sqrt{s}=1.96$~TeV are reported in Refs.~\cite{bib:dilepton,bib:bkin,bib:secvtx,bib:slt,bib:kin} and use 
different techniques and top decay channels. The measurement described here 
analyzes more data than the above, and uses a jet probability $b$-tagging algorithm.
A feature of this algorithm is that $b$-tagging is based on a continuous probability
function rather than on a discrete object such as a secondary vertex. Potentially,
this tagger can also be used to statistically separate $b$ and $c$ heavy flavor
contributions.

The organization of this paper is as follows. Section~\ref{sec:cdf}
reviews the detector systems relevant to this analysis. In
Section~\ref{sec:data}, we describe the data sample and event reconstruction. 
The $b$-tagging algorithm and its efficiency and misidentification (``fake") rate 
are discussed in Section~\ref{sec:jp}. 
Section~\ref{sec:evt_sel} describes the event selection. The estimate
of the different backgrounds is presented in Section~\ref{sec:bkg}. 
The $t\bar{t}$ event acceptance and  
tagging efficiency are derived in Section~\ref{sec:acc}. 
The $t\bar{t}$ production cross section measurements in single and
double tagged events are reported in Sections~\ref{sec:single_jp},
and~\ref{sec:double_jp}, respectively. Finally,
the conclusions are presented in Section~\ref{sec:concl}.

\section{\label{sec:cdf}The CDF II Detector}
The {\CDFII} detector uses a cylindrical coordinate system with
the $z$ coordinate along the proton direction, the azimuthal angle
$\phi$, and the polar angle $\theta$ usually expressed in terms of the
pseudo-rapidity $\eta$ = -ln[tan($\theta$/2)]. The rectangular
coordinates $x$ and $y$ point radially outward and vertically upward
from the Tevatron ring, respectively. The detector has been described in detail
elsewhere~\cite{bib:cdf_det}. In this section, we give a brief description of 
the parts relevant for the analysis. 
 
Tracking systems are essential to trigger on and identify high momentum 
charged particles such as electrons and muons. The charged particle tracking 
detectors are contained in a superconducting solenoid which generates a 
magnetic field of 1.4~$\Tesla$, oriented parallel to the proton beam direction.
The Central Outer Tracker ({{\COT})~\cite{bib:cdf_cot} is a 3.1~$\m$ long open 
cell drift chamber which performs up to 96 track position measurements in the 
region between 
0.40~$\m$ and 1.37~$\m$ from the beam axis. Sense wires are arranged in 8 
alternating axial and $\pm$ 2$^\circ$ stereo superlayers with 12 wires each. The 
position resolution of a single drift time measurement is approximately 
140~$\microns$. For high momentum tracks, the {\COT} transverse momentum 
resolution is $\sigma_{p_T}/p^{2}_{T}\, \approx \, 0.0017~\Gevc$. 

Inside the inner radius of the {\COT}, a five layer doubled-sided silicon 
microstrip detector ({\SVX})~\cite{bib:cdf_svx2} covers the region between 
2.5~$\cm$ to 11~$\cm$ from the beam axis. Three separate {\SVX} 
barrel modules along the beamline cover a length of 96~$\cm$, approximately 
90\% of the luminous beam intersection region. Three of the five layers 
combine an $r$-$\phi$ measurement on one side and a 90$^\circ$ stereo 
measurement on the other, and the remaining two layers combine $r$-$\phi$ with 
a small stereo angle of $\pm$1.2$^\circ$. Silicon microstrips have a pitch
of 60 to 65~$\microns$ depending on the layer. 
Three additional Intermediate Silicon Layers 
({\ISL})~\cite{bib:isl} at radii between 19~$\cm$ and 30~$\cm$ in the 
central region link tracks in the {\COT} to hits in {\SVX}.

Electromagnetic ({\EM}) and hadronic ({\HAD}) 
sampling calorimeters~\cite{bib:cem,bib:pem,bib:chad} surround the tracking 
system and measure the energy flow of interacting particles in the 
pseudo-rapidity range $|\eta| <$~3.64. The {\EM} and {\HAD}
calorimeters are lead-scintillator and iron-scintillator sampling devices, 
respectively. They are segmented into projective towers, each one covering a 
small range in pseudo-rapidity and azimuth. Most towers cover 15 degrees in 
$\phi$ and 0.10 to 0.13 units in pseudo-rapidity. 
Proportional chambers ({\CES}) measure the transverse profile of {\EM} showers 
at a depth corresponding to the shower maximum for electrons.
Electrons are reconstructed in the central electromagnetic calorimeter 
({\CEM}) with a transverse energy precision 
$\sigma(E_T)/E_{T} = 13.5\% / \sqrt{E_{T}/ \Gev}\oplus 2\%$~\cite{bib:cem}. 
Jets are identified as a group of electromagnetic and hadronic calorimeter 
towers with an energy resolution of approximately 0.1$\cdot E_{T}$ + 1.0~$\Gev$~\cite{bib:jet_e_res}.

The muon system is located outside of the calorimeters. Four layers of
planar drift chambers ({\CMU})~\cite{bib:cmu} detect muons
with $p_T > 1.4~\Gevc$ that penetrate the five absorption lengths of
calorimeter steel in the central region of $|\eta| <$~0.6. An additional four 
layers of planar drift chambers
({\CMP})~\cite{bib:cmp} located behind 0.6~$\m$ of steel outside the 
magnet return yoke detect muons with $p_T > 2.0~\Gevc$. When a track is linked 
to both {\CMU} and {\CMP}, it is called a {\CMUP} 
muon. The Central Muon Extension detector ({\CMX}),
arranged in a conical geometry, provides muon detection in the region
0.6~$<\,|\eta|\,<$~1.0 with four to eight layers of drift chambers,
depending on the polar angle. All the muon chambers measure the
azimuthal coordinates of hits via a drift time
measurement. The {\CMU} and {\CMX} also measure the longitudinal coordinate, $z$.

The beam luminosity is determined by using gas Cherenkov 
counters~\cite{bib:lumi} located in the region 3.7~$ <|\eta| <$~4.7 which 
measure the average number of inelastic $p\bar{p}$  collisions per bunch
crossing. The total uncertainty on the luminosity is
5.9\%, where 4.4\% comes from the acceptance and operation of the
luminosity monitor and 4.0\% from the calculation of the inelastic
$p\bar{p}$ cross section~\cite{bib:lumi_sys}.

\section{\label{sec:data}Data Sample and Event Reconstruction}
The data used in this analysis are from $p\bar{p}$ collisions at a
center-of-mass energy of $\sqrt{s}\, =$ 1.96~$\Tev$ recorded by the 
{\CDFII} detector between March 2002 and September 2004. 
The data sample has been collected by triggers based on the selection of 
a high transverse momentum lepton (electron or muon). The total integrated
luminosity is 318~$\pb^{-1}$ for {\CEM} electron and {\CMUP} muon
candidates, and 305~$\pb^{-1}$ for {\CMX} muon candidates.
Briefly, we discuss the trigger and lepton identification requirements, the 
reconstruction of jets, and the missing transverse energy, $\met$. 

{\CDF} has a three-level trigger system to filter events from a 2.5~$\mhz$
beam crossing rate down to 60~$\hz$ for permanent storage. The first two
levels of triggers are special purpose hardware and the third consists
of a farm of computers. 

The first trigger level ({\LUNO}) reconstructs charged particle tracks in
the {\COT} $r$-$\phi$ projection using a hardware track processor
called the Extremely Fast Tracker ({\XFT})~\cite{bib:xft}. The {\LUNO}
electron trigger requires a {\XFT} track with $p_{T}\,>$ 8~$\Gevc$ 
matched
to an {\EM} calorimeter tower with $E_T\,>$ 8~$\Gev$ and with a ratio of
hadronic-to-electromagnetic energy less than 0.125. The {\LUNO} muon
trigger requires an {\XFT} track with $p_{T}\,>$ 4~$\Gevc$ matched to 
a muon
track segment with $p_{T}\,>$ 6~$\Gevc$ from the {\CMU} and {\CMP} chambers
or a track with $p_{T}\,>8$~$\Gevc$ matched to a muon track segment with
$p_{T}\,>$ 6~$\Gevc$ in the {\CMX} chambers.

The second level ({\LDUE}) electron trigger requires the {\XFT} track 
found at {\LUNO} to be matched to a cluster of energy in the central {\EM} 
calorimeter with $E_{T}\,>$ 16~$\Gev$. The cluster adds the energy of the 
neighboring trigger towers with $E_{T}\,>$ 7.5~$\Gev$ to the original {\LUNO} 
trigger tower. A trigger tower consists of two calorimeter towers.
The {\LDUE} muon trigger accepts events passing {\LUNO}.

The third trigger level ({\LTRE}) is a farm of Linux computers which perform
on-line event reconstruction, including 3D charged particle
reconstruction. The {\LTRE} electron trigger requires a track with
$p_{T}\,>$ 9~$\Gevc$ matched to an energy cluster of three adjacent towers
in pseudo-rapidity in the central {\EM} calorimeter with $E_{T}\,>$ 18~$\Gev$, consistent with the shower profile expected from test beam electrons. 
The {\LTRE} muon trigger requires a track
with $p_{T}\,>$ 18~$\Gevc$ matched to a track segment in the muon
chambers within 10~$\cm$ in the $r-\phi$ view and, for {\CMU} and {\CMX} muons
only, within 20~$\cm$ in the $z$ view. The efficiency of these triggers is 
measured using $W^{\pm} \rightarrow e^{\pm}\nu$ and $Z\rightarrow \mu^+\mu^-$ 
data (the method is described in Ref.~\cite{bib:eff_tri}) and is found to be (96.2 $\pm$ 0.6)\% for 
{\CEM} electrons, and (90.8 $\pm$ 0.5)\% and (96.5 $\pm$ 0.4)\% 
for {\CMUP} and {\CMX} muons respectively, for electrons and muons passing 
through the fiducial volume of these detectors.
    
\subsection{Track and Primary Vertex Reconstruction}\label{s:trk_q}
The trajectories of charged particles are found (in a first
approximation) as a series of segments in the axial superlayers of the
{\COT}. Two complementary algorithms associate the segments lying on a
co\-mmon circle to define an axial track. Segments in the stereo layers are 
associated with the axial tracks to reconstruct 3D tracks. For muons and 
electrons used in this analysis, {\COT} tracks are required to have at least 3
axial and 2 stereo segments with at least 5 hits per superlayer.
The efficiency for finding isolated high momentum {\COT} tracks in the {\COT}
fiducial volume with
$p_{T}\,>$ 10~$\Gevc$ is measured using electrons from 
$W^{\pm}\rightarrow e^{\pm}\nu$ events and is found to be (98.3 $\pm$ 0.1)\%. 
Silicon hit information is added to reconstructed {\COT} tracks using 
an ``outside-in'' tra\-cking algorithm. The {\COT} tracks 
are extrapolated to the silicon detector and the track is refit using the
information from the silicon measurements. The initial track parameters
provide a width for a search region in a given layer. For each
candidate hit in that layer, the track is refit and used to define the
search region into the next layer. The search uses the two best
candidate hits in each layer to generate a small tree of final track
candidates, and the one with the best fit $\chi^{2}$ is selected. 
The efficiency to associate at least three silicon hits with an isolated
{\COT} track is found to be (91 $\pm$ 1)\%.

The primary vertex location for a given event is found by fitting
well-measured tracks to a common point of origin. 
At high luminosities, more than one collision can occur on a given
bunch crossing. For a luminosity of $\sim$10$^{32}~\cm^{-2}$s$^{-1}$, there are $\sim$2.3 interactions per bunch crossing. The luminous region is long, with $\sigma_z$ = 29~$\cm$; 
therefore the primary vertices of each collision are  typically
separate in $z$. The first estimate of the primary
vertices ($x_V$, $y_{V}$, $z_{V}$) is binned in the $z$ coordinate,
and the $z$ position of each vertex is then
calculated from the weighted average of the $z$ coordinate of all
tracks within 1~$\cm$ of the first iteration vertex, with a typical
resolution of 100~$\mu$m. The primary vertex is determined event by event
by an iterative algorithm which uses tracks around a seed vertex, defined as above, 
to form a new vertex. The $\chi^2$ for all tracks relative to the 
new vertex is calculated, tracks with bad $\chi^2$ are removed, and the cycle 
is repeated until all tracks have a good $\chi^2$. 
The locus of all primary vertices defines the beamline, the position of the
luminous region of the beam-beam collisions through the detector. A
linear fit to ($x_V$, $y_V$) vs. $z_V$ yields the beamline for each stable
ru\-nning period. The beamline  is used as a constraint to refine the knowledge 
of the primary vertex in a given event. The transverse beam
cross section is circular, with a rms width of $\approx\,30~\microns$ at $z$
= 0, rising to $\approx$ 50 - 60~$\microns$ at $|z|$ = 40~$\cm$. The beam is
not necessarily parallel nor centered in the detector.

\subsection{Electron Identification}\label{s:ele_id}
Electron reconstruction begins with a track with
$p_T>9~\Gevc$ that extrapolates to a cluster of three {\CEM} towers adjacent in
pseudo-rapidity with a total $E_T>20~\Gev$. Several cuts are successively applied 
in order to improve the purity of the electron selection, as summarized in
Table~\ref{t:elecut}. Electron candidates passing these requirements are
called tight electrons.

\begin{table*}[!htb]
\caption{Selection requirements for tight electrons.}
\label{t:elecut}
\begin{center}
\begin{tabular}{l@{\hspace{1cm}}c}
\hline
\hline
Electron Variable & Cut \\
\hline
 $p_T$   &   $\geq 10~\Gevc$ \\
 $E_T$   &   $\geq 20~\Gev$ \\
 {\COT} Axial  Segments   &   $\geq 3$  \\
 {\COT} Stereo Segments   &   $\geq 2$  \\
 Hits for Each {\COT} Segment   &   $\geq 5$  \\
 $E_{HAD}/E_{EM}$   &   $\leq 0.055+0.00045\times E$  \\
 $E/p$   &   $\leq 2.0$ unless $p_T\geq 50~\Gevc$ \\
 Isolation   &   $\leq 0.1$  \\
 $L_{shr}$   &   $\leq 0.2$  \\
 {\CES} $|\Delta z|$   &   $\leq 3.0~\cm$  \\
 {\CES} $Q\times |\Delta x|$   &   $-3.0\leq Q\times |\Delta x|\leq 1.5~\cm$   \\
 {\CES} $\chi^2_{strip}$  &   $\leq 10$  \\
 Photon Conversions   &   Veto if $D\leq 0.2~\cm$ and $\Delta \cot(\theta )<0.04$  \\
\hline
\hline
\end{tabular}
\end{center}
\end{table*}

The ratios between the hadronic and the electromagnetic 
cluster energies $E_{HAD}/E_{EM}$ and between the cluster energy and the track 
momentum $E/p$ are required to be consistent with an electron's energy 
deposition in the calorimeters. The cluster is further required 
to be isolated, the isolation $I$ being defined as the ratio of the addi\-tio\-nal
transverse energy in a cone of radius 
$R=\sqrt{(\Delta\phi)^2+(\Delta\eta)^2}=0.4$
around the cluster to the transverse energy of the cluster itself.

The position of the electromagnetic shower measured by the {\CES} detector is 
used
to define matching requirements between the extrapolated track and the cluster
in the {\CES} $x$ and $z$ local coordinates. In particular, a charge dependent 
cut in the 
$x$ position is applied to take into account the different flow of
energy deposited by bremsstrahlung photons emitted by an electron or a 
positron. In addition, the {\CES} provides electron identification 
through the observed shower shape. The {\CES} shower shape is fitted in the z 
view to the distribution expected for an electron, and the chisquare 
probability for the fit, $\chi^2_{strip}$, is used as a cut on the shower profile. 
Finally, the sharing of energy between adjacent calorimeter towers is 
quantified by the lateral shower profile $L_{shr}$, which 
measures how close the energy distribution in the {\CEM} towers adjacent to 
the cluster seed is to the electron hypothesis.

Electrons from photon conversions throughout the detector material are 
vetoed by rejecting electron candidates if an oppositely charged track with
a small distance of closest approach ($D$) is found. This analysis is 
sensitive to any loss in efficiency from the misidentification of an electron 
from the $W$ boson decay as a photon conversion. Therefore, in order to avoid loss of efficiency, 
the veto is not applied to 
events consistent with electrons radiating a photon that subsequently converts. 
The performance of this algorithm to identify electrons from photon conversions 
is estimated to be (72.6 $\pm$ 0.1)\%~\cite{bib:kin}, where the uncertainty 
covers both statistical and systematic.

The efficiency of the electron selection on $t\bar t$ events is determined 
by means of Monte Carlo simulation. Studies of $Z\rightarrow e^+e^-$ processes
show that a data to Monte Carlo simulation 
scale factor of (99.6 $^{+0.4}_{-0.5}$)\% is needed 
to correct the simulation predictions for the efficiency for {\CEM} electron 
identification.

Other electron categories are defined. Candidate electrons passing all the 
above requirements except for the isolation cut are called loose electrons. 
Tracks matched to an energy deposit in the plug calorimeter ($1.2<|\eta|<2.0$) 
are called plug electrons. 

\subsection{Muon Identification}\label{s:muon_id}

Muon identification starts by requiring an isolated, high momentum {\COT} 
track that extrapolates to a track segment in the muon chambers. Several 
additional 
requirements are imposed in order to minimize contamination from hadrons 
punching through the calorimeter, decays in flight of charged hadrons 
and cosmic rays. 
Table~\ref{t:muocut} lists the selection requirements for candidate muons.
Muon candidates passing these cuts are called tight muons.

\begin{table*}[!htb]
\caption{Selection requirements for tight muons.}
\label{t:muocut}
\begin{center}
\begin{tabular}{l@{\hspace{1.4cm}}c}
\hline
\hline
Muon Variable & Cut \\
\hline
 $p_T$       &   $\geq 20~\Gevc$  \\
 {\COT} Axial  Segments   &   $\geq 3$  \\
 {\COT} Stereo Segments   &   $\geq 2$  \\
 Hits for Each {\COT} Segment   &   $\geq 5$  \\
 $|d_0|$   &   $\leq 0.2~\cm$ if no silicon hits\\
 $|d_0|$   &   $\leq 0.02~\cm$ if silicon hits\\
 $E_{HAD}$   & $\leq$  max$(6, 6+0.0280(p-100))~\Gev$ \\
 $E_{EM}$    & $\leq$  max$(2, 2+0.0115(p-100))~\Gev$ \\
 {\CMU} $|\Delta x|$   &   $\leq 3.0~\cm$ \\
 {\CMP} $|\Delta x|$   &   $\leq 5.0~\cm$ \\
 {\CMX} $|\Delta x|$   &   $\leq 6.0~\cm$ \\
 Isolation   &   $\leq 0.1$  \\
 Cosmic Rays   &   Veto  \\
\hline
\hline
\end{tabular}
\end{center}
\end{table*}

The {\COT} track must have $p_T \geq 20~\Gevc$, and at least 3 axial and 2 
stereo segments with a minimum of 5 hits per segment. The distance of closest 
a\-ppro\-ach of the track to the beamline in the transverse plane, $d_0$, must be 
small in order to select prompt muons (coming from the interaction primary 
vertex) and reject cosmics and in-flight decays. 
The energy deposition in the {\EM} and {\HAD} calorimeters, $E_{\rm EM}$ and 
$E_{\rm HAD}$, must be small as expected for the pa\-ssa\-ge of a minimum ionizing 
particle. 
The distance between the extrapolated {\COT} track and the track segment in 
the muon chambers, $\Delta x$, must be small in order to ensure a good match. 
If a track is matched to a {\CMU} segment, a matching {\CMP} segment is also 
required, and vice versa. 
Isolation is defined as the ratio between any additional transverse energy in 
a cone of radius $R = 0.4$ around the track direction and the muon $p_T$, and 
it is required to be smaller than 0.1. Cosmic rays are efficiently identified 
and rejected
through their asynchronous track timing re\-la\-tive to the beam crossing time and 
their incoming and outgoing back-to-back track topology.

Studies of $Z\rightarrow \mu^+\mu^-$ processes show that a data to Monte Carlo 
simulation scale factor of (87.4 $\pm$ 0.9)\% ((98.9 $\pm$ 0.6)\%) is needed to 
correct the simulation predictions for the {\CMUP} ({\CMX}) muon 
identification efficiency.

As for the electrons, candidate muons passing all the cuts except the 
isolation cut are called loose muons. A track matched to a {\CMU} or a {\CMP} 
segment only, which passes all the other cuts including isolation, is also 
accepted as a loose muon.

\subsection{Jet Reconstruction and Corrections}\label{s:jet_reco}
The jets used in this analysis are reconstructed from calorimeter
towers using a cone algorithm~\cite{bib:jet_rec} with a radius $R \leq$
0.4, for which the $E_T$ of each tower is calculated with respect to the
$z$ coordinate of the event. The calorimeter towers belonging to any
electron candidate are not used by the jet clustering algorithm. The
energy of the jets is corrected~\cite{bib:jet_e_corr} for the pseudo-rapidity dependence of the calorimeter response, the 
calorimeter time dependence, and extra $E_T$ from any multiple interactions.

By definition, tight jets have corrected $E_{T} \geq$ 15~$\Gev$ and 
detector $|\eta| <$~2.0, whereas loose jets have corrected $E_{T} \geq$ 8~$\Gev$ and detector $|\eta| <$~2.0. Detector $\eta$ is the pseudo-rapidity of
the jet calculated with respect to the center of the detector.

\subsection{Missing Transverse Energy Reconstruction}\label{s:met_reco}
The presence of neutrinos in an event is inferred by an imbalance of
transverse energy in the detector. The missing transverse energy, $\met$, is
defined as the magnitude of $-\sum_i
[E_{T,i}cos(\phi_{i}),E_{T,i}sin(\phi_{i})]$, where $E_{T,i}$ is the
transverse energy of the calorimeter tower $i$ calculated with respect
to the $z$ coordinate of the event, $\phi_{i}$ is its azimuthal angle,
and the sum is over all calorimeter towers. The 
$\met$ is corrected by subtracting the transverse momentum of the
muon track and adding back the transverse energy in the
calorimeter towers traversed by the muon. Because the $\met$ calculation uses all
calorimeter towers, the $\met$ vector is adjusted for the effect of the jet corrections
for all jets with $E_{T}\,\geq$ 8~$\Gev$ and $|\eta|\,<$~2.5. 
 
\subsection{Monte Carlo Samples and Detector Simulation}\label{s:mcsamples}
The understanding of acceptances, efficiencies and backgrounds relies on 
detailed simulation of physics processes and detector response.

The detector acceptance for $t\bar{t}$ events is modeled using 
{\PYTHIAv}~\cite{bib:pythia} and {\HERWIGv}~\cite{bib:herwig}. This analysis
uses the former for the final cross section estimate and the latter to estimate
the systematics due to differences in the modeling of $t\bar{t}$
production and decay.  
These generators employ leading order matrix elements for the hard parton 
scattering, followed by parton showering to simulate gluon radiation and 
fragmentation effects. The generators are used with the {\CTEQcL} parton 
distribution functions~\cite{bib:cteq5l}. Decays of $b$ and $c$ hadrons are 
modeled using {\QQv}~\cite{bib:qqv9.1}. Estimates of backgrounds from 
$W$ bosons produced in association with jets are derived using the 
{\ALPGEN} generator~\cite{bib:alpgen} with parton showering provided by 
{\HERWIG}. The background from electroweak processes and single top 
production is studied using {\PYTHIA}.

The {\CDFII} detector simulation reproduces the response of the 
detector and uses the same detector geometry database as the event 
reconstruction.
Particle interactions through matter are performed with 
{\GEANTt}~\cite{bib:geant3}. Charge deposition in the silicon detectors is 
calculated using a simple geometrical model based on the path length of the 
ionizing particle and an unrestricted Landau distribution. The drift model for 
the {\COT} uses a parametrization of a {\GARFIELD}~\cite{bib:garfield} 
simulation with
parameters tuned to match {\COT} collider data~\cite{bib:cdf_cot}. The
calorimeter simulation uses the {\GFLASH}~\cite{bib:gflash} parametrization
package interfaced with {\GEANTt}. The {\GFLASH} parameters are tuned 
to test beam data for electrons and high-$p_{T}$ pions and they are checked by 
comparing the calorimeter energy of isolated tracks in collision data to their 
momentum as measured in the {\COT}. More details on the 
{\CDFII} simulation can be found in Ref.~\cite{bib:cdf_sim}.

\section{\label{sec:jp} \boldmath Jet Probability $b$-Tagging Algorithm}
The {\JP} $b$-tagging algorithm~\cite{bib:jet_probab} 
is used to determine whether a jet has been produced from the hadronization 
process of a light parton or a heavy 
parton. The latter result in long-lived hadrons whose decay gives rise to tracks displaced from the primary interaction vertex. This algorithm uses  
tracks a\-sso\-cia\-ted with a jet to determine 
the probability for these to come from the primary vertex of the interaction.
The calculation of the probability is based on the impact 
parameters ($d_0$) of the tracks in the jet and their uncertainties.
The impact parameter is assigned a positive or negative sign depending on 
the position of the track's point of closest approach to the primary
vertex with respect to the jet direction, as shown in Fig.~\ref{f:signIP1}. 
By construction, the probability for tracks originating from the primary vertex is 
uniformly distributed from 0 to 1.
For a jet coming from heavy flavor hadronization, the distribution peaks at 0, 
due to tracks from long lived particles that have a large impact parameter 
with respect to the primary vertex.  
\begin{figure}[!htb]
\begin{center}
    \includegraphics[width=8.0cm,clip=]{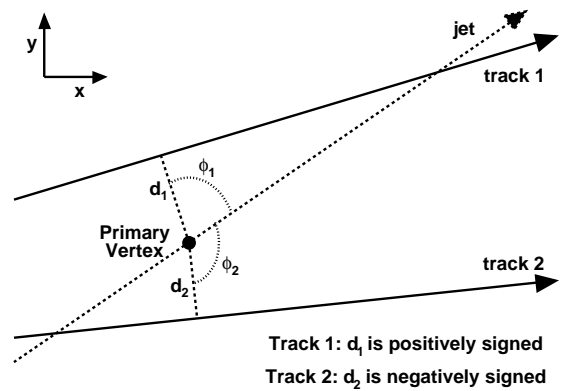}
\caption{\label{f:signIP1} The sign of the impact parameter of a track. The 
impact parameter is positive (negative) if the angle $\phi$ between the jet 
axis and the line connecting the primary vertex and the track's point of 
closest approach to the primary vertex itself is smaller (bigger) than 
$\pi/2$.} 
\end{center}
\end{figure}

The particles in a jet coming from a light parton originate at the primary 
vertex, but these tracks are reconstructed with a non-zero impact parameter 
due to the finite tracking resolution. They have an equal probability of being 
positively or negatively signed.
Jets which originate from a heavy parton 
contain long lived hadrons giving rise to tracks displaced in the 
jet direction, which preferentially populate the positive side of the
signed impact parameter distribution.
The width of the negative impact parameter distribution is solely 
due to the tracking detector resolution, beam spot size, and multiple 
scattering.

The tracking resolution can be extracted from the data by fitting the negative
side of the signed impact parameter distribution of tracks from
prompt jets, which are the
dominant component of inclusive jet data. Tracks are divided into 72 different
categories according to the number and quality of {\SVX} hits, detector 
$\eta$ and $p_T$. To minimize the contribution from
badly measured tracks with a large reconstructed impact parameter, the
signed impact parameter significance, $S_{d_{0}}$ (ratio of the impact
parameter to its uncertainty), is parameterized for each track category. 
Tracks are fitted to a helix, and the impact pa\-ra\-me\-ter is corrected for
beam offsets in order to take into account any displacement of the
primary vertex from the nominal position. The uncertainty in the impact
parameter is given by the error propagation of the uncertainties in the
fit and in the beam offset correction. We parameterize the impact parameter significance for tracks satisfying the 
quality criteria listed in Table~\ref{tab:track_qua} that are associated with 
jets with $E_T >7~\Gev$ and $|\eta| <$~2.5. These tracks must 
have $p_T>0.5~\Gevc$,
impact parameter less than 0.1~cm (in order to reject long lived $K$'s and 
$\Lambda$'s), three to five hits on different axial layers of the {\SVX}, at 
least 20 (17) hits in the {\COT} axial (stereo) layers, and the $z$ position of 
the track must be within 5~cm of the event primary vertex.
Tracks passing this selection are called jet probability tracks.
The $|d_{0}|$ is measured with respect to the primary vertex. The event is 
required to have a primary vertex, and the vertex with highest sum of 
transverse 
momentum of all tracks is chosen in events which have more than one vertex.

\begin{table}[!htb]
\begin{center}
\caption{\label{tab:track_qua} Selection criteria for tracks used by the 
{\JP} algorithm.}
\begin{tabular}{l@{\hspace{1.5cm}}c} 
\hline
\hline
Variable & Cut \\
\hline
$p_T$ &$>0.5\, \Gevc$ \\
$|d_{0}|$ & $<$ 0.1 $\cm$\\
$N_{\mbox{{\SVX} axial}}$ & $\geq$ 3 and $\leq$ 5 \\
$N_{\mbox{{\COT} axial}}$ & $\geq$ 20 \\
$N_{\mbox{{\COT} stereo}}$ & $\geq$ 17 \\
$|z_{trk}\, -\, z_{pv}|$& $<$ 5 $\cm$ \\ 
\hline
\hline
\end{tabular} 
\end{center}
\end{table}

Figure~\ref{fig:sigS} shows the distribution of the impact parameter 
significance of tracks in an inclusive jet sample for one of the track 
categories, namely tracks with at least 5 good {\SVX} hits, $p_T>5~\Gevc$ and 
$|\eta|<0.6$. 
The negative side of this distribution is fitted with a function 
$R(S)$ called the resolution function, which is 
\begin{figure}[!htb]
\begin{center}
    \includegraphics[width=8.7cm,clip=]{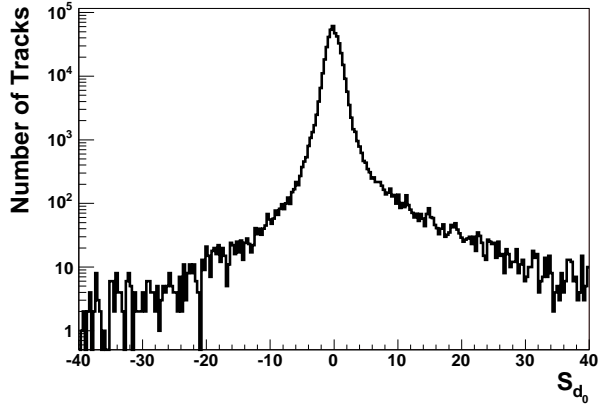}
\caption{\label{fig:sigS} Distribution of the impact parameter 
significance for tracks in an inclusive jet sample with at least 5 good 
{\SVX} hits, $p_T>5~\Gevc$, and $|\eta|<0.6$.}
\end{center}
\end{figure}
used to determine the probability, $P_{tr}(S_{d_0})$, that the impact 
parameter significance of a given track is due to the detector resolution, 
defined as: 
\begin{equation}
  P_{tr}(S_{d_0}) = \frac{\int^{-|S_{d_0}|}_{-\infty} R(S)dS}{\int^{0}_{-\infty}R(S)dS}.
\end{equation}
The $S_{d_0}$ distribution peaks at zero and
falls quickly with increasing absolute value of $|S_{d_0}|$, but the tails 
are rather long. In order to improve the statistics and obtain a better fit
in the tail, we use non-linear bins by transforming it to 
$X=\ln(|S^-_{d_0}|)$, where the minus sign indicates that only the 
negative part
of $S_{d_0}$ is used. Figure~\ref{fig:sigSfit} shows the result of such a 
fit, together with the fit residues defined as (data-fit)/uncertainty, where 
the uncertainty is taken as the statistical uncertainty on each data point.
\begin{figure}[!htb]
\begin{center}
    \includegraphics[width=8.7cm,clip=]{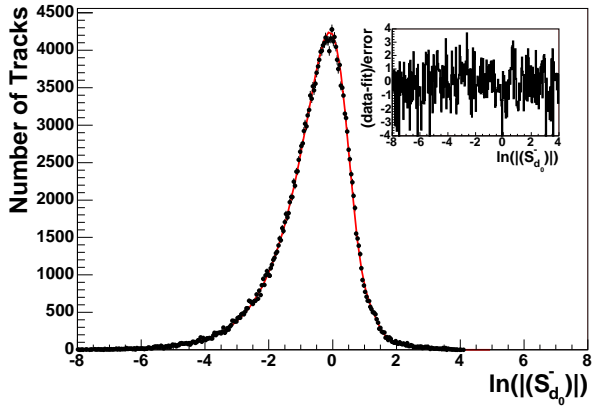}
\caption{\label{fig:sigSfit} Fit to the transformed impact 
parameter significance, $\ln(|S^-_{d_0}|)$, where only the negative side 
of the 
$S_{d_0}$ distribution is used. The resolution function is chosen as the 
convolution of four Gaussians. The inset shows the residues of the fit, 
(data-fit)/uncertainty.}
\end{center}
\end{figure}
A resolution function parameterized as the convolution of four Gaussians with 
means at zero is found to fit well all distributions for all 72 track   
categories:
\begin{equation}
R(S) = \sum_{i=1}^{4}\frac{p_i}{\sqrt{2\pi}\sigma_i}e^{-S^2/2\sigma^2_i}.
\end{equation}
After the transformation to a logarithmic axis, the resolution function becomes
\begin{equation}
R(X) = \sum_{i=1}^{4}\frac{p_i}{\sqrt{2\pi}\sigma_i}e^{(X-\frac{e^{2X}}{2\sigma^2_i})}
\end{equation}
and $R(X)$ is used to fit the transformed 
$X = \ln(|S^-_{d_0}|)$ distribution.

The jet probability $P_J$ that a jet is consistent with a zero lifetime 
hypothesis is defined as
\begin{equation}
   {{P_J}} = \prod \times \sum_{k=0}^{N_{trk}-1}{ \frac{(-ln\prod)^k}{k!}},
\end{equation}
where 
\begin{equation}
   \prod = \prod_{l=1}^{N_{trk}}P_{tr} 
\end{equation}
and $N_{trk}$ is the number of jet probability tracks with positive impact 
parameter.
Jets are required to have at least two jet probability tracks with positive impact parameter to be taggable.
Both of these distributions should be uniformly distributed in the interval [0-1] for jets
having only prompt tracks.
Tracks with negative impact parameter are used to define a ¨negative¨ $P_J$,
which is used to check the algorithm and to estimate the misidentification rate.
We define positive (negative) tagged jets as those jets whose positive 
(negative) $P_J$ is less than a cutoff, where we use 1\% (main result) and 5\% (cross check).
Positive tagged jets are expected to be enriched in heavy flavor.
The 1\% cut was used in previous publications~\cite{bib:jet_probab} and has 
similar performance to the secondary vertex tagger~\cite{bib:secvtx}, while 
the loose (less restrictive) 5\% cut was chosen  near the point where the $P_J$ distribution becomes flat 
(see Fig.~\ref{f:jp_dist}). Further gain in $t\bar{t}$ selection efficiency 
resulting from a looser $P_J$ cut is accompanied by an increase in background from light jets misidentified as
heavy flavor (mistags). For comparison, both the 1\% and 5\% numbers and figures are presented together throughout the paper.

A feature of this algorithm is that the $b$-tagging is performed using a
continuous variable instead of a discrete object like a reconstructed
secondary vertex. It therefore provides a variable that allows one to move 
continuously along the 
efficiency curve and to select the optimal signal to background point for a 
specific analysis. Furthermore, the ability to adjust the $P_J$ cut is a valuable
tool to understand the heavy flavor content of the sample.
Potentially~\cite{bib:jet_probab}, this method can be used to statistically separate
$b$ and $c$ heavy flavor contributions.
This feature is illustrated in the left plot in Fig.~\ref{f:jp_dist_mcdata}, where the jet 
probability distributions for $b$, $c$ and light jets are shown. Monte
Carlo simulated $2\rightarrow 2$ parton events are used as des\-cri\-bed 
in Section~\ref{sec:jp_eff}. 
In the right plot, we show the jet probability distributions observed in two
different data sets of jets. The first sample is enriched in heavy flavor content
by requiring the jets to contain a soft momentum electron; here, events are triggered 
on low $p_T$ inclusive electrons (see Section~\ref{sec:jp_eff}). 
The second set consists of generic QCD jets selected by requiring events with at least one 
jet with $E_T>50~\Gev$ (the Jet50 sample).

\begin{figure*}[!htb]
\begin{center}
    \includegraphics[width=0.49\textwidth,clip=]{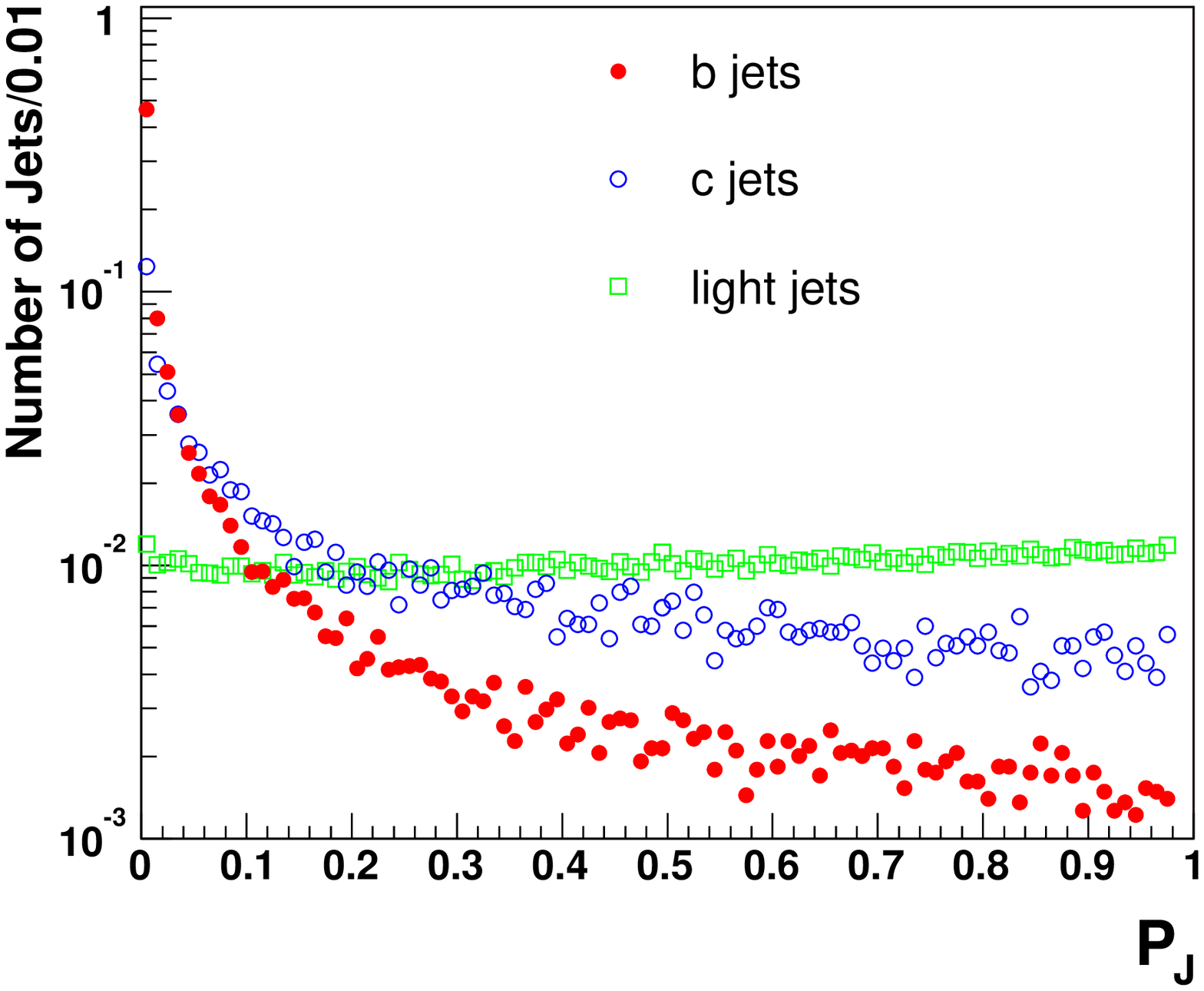}
    \includegraphics[width=0.49\textwidth,clip=]{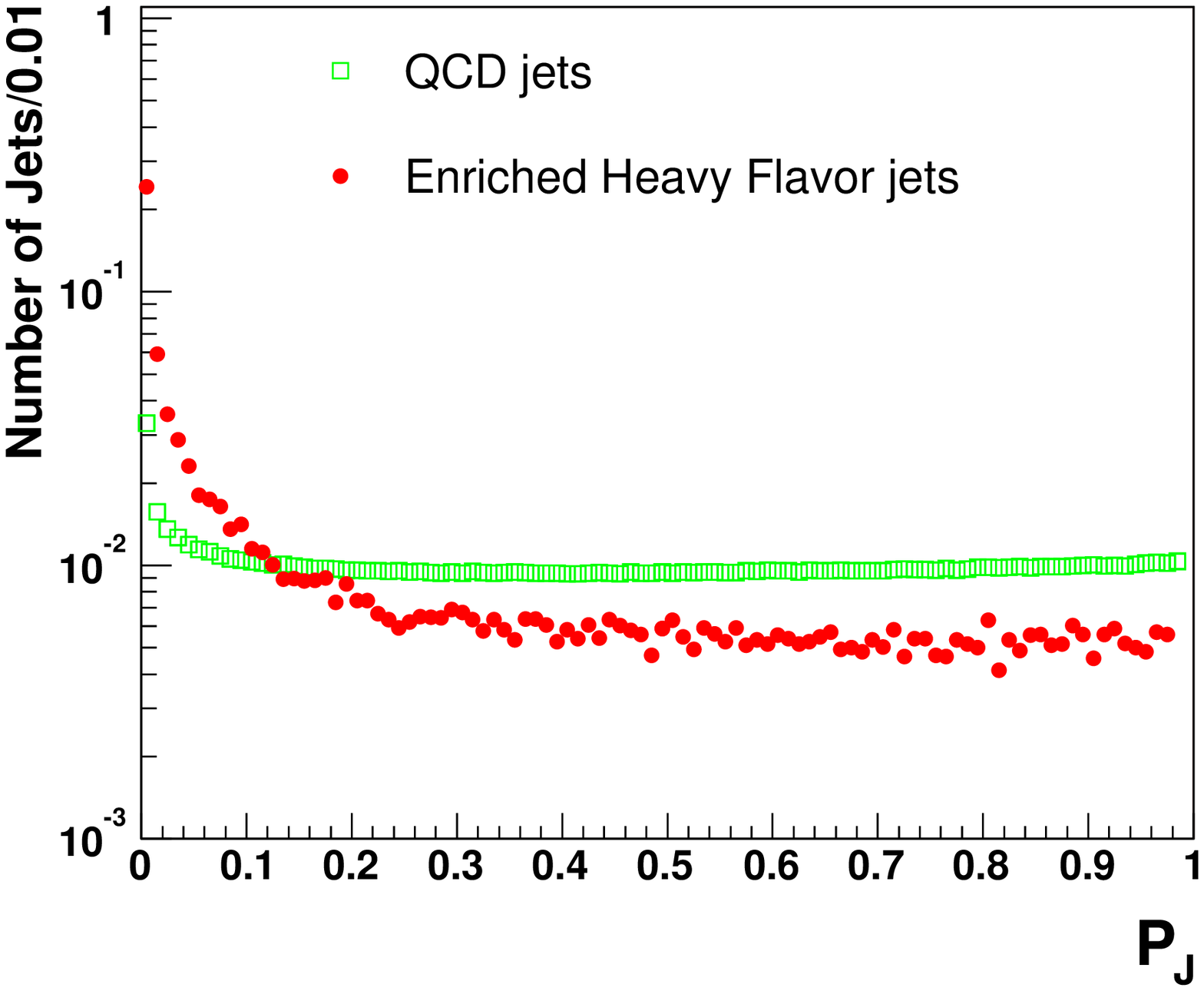}
\caption{\label{f:jp_dist_mcdata} Left: jet probability distributions for jets matched 
to $b$ (full circles), $c$ (empty circles) and light (empty squares) quarks
in Monte Carlo simulated events. Right: jet probability distributions for electron jets 
in inclusive electron data (full circles) and for generic QCD jets in Jet50 data (empty squares).} 
\end{center}
\end{figure*}

In this section we discuss the $b$-tagging algorithm itself, independently of the
other details of this ana\-ly\-sis.

\subsection{\label{sec:jp_eff}Measurement of the Tagging Efficiency
for Heavy Quark Jets}

The method used to measure the {\JP} tagging efficiency for 
heavy flavor jets is des\-cri\-bed in detail in Ref.~\cite{bib:secvtx}. The ideal events
to study this efficiency are $b\bar{b}$ dijet events.
We use a calibration data sample of jets whose heavy flavor fraction can be 
measured: a sample triggered on low $p_T$ inclusive electrons which is 
enriched in semileptonic decays of bottom and charm hadrons. 
The tagging efficiency is also measured for simulated jets by using a 
Monte Carlo sample similar to the calibration sample. We use 
{\HERWIG} to generate 2$\rightarrow$2 parton events, which are passed 
through a filter requiring an electron with $p_T > 7~\Gevc$ and 
$|\eta| < 1.3$. Events passing this filter are processed using the detector 
simulation des\-cri\-bed in Section~\ref{s:mcsamples}.
Electrons are identified using a selection similar to that described in 
Section~\ref{s:ele_id}, except that they are required to be non-isolated and 
have a lower energy threshold ($E_T > 9~\Gev$ and track $p_T > 8~\Gevc$).
The heavy flavor content of the sample is further enhanced by requiring two
jets in the event, an ``electron jet'', presumed to contain the decay products
of a heavy flavor hadron, and an ``away jet''. The electron jet must have 
$E_T > 15~\Gev$ and be within 0.4 of the electron direction in $\eta$-$\phi$ 
space.
The away jet is required to have $E_T > 15~\Gev$ and $|\eta| < 1.5$, and it 
must be approximately back-to-back with the electron jet 
($\Delta\phi_{e-j} > 2$~rad).
If the fraction of electron jets containing heavy flavor for which the away 
jet is tagged ($F_{HF}^a$) is known, and if there were no prompt jets
misidentified as $b$-jets, the efficiency to tag a
heavy flavor jet containing an electron would be given by
\begin{eqnarray}
\epsilon = \frac{N_{a+}^{e+}}{N_{a+}} \cdot \frac{1}{F_{HF}^a}, 
\end{eqnarray}
where $N_{a+}^{e+}$ is the number of events for which both the electron jet 
and the away jet are positively tagged, and $N_{a+}$ is the total number of 
events for which the away jet is positively tagged. Since light jets can be 
tagged as well, we correct for this effect by subtrac\-ting the number of 
negative tags. We define the po\-si\-tive tag excess for events with  a positive 
or negative tag in the away jet as
\begin{eqnarray}
\Delta_{a+} = N_{a+}^{e+} - N_{a+}^{e-}     \\
\Delta_{a-} = N_{a-}^{e+} - N_{a-}^{e-}    
\end{eqnarray}
where, for example, $N_{a+}^{e-}$ is the number of events where the electron 
jet is negatively tagged and the away jet is positively tagged.
The tagging efficiency for heavy flavor jets containing an electron is then 
given by
\begin{equation}
\epsilon\,=\frac{\Delta_{a+} - \Delta_{a-}}{(N_{a+}\,-\,N_{a-})}\cdot
\frac{1}{F_{HF}^{a}}.
\label{eq:epsilon}
\end{equation}
Since events with an electron jet and a tagged away jet are mostly due to heavy flavor 
pair production, one expects 
$F_{HF}^{a}$ to be close to unity. This number is less than 1.0 due to events 
in which the away jet is mistagged or contains heavy flavor due to gluon 
splitting or flavor excitation, and the electron is either a jet misidentified as 
electron or part of a photon conversion pair. If $P$ denotes the probability to 
positively tag the away jet in an event where the electron jet is a light jet, 
then $F_{HF}^a$ is given by
\begin{equation}
F_{HF}^a = 1 - P(1-F_{HF}),
\end{equation}
where $F_{HF} = F_b+F_c = F_b \cdot (1+F_{c/b})$ denotes the total heavy flavor
fraction of electron jets. Here $F_b$ and $F_c$ are the total $b$ and $c$ 
fractions of electron jets, respectively, and $F_{c/b}$ is the $c$ to $b$ 
fraction ratio. 
We estimate $P$ using identified conversions as
\begin{equation}
P =  \frac{\frac{N_{c}^{a+}-N_{c}^{a-}}{N_{a+}-N_{a-}} - \epsilon_c^{'}}{\frac{N_{c}}{N}\,-\,\epsilon_c^{'}},
\end{equation}
where N is the number of events passing the selection, 
$\epsilon_c^{'} = \frac{N_c^{e+}-N_c^{e-}}{N_{e+}-N_{e-}}$, and the $c$ 
subscript refers to events where the electron was identified as a conversion. 
A full derivation of this expression can be found in Ref.~\cite{bib:secvtx}. 
Two methods are used to measure the $b$-fraction, $F_b$, of the electron jets. 
The first method is to reconstruct $D^0 \rightarrow K^-\pi^+$ decays within 
the electron jet and use the invariant mass sidebands to subtract background. 
The second method involves searching for secondary muons within the electron 
jet resulting from cascade $b \rightarrow c \rightarrow l \nu q$ decays using the 
same-sign rate to estimate the background. The contribution from 
charm, $F_{c/b}$, is determined from Monte Carlo simulation to be $F_{c/b}=0.61 \pm 0.10$.
For inclusive electron data we measure $F_{HF} = 0.259 \pm 0.064$ and
$F_{HF}^a = 0.71 \pm 0.05$.

The efficiencies to tag a taggable heavy flavor jet with $E_T > 15~\Gev$ in 
data are summarized in Table~\ref{tab:tageff1} for $P_J<$ 1\% and 5\%.
The ratio of data efficiency to Monte Carlo simulation efficiency is called the 
tagging scale factor ($SF$). The uncertainties shown are statistical and systematic,
which are des\-cribed below.

\begin{table}[!htb]
\begin{center}
\caption{\label{tab:tageff1} Efficiency to tag a taggable heavy flavor 
electron jet in data and the tagging scale factor ($SF$) for {\JP} cuts of 
1\% and 5\%.}
\begin{tabular}{l@{\hspace{1.2cm}}c@{\hspace{1.2cm}}c}
\hline
\hline
  &  $P_J<$ 1\% & $P_J<$ 5\% \\
\hline
$\epsilon$ (Data) &  0.258 $\pm$ 0.011 & 0.334 $\pm$ 0.016 \\
$SF$              &  0.817 $\pm$ 0.07  & 0.852 $\pm$ 0.072 \\ 
\hline
\hline
\end{tabular}
\end{center}
\end{table}

\begin{figure*}[!htb]
\begin{center}
\includegraphics[width=8.3cm,height=7.0cm,clip=]{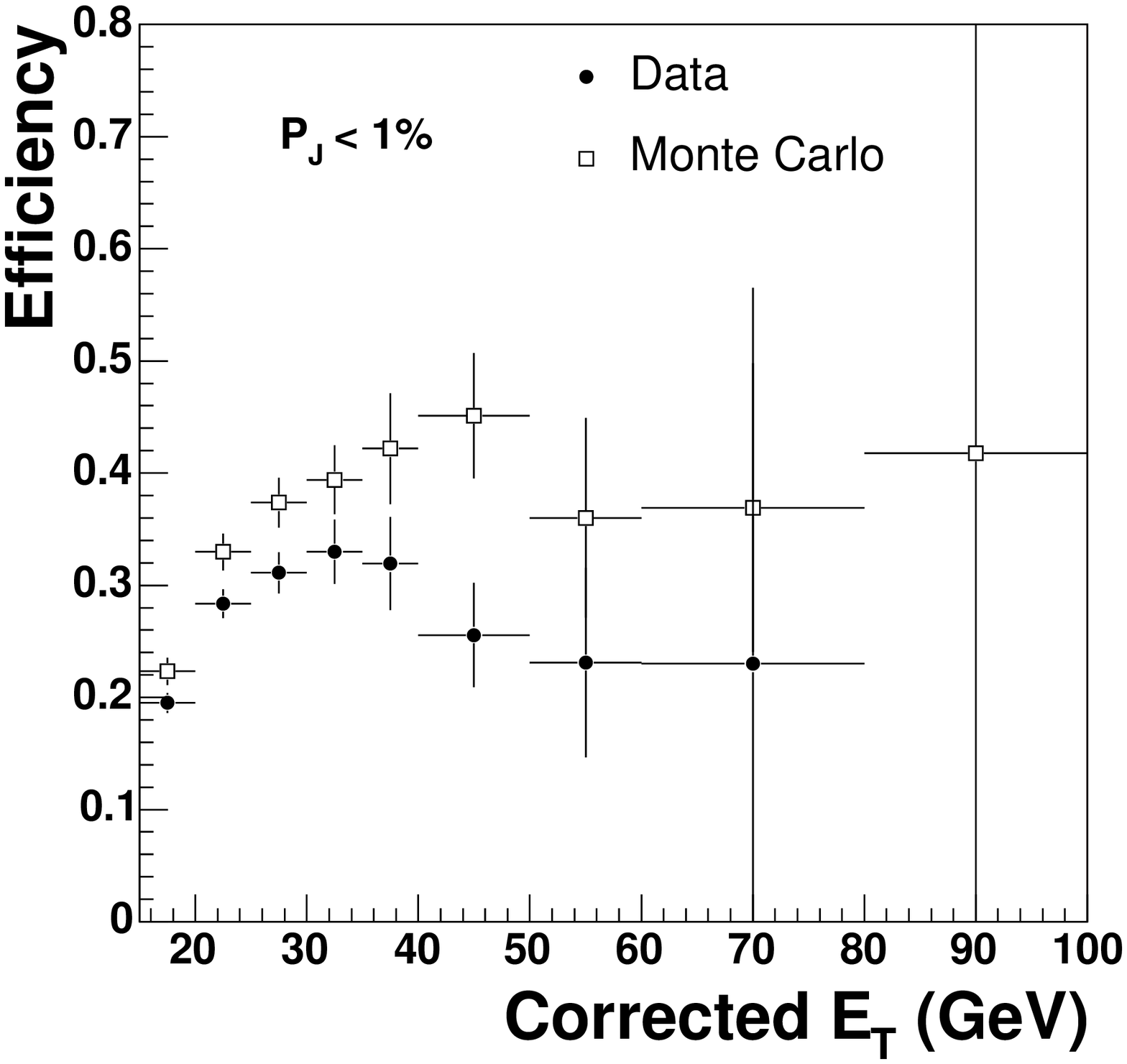}
\includegraphics[width=8.3cm,height=7.0cm,clip=]{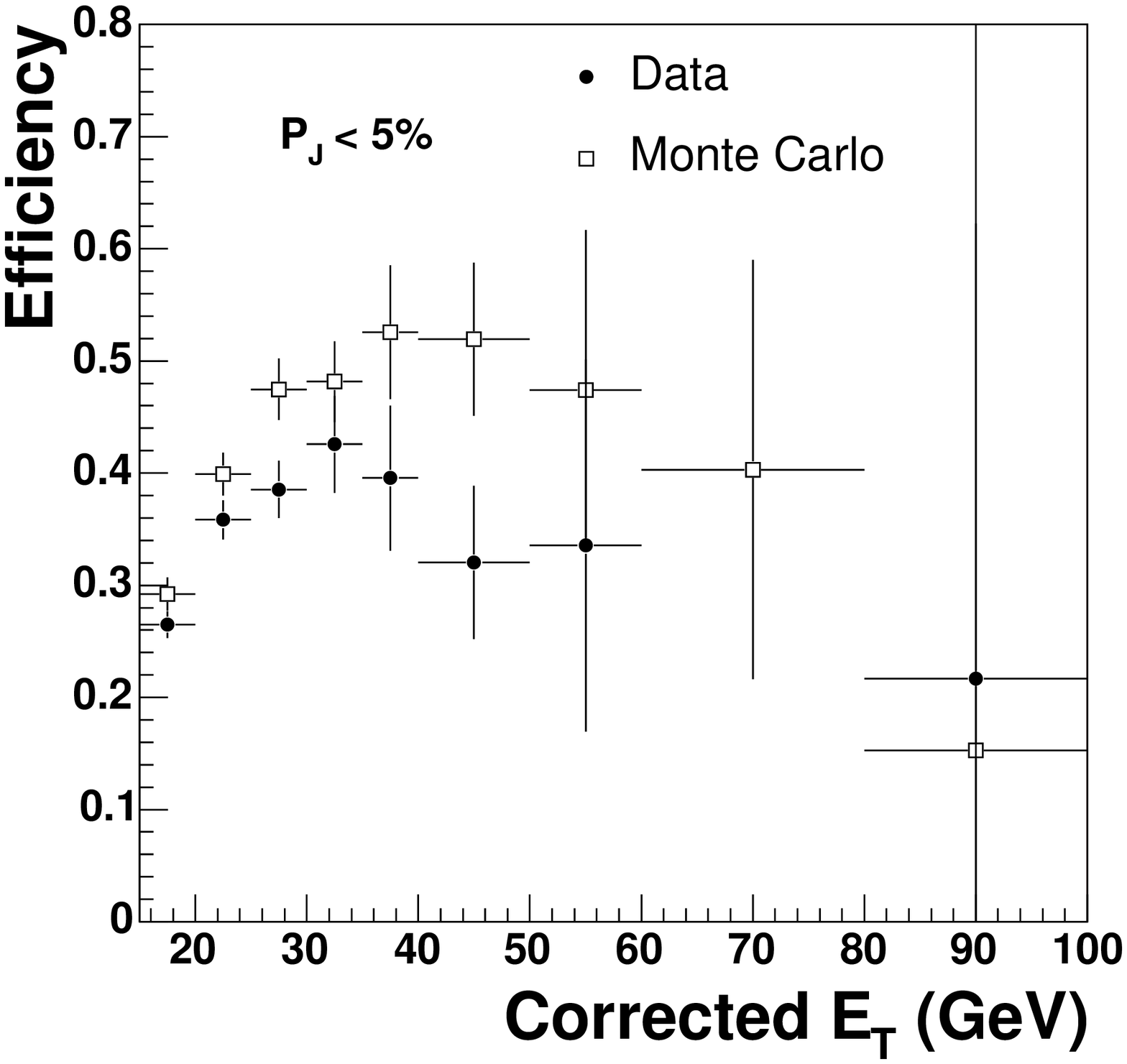}
  \caption{\label{f:jpeff111} Efficiency to tag a heavy flavor jet as a 
function of corrected jet $E_{T}$ in data and Monte Carlo simulation for 
1\% (left) and 5\% (right) {\JP} cut values.}
\end{center}
\end{figure*}
\begin{figure*}[!htb]
  \includegraphics[width=8.3cm,height=7.0cm,clip=]{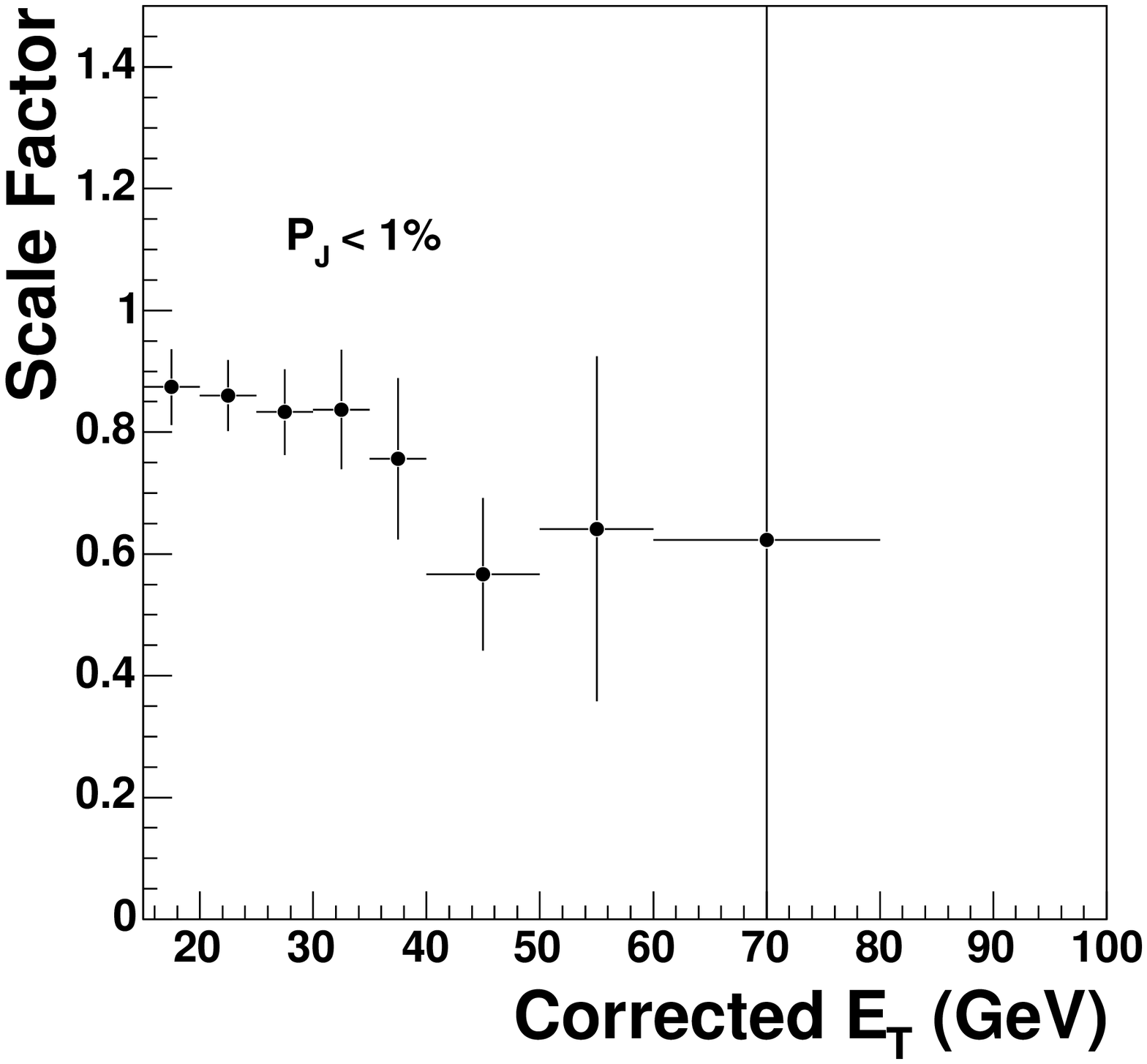}
  \includegraphics[width=8.3cm,height=7.0cm,clip=]{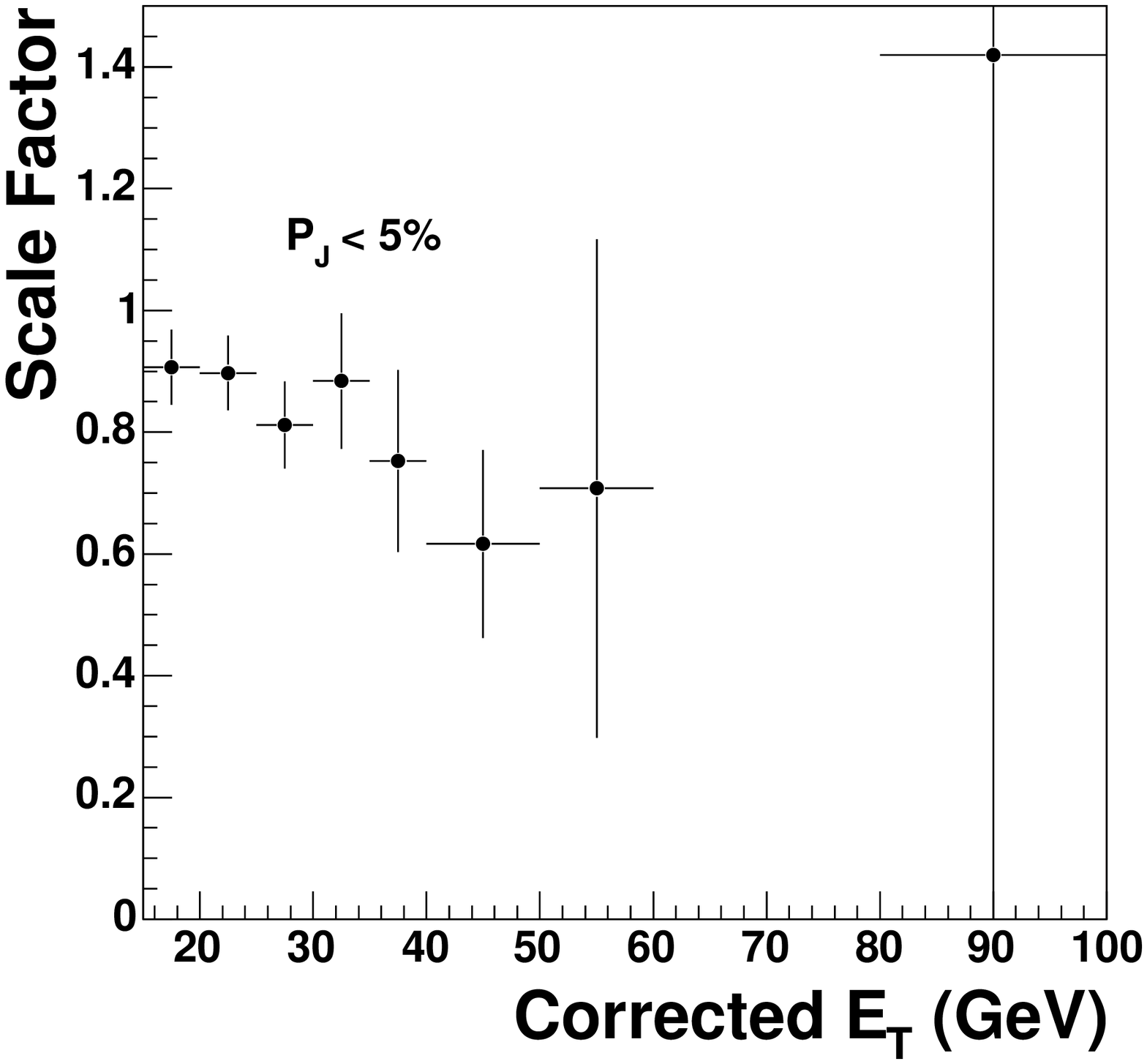}
  \caption{\label{f:jpsf111}  The scale factor ($SF$) as a function of corrected jet $E_{T}$ 
for 1\% (left) and 5\% (right) {\JP} cut values.}
\end{figure*}

It is crucial to understand the tagging efficiency and scale factor dependences
on the jet $E_T$ in order to characterize the {\JP}
algorithm performance. The $E_T$ dependence observed in the inclusive electron 
sample is shown in Fig.~\ref{f:jpeff111} 
and~\ref{f:jpsf111} for the tagging efficiency and scale factor respectively.
Due to the lack of statistics at high jet $E_T$, we repeat the study using 
two samples of high energy jets selected by requiring events with at least one 
jet with $E_T>20~\Gev$ (the Jet20 sample) or with $E_T>50~\Gev$ (the Jet50 
sample). 
The absolute value of the $SF$ can not be extracted because of the 
unknown content of heavy flavor in these samples. However, since the 
variations of heavy flavor fraction are relatively small over a large range of 
$E_T$, we can still estimate the $E_T$ dependence of the scale factor from the
$E_T$ dependence of the ratio of positive tag excess between data and Monte Carlo 
simulation.
\begin{table*}[!htb]
\begin{center}
\caption{\label{t:etdepsf} Summary of the scale factor $vs$. $E_{T}$ slope 
measurements in various samples.}
\begin{tabular}{l@{\hspace{2cm}}c@{\hspace{2cm}}c}
\hline
\hline
Sample & $P_J<$ 1\% & $P_J<$ 5\% \\
\hline
Inclusive Electron      & -0.0082 $\pm$ 0.0037 & -0.0081 $\pm$ 0.0044\\   
Jet 20           & -0.0008 $\pm$ 0.0019 & -0.0028 $\pm$ 0.0024\\  
Jet 50           &  0.0005 $\pm$ 0.0008 &  0.0005 $\pm$ 0.0009\\  
\hline
Weighted Average & -0.00002 $\pm$ 0.00070 & -0.00020 $\pm$ 0.00072\\  
\hline
\hline
\end{tabular}
\end{center}
\end{table*}
Table~\ref{t:etdepsf} shows the results of a linear fit of the tagging scale 
factor to the jet $E_T$ in the inclusive electron, Jet20 and Jet50 samples.
The combined estimate of the slopes is found to be consistent with a flat
$E_T$ dependence of the scale factor both when a $P_J$ cut of 1\% and 5\% is 
applied. Based on these results, we conclude that the scale factor measured in 
the inclusive electron sample is valid at any $E_{T}$. 

Different sources of systematic uncertainty in the determination of $SF$ have 
been considered. 
An uncertainty on the value of $F_b$, determined from the rate of 
$D^0\rightarrow K\pi$ decays, comes from the branching ratio 
$BR$($B\rightarrow l\nu D^0X$). A factor $1.131\pm 0.070$ is used to normalize 
the Monte Carlo simulation prediction to the PDG~\cite{bib:pdg} value. 
The uncertainty includes both the PDG branching 
ratio uncertainty and the Monte Carlo simulation statistical error. Another uncertainty on 
$F_b$ comes from the difference in $D^0$ reconstruction efficiency, 
$\epsilon_{D^0}$, between data and Monte Carlo simulation. 
This uncertainty is derived by 
studying the efficiency of reconstruction for simulated $D^0\rightarrow K\pi$ 
decays embedded into data events, and is found to be 10\%. There is an additional 
uncertainty due to the assumption of symmetry between negative tags and positive mistags 
implicit in the derivation of Equation~\ref{eq:epsilon}. 
The effect of a mistag asymmetry is 
estimated by scaling the subtracted negative tags by di\-ffe\-rent factors 
ranging from 0 to 2 (0.4 to 1.4) for 1\% (5\%) {\JP} cuts, and an 
uncertainty of 7\% is conservatively derived on the tagging scale factor 
due to this effect.
Final estimates for {\JP} tagging efficiencies and scale 
factors are summarized in Table~\ref{tab:tageff1}. 

We do not measure the tagging scale factor for $c$ jets. We assume a common 
scale factor for jets from $b$ and $c$ quarks and we increase the uncertainty 
for a $c$ quark scale factor by 100\% to take into account additional 
uncertainties due to this assumption.




\subsection{\label{sec:jp_mistag}Measurement of the Mistag Rate}
An important ingredient of any analysis which uses heavy flavor tagging is the
background from light quark or gluon jets incorrectly tagged as heavy flavor.
The probability of (positively) tagging a light jet (the ``mistag rate") is closely 
related to the negative tag rate. We remind the reader that a positive (negative) $P_J$ 
is calculated using positively (negatively) signed impact parameter tracks, and a jet 
which has positive (negative) $P_J$ smaller than a certain cut is said to be 
positively (negatively) tagged. It is assumed that the negative tags are due to detector
resolution effects only, while the positive tag rate has an additional   
contribution from real heavy flavor in the jets. Under this assumption, the
mistag rate is equal to the negative tag rate, although in reality there is
also a small contribution from heavy flavor jets to the negative tag
rate and there are contributions from $K$'s, $\Lambda$'s and nuclear
interactions with the detector material to the positive tag rate.    
These effects are considered later in Section~\ref{sec:misasy}.

Since the tag rate has a considerable dependence on jet kinematics, it 
is parameterized as a 6 dimensional tag rate matrix, or look-up table, 
of the transverse energy 
$E_T$ of the jet, the number of jet probability tracks in the jet $N_{trk}$, 
the sum of the transverse e\-ner\-gy of all jets in the event $\sum E_{T}^{jet}$, 
the $\eta$ of the jet computed with respect to the center of the detector, the 
$z$ vertex position $z_{vtx}$ and the $\phi$ of the jet. The tag rates are 
obtained from four inclusive jet samples selected 
by requiring the $E_T$ of the most energetic jet in the event to be greater 
than 20, 50, 70 or 100~$\Gev$ respectively. For a 1\% (5\%) $P_J$ cut, the overall 
ne\-ga\-tive tag rate is (1.22 $\pm$ 0.08)\% ((5.30 $\pm$ 0.25)\%), while 
the overall positive tag rate is (3.54 $\pm$ 0.18)\% ((9.20 $\pm$ 0.26)\%).
Overall tag rates depend on the sample, which is why the tag rates
are parameterized as a function of different variables.
Figure~\ref{f:prmistag} shows the negative tag rates for $P_J<$ 1\% and $P_J<$ 5\% 
as a function of 
the jet $E_T$ and pseudo-rapidity. The bands represent the total uncertainty. 
\begin{figure}[!htb]
\centering
  \includegraphics[width=8.7cm,clip=]{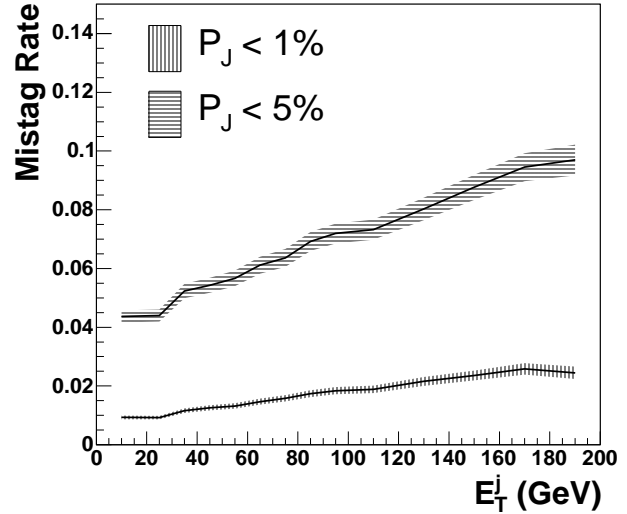}
  \includegraphics[width=8.7cm,clip=]{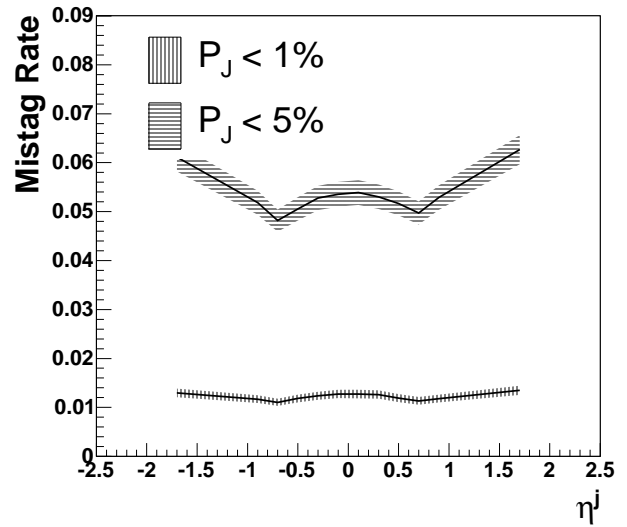}
  \caption{Mistag rate for {\JP} cuts
of 1\% and 5\% as a function of jet $E_T$ (top) and jet pseudo-rapidity (bottom) in inclusive 
jet data sample. The bands represent the statistical and systematic 
uncertainties added in quadrature.}
  \label{f:prmistag}
\end{figure}

We estimate the systematic uncertainties by comparing the observed and 
predicted tag rates in diffe\-rent data samples. We apply tag rate matrices, 
constructed using different inclusive jet subsamples, to different datasets. 
Results are shown in Table~\ref{tab:matrix} for a $P_J$ cut of 1\%. 
The largest deviation between observed and predicted tag rates across
the different jet samples is taken as the systematic uncertainty due to
the sample dependence of the matrix.
In order to account for any possible bias due to the trigger selection,
we apply the matrix separately to trigger and non-trigger jets.
A trigger jet is defined as the jet closest in $\eta$-$\phi$ space to the
level 2 cluster that fired the trigger.
We also apply a matrix built with a high statistics sample of Jet20 events, 
to the Jet50 sample which has several jets below the trigger threshold of 
50~$\Gev$. Also considered is a sample selected by requiring at least four jets
with $E_T >15~\Gev$ and $\sum E_T^{jet}>125~\Gev$. 
These events are expected to give a reasonable estimate of the systematic
uncertainty because of the higher jet and track multiplicities. 
Furthermore, this sample is not used to build the matrix, making 
it sensitive to any additional sources of systematic uncertainty.
Fi\-gu\-re~\ref{f:tagrates} compares the observed and predicted tag rates in the
$\sum E_T^{jet}$ sample as a function of jet $E_{T}$.
The total systematic uncertainty on the overall tag rates is conservatively
taken as the sum in quadrature of the $\sum E_T^{jet}$, Jet20 to Jet50,
and the largest of the trigger and sample contributions.
Table~\ref{tab:matrix_sys} summarizes the relative uncertainties on the
overall tag rates for $P_J$ cuts of 1\% and 5\%. The total relative uncertainty 
is 5.0\% (2.8\%) for positive tag rate and 6.7\% (4.7\%) for negative tag rate 
for a $P_J$ cut of 1\% (5\%).
\begin{table*}[!htb]
\caption{ \label{tab:matrix} Ratios of observed to predicted rates 
of positive and negative tags when a $P_J$ cut of 1\% is applied. 
The first column specifies the sample used to build the
matrix, while the second column reports the sample used to compute the rates.
All(even)[odd] means that all(only even event number)[only odd event number]
events are used. The errors shown are statistical only.}
\begin{center}
\begin{tabular}{l@{\hspace{0.5cm}}c@{\hspace{0.5cm}}c@{\hspace{0.5cm}}c}
\hline
\hline
Matrix        & Sample               & Obs./Pred. Pos. Tag Rate Ratio & Obs./Pred. Neg. Tag Rate Ratio \\
\hline
Inc. Jet Even  &  Inc. Jet Odd        & 0.997$\pm$0.002       & 0.999$\pm$0.003       \\
Inc. Jet Even  &  Jet20 Odd           & 0.987$\pm$0.003       & 0.970$\pm$0.006       \\
Inc. Jet Even  &  Jet50 Odd           & 0.991$\pm$0.003       & 0.998$\pm$0.006       \\
Inc. Jet Even  &  Jet70 Odd           & 0.997$\pm$0.004       & 0.996$\pm$0.006       \\
Inc. Jet Even  &  Jet100 Odd          & 0.989$\pm$0.003       & 1.029$\pm$0.005       \\
Jet20 All      &  Jet50 All           & 1.020$\pm$0.003       & 1.044$\pm$0.008       \\
Inc. Jet Even  &  Trig. Jet  Odd      & 0.976$\pm$0.002       & 0.978$\pm$0.004       \\
Inc. Jet Even  &  Non trig. Jet Odd   & 1.028$\pm$0.003       & 1.028$\pm$0.005       \\
Inc. Jet All   &  $\sum E_{T}^{jet}$ All      & 1.037$\pm$0.002       & 0.966$\pm$0.003       \\
\hline
\hline
\end{tabular}
\end{center}
\end{table*} 

\begin{figure}[!htb]
\begin{center}
  \includegraphics[width=8.8cm,clip=]{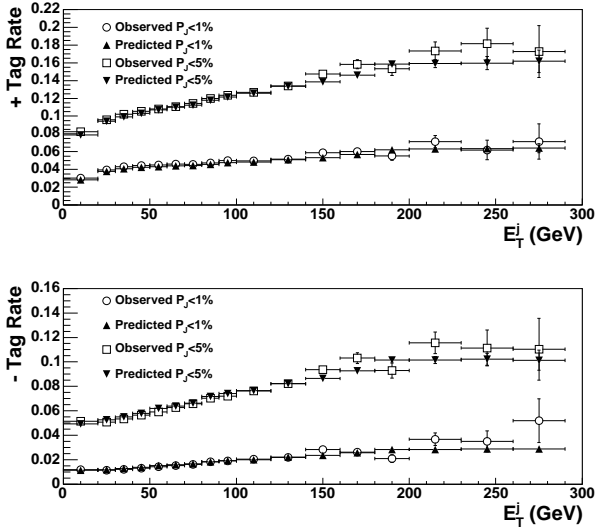}
  \caption{\label{f:tagrates} Observed positive (top) and negative (bottom) tag rates as a 
function of $E_{T}$ for events in $\sum E_{T}$ data $vs$. prediction from the 
matrix built with events in inclusive jet data. The two curves 
correspond to $P_J$ cuts of 1\% and 5\%.}
\end{center}
\end{figure}

\begin{table*}[!htb]
\begin{center}
\caption{ \label{tab:matrix_sys} Total relative uncertainties on the overall 
positive and negative tag rates. Different tag rate matrices are applied to 
orthogonal samples of jets as shown in Table~\ref{tab:matrix}, and the 
total uncertainty is taken as the sum in quadrature of the most relevant
contributions. Sample refers to the largest uncertainty from lines
2 to 5 in Table~\ref{tab:matrix}.}
\begin{tabular}{c@{\hspace{1cm}}c@{\hspace{1.1cm}}c@{\hspace{1.1cm}}c@{\hspace{1.1cm}}c@{\hspace{1.1cm}}c@{\hspace{1.1cm}}c}
\hline
\hline
$P_J$ cut   & Statistical   & Trigger      & $\sum_{j}$        &  Jet20$\rightarrow$Jet50 & Sample   & TOTAL  \\
\hline
Pos. $P_J<$ 1\%  & 0.11\%  & 2.4\%        & 3.7\%        & 2.0\%          & 1.3\%    & 5.0\%  \\
Neg. $P_J<$ 1\%  & 0.25\%  & 2.2\%        & 3.4\%        & 4.4\%          & 3.0\%    & 6.7\%  \\
Pos. $P_J<$ 5\%  & 0.07\%  & 1.5\%        & 1.2\%        & 1.6\%          & 1.2\%    & 2.8\%  \\
Neg. $P_J<$ 5\%  & 0.09\%  & 1.3\%        & 2.4\%        & 3.1\%          & 2.2\%    & 4.7\%  \\
\hline
\hline
\end{tabular}
\end{center}
\end{table*}

\subsubsection{Mistag Asymmetry}\label{sec:misasy}
The rate of negatively tagged jets does not reflect the rate of positive 
mistags of light jets because of residual lifetime effects from $\Lambda$'s 
and K's or interactions with the detector material. 
Corrections for these effects are determined 
by studying the flavor composition of tagged jets in data.

The set of jet probability tracks inside a tagged jet is used to build a 
variable sensitive to the flavor content of the jet itself. The relative
contributions from heavy and light partons to data are determined by fitting 
the distribution of this variable for tagged jets in data to Monte Carlo simulation
templates for $b$, $c$ and light jets. For data, a sample selected by 
requiring a jet with $E_{T}>$ 50~$\Gev$ at the trigger level is used.
For Monte Carlo simulation distributions, {\HERWIG} is used to generate 
$2\rightarrow 2$ processes with an outgoing parton $p_{T}>40~\Gevc$.
We perform the fit using six different variables,
the maximum impact parameter $d_0$ of the tracks in the jet, 
the maximum impact parameter significance $S_{d_{0}}$ of the tracks in the 
jet, 
the mass of the system of tracks with $|d_{0}|\,> 0.01$~$\cm$ and $S_{d_0}>2$, 
and the transverse momentum ($P_T^{rel}$) with respect to the jet direction 
of the system of tracks with $|d_{0}|\,> 0.01$ $\cm$ and $S_{d_0}>2$. 

The fit is made more robust by fitting the positive excess only, for which the
distributions for negative tags are subtracted from the positive
side. This removes contribution to the mistags due to detector resolution,
which could be simulated poorly. The number of negative tags
obtained for $b$, $c$ and light jets in Monte Carlo simulation is normalized to
the total number of negative tags found in data. From
the fit, the fractions of $b$, $c$ and light jets in data are
obtained; thus the ratio of positive to negative tags from light jets, $\beta$.

Figure~\ref{fig:misasy1} shows the result of the fit of the positive tag excess
in data to Monte Carlo templates of the maximum impact parameter of the tracks
contained within $b$, $c$ and light tagged jets. A $P_J$ cut of 1\% is used.
It should be noted that the $c/b$ ratio gets a contribution from the $c/b$ tagging
efficiency ratio of about 0.2. The observed rise of light jets is the result of the
fact that tags for light jets are usually due to one large impact parameter track. 
Table~\ref{t:misasy_res} summarizes the results of the mistag asymmetry 
measurement with the six variables chosen for $P_J$ $<$ 1\% and 5\%.
As a final estimate of the mistag asymmetry, we take the average of the six 
measurements and assign the maximum difference between the average and each 
single determination as the uncertainty. The results are $1.56 \pm 0.14$ and
$1.27 \pm 0.17$ for $P_J$ cuts of 1\% and 5\%, respectively. The asymmetry is 
caused by secondary interactions with the detector material and residual 
lifetime effects from $K$'s and $\Lambda$'s, giving an excess of positive 
mistags. We study the expected contribution of $K$'s and $\Lambda$'s 
decays to the mistag asymmetry in Monte Carlo simulated events. We find the ratio
of positively to negatively tagged light jets to be $1.55\pm 0.11$ 
($1.21\pm 0.04$) for a $P_J$ cut of 1\% (5\%). Uncertainties are statistical
only. These results are in good agreement with our measurements on data
and suggest $K$'s and $\Lambda$'s to be the main source of mistag asymmetry.
The negative tag rates measured have therefore to be scaled up by the 
asymmetry  factor in order to obtain an accurate estimate of the positive 
mistag rate.
We repeat the measurement in bins of jet transverse energy to study
the dependence of the mistag asymmetry on the jet $E_T$. Results are shown in 
Fig.~\ref{f:mistag_jetet}. The asymmetry exhibits a small dependence with 
jet $E_T$ which is taken into account to estimate the mistag background.
\begin{figure}[!htb]
   \centering
    \includegraphics[width=0.52\textwidth,clip=]{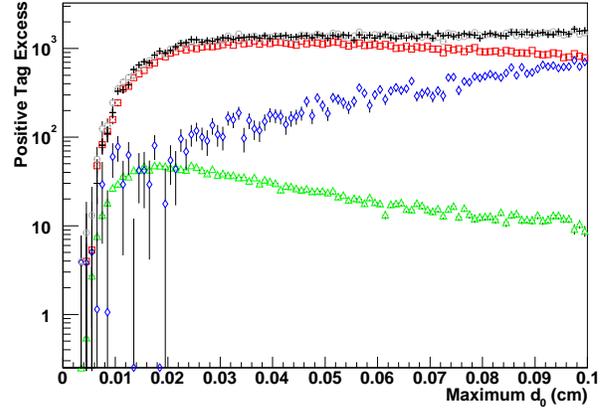}
\caption{\label{fig:misasy1} Result of the fit of the positive tag excess in
{\scshape{Jet50}} data as a function of the  maximum impact parameter
$d_0$ of jet probability tracks inside the tagged jets. A $P_J$ cut of 1\% is 
used. Observed data are the crosses, while the fitted component
from $b$, $c$ and light jets are the squares, triangles and diamonds,
respectively. The circles are the sum of the three fitted components.}
\end{figure}                   

\begin{table*}[!htb]
\begin{center}
\caption{\label{t:misasy_res}Mistag asymmetry measured in Jet50 data for $P_J$ 
cuts at 1\% and 5\%. The quoted uncertainties are derived from the uncertainty
in the fits. The uncertainty used for the average is the maximum difference between 
the average and each measurement.}
\begin{tabular}{l@{\hspace{1.3cm}}c@{\hspace{1.3cm}}c}
\hline
\hline
  Fitted variable & $\beta$ ($P_J<$ 1\%) & $\beta$ ($P_J<$ 5\%) \\
\hline
  Maximum $d_0$ & 1.64 $\pm$ 0.02 & 1.37 $\pm$ 0.02 \\
  Maximum $S_{d_0}$ & 1.56 $\pm$ 0.03 & 1.10 $\pm$ 0.02 \\
  Mass of the system of tracks with $|d_0|>0.01~cm$ & 1.51 $\pm$ 0.04 & 1.30 $\pm$ 0.02 \\
  Mass of the system of tracks with $S_{d_0}>2$ & 1.43 $\pm$ 0.03 & 1.20 $\pm$ 0.02 \\
  P$_T^{rel.}$ of the system of tracks with $|d_0|>0.01~cm$ & 1.67 $\pm$ 0.03 & 1.32 $\pm$ 0.02 \\
  $P_T^{rel.}$ of the system of tracks with $S_{d_0}>2$ & 1.57 $\pm$ 0.02 & 1.30 $\pm$ 0.02 \\
\hline
  Average & 1.56 $\pm$ 0.14 & 1.27 $\pm$ 0.17 \\
\hline
\hline
\end{tabular}
\end{center}
\end{table*}

\begin{figure}[!htb]
   \centering
    \includegraphics[width=0.49\textwidth,clip=]{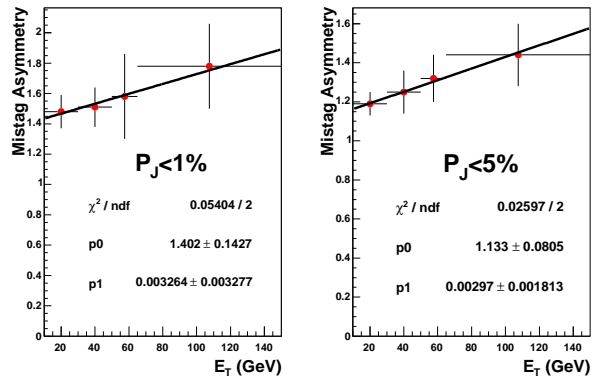}
\caption{Mistag asymmetry as a function of the jet transverse energy for 
         $P_J<$ 1\% (left plot) and $P_J<$ 5\% (right plot).}
\label{f:mistag_jetet}
\end{figure}

\subsection[Jet Probability Performance on $t\bar t$ Events]{Jet Probability Performance on \boldmath{$t\bar t$} Events}
We study the performance of the {\JP} algorithm by computing the efficiency to 
tag a $b$ jet in {\PYTHIA} Monte Carlo $t\bar t$ events generated with a top 
mass = 178~$\Gevcc$.
Results are shown in Fig.~\ref{f:preff} as a function of the transverse 
energy $E_T$ and of the pseudo-rapidity $\eta$ of the jets for $P_J$ cuts of 
1\% and 5\%. Jets are matched to $b$ quarks (by requiring $\Delta R < 0.4$ between
the reconstructed jet and the $b$ quark) and the tagging $SF$ is 
applied to the resul\-ting efficiency. We also measure the average efficiency to 
tag a $b$ or a $c$ jet in $t\bar t$ events passing the kinematic event 
selection described in Section~\ref{sec:evt_sel}. 
Results are shown in Table~\ref{t:bceff} before and after applying the tagging 
$SF$. The scaled per-jet efficiencies, together with the mistag matrix,
are used to determine the efficiency to tag at least $n$ jets per $t\bar t$ 
event, as described in Section~\ref{sec:acc}. Although the tagging 
requirement results in some loss of efficiency for the $t\bar t$ signal, it
significantly increases the signal-to-background ratio by heavily suppressing 
the dominant $W$+jets background.

\begin{figure}[!htb]
\centering
  \includegraphics[width=0.5\textwidth,clip=]{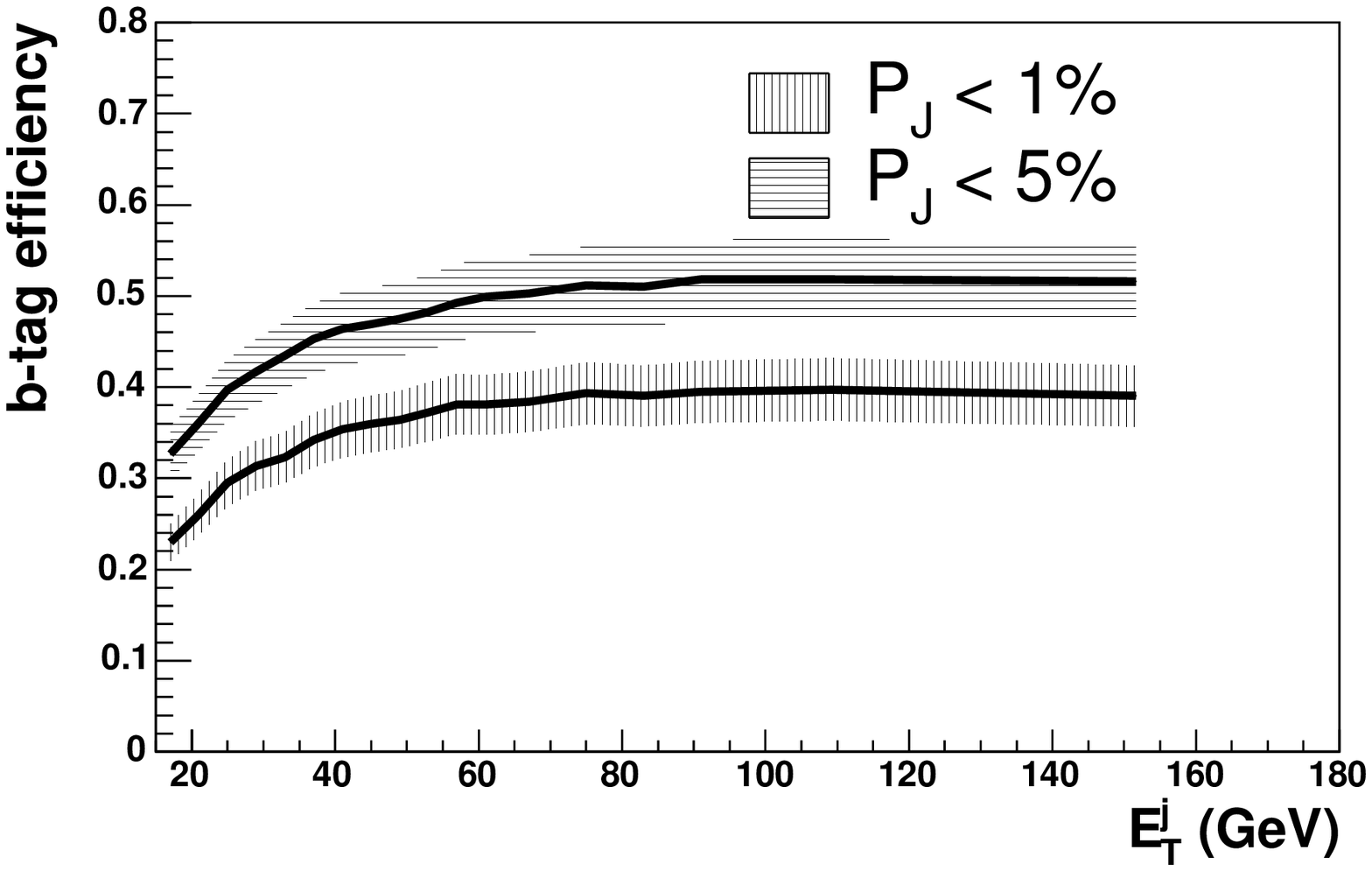}
  \includegraphics[width=0.5\textwidth,clip=]{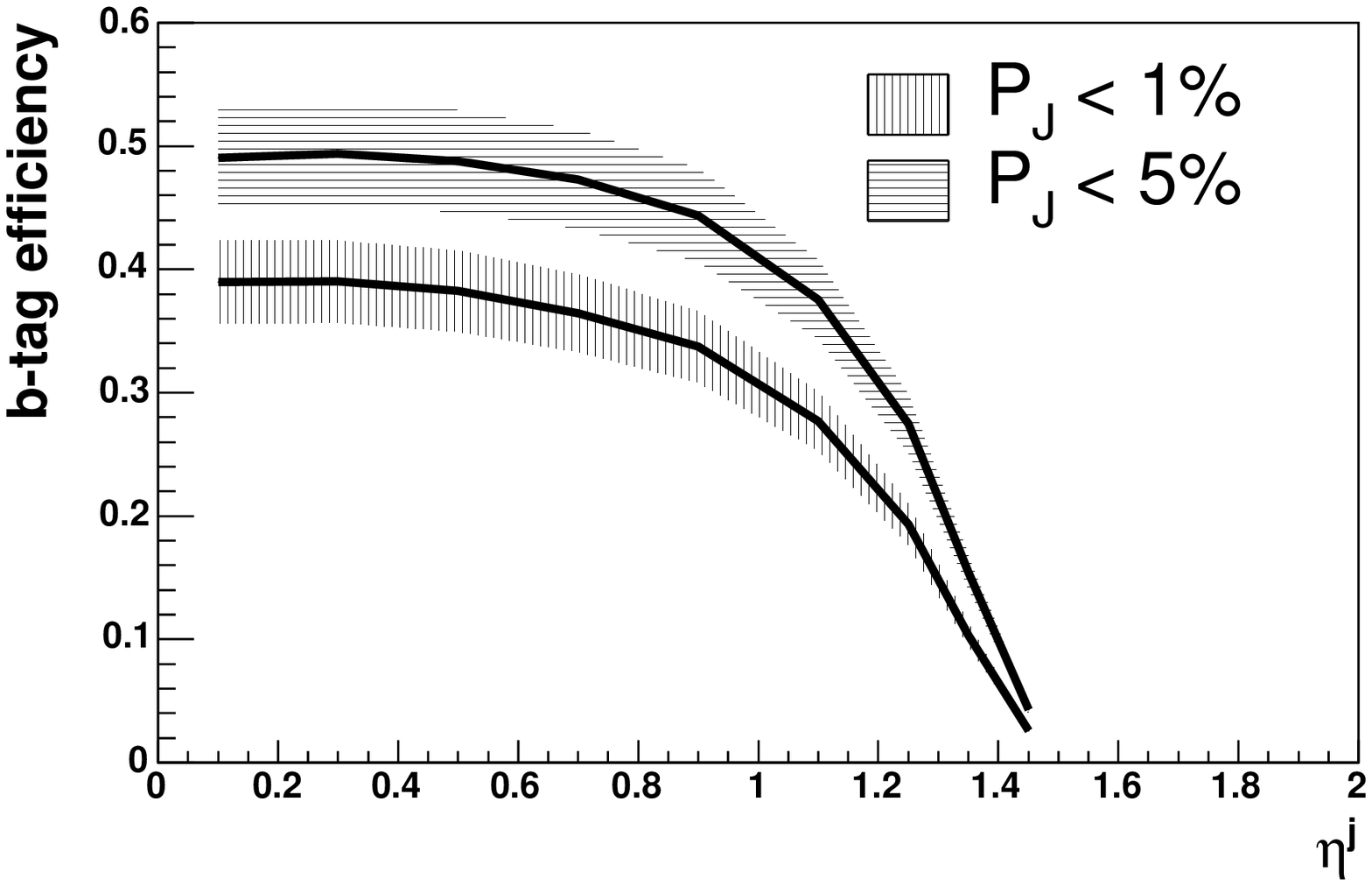}
  \caption{Efficiency to tag $b$ jets in $t\bar t$ Monte Carlo simulated events 
as a function of jet $E_T$ (top) and jet $\eta$ (bottom) for 
two different $P_J$ cuts. The efficiency is obtained by multiplying the tag rate 
for jets matched to $b$ quarks in the Monte Carlo simulation by the appropriate tagging 
scale factor. The bands represent the systematic uncertainty on the scale 
factors.}
  \label{f:preff}
\end{figure}

\begin{table*}[!htb]
\begin{center}
\caption{\label{t:bceff}Tagging efficiencies for $b$ and $c$ jets in $t\bar t$ events ($m_t=178~\Gevcc$) before (raw efficiency) and after (scaled efficiency) applying the tagging scale factor.}
\begin{tabular}{l@{\hspace{1.3cm}}c@{\hspace{1.3cm}}c@{\hspace{1.3cm}}c@{\hspace{1.3cm}}c}
\hline
\hline
 & \multicolumn{2}{c}{{ b jets}} & \multicolumn{2}{c}{{ c jets}} \\
\hline
 & Raw Eff. (\%) & Scaled Eff. (\%)  & Raw Eff. (\%)  & Scaled Eff. (\%)  \\
\hline
 $P_J<$ 1\% & $43.2\pm 0.1$  & $35.3\pm 3.0$  & $ 9.6\pm 0.2$  & $ 7.8\pm 0.7$ \\
 $P_J<$ 5\% & $54.6\pm 0.1$  & $46.5\pm 3.9$  & $20.3\pm 0.2$  & $17.3\pm 1.5$ \\
\hline
\hline
\end{tabular}
\end{center}
\end{table*}

\section{\label{sec:evt_sel}Event Selection}
Top quark events in the lepton+jets channel are characterized by the presence 
of an electron or muon with high transverse energy, large missing transverse energy 
and four high energy jets, two of which are $b$ jets.
The basic pretag selection requires one tight electron or muon, 
$\met > 20~\Gev$ and jets with corrected $E_{T} > 15~\Gev$ and $|\eta| < 2$. 

In order to select a lepton+jets sample completely disjoint from the top
dilepton sample ($t\bar t \rightarrow l^{+}l^{-}\nu\bar{\nu}q\bar q$), we 
reject events with an extra lepton that passes the loose requirements. Events consistent with 
$Z \rightarrow l^{+}l^{-}$ are removed if a tight lepton and a second object 
form an invariant mass within the range [76, 106]~$\Gevcc$. 
If the tight lepton is an electron, the second object may be an isolated 
electromagnetic object, a jet with electromagnetic fraction greater than 0.95 
or an oppositely-signed isolated track. If the tight lepton is a muon, the 
second object may be an isolated muon or an opposite-signed isolated track.

The event vertex $z$ position is used to cluster jets and to ensure leptons 
and jets come from the same interaction. If more than one primary vertex is 
reconstructed in the event, the vertex closest to the lepton track is selected 
as event vertex.
Events are rejected if the $z$ of the lepton track is farther than 5~cm from the $z$ 
of the event vertex.
The vertex $z$ position is required to be within 60~cm of the center of the
detector in order to ensure good event reconstruction in the
projective tower geometry of the {\CDF} detector.
The efficiency of this requirement is measured using minimum bias data and 
found to be (95.1 $\pm$ 0.3)\%.
For consistency with the $b$-tagging algorithm, events are also rejected if the
$z$ of the vertex with highest $\sum p_T$ of all tracks is farther than 5~cm 
from the event vertex $z$. The efficiency of this requirement is (98 $\pm$ 
2)\%, where the 2\% error accounts for the uncertainty in the simulation
of multiple interactions.

The events selected by the above criteria are dominated by {\QCD} production 
of $W$ bosons in
association with jets. 
Figure~\ref{f:jp_dist} shows the $P_J$ distribution for taggable jets in this sample.
In order to improve the signal to background ratio for 
$t\bar t$ events, we require at least one jet in the event to be tagged as a 
$b$ jet. A $t\bar t$ event is expected to have four jets in the final state,
but due to gluon radiation, jet merging, and inefficiencies in jet
reconstruction, this number can eventually be different.
We therefore use the tagged events with three or more jets to define our 
signal sample, while the events with one and two jets are used as 
a control sample. 

\begin{figure}[!htb]
\begin{center}
    \includegraphics[width=8.7cm,clip=]{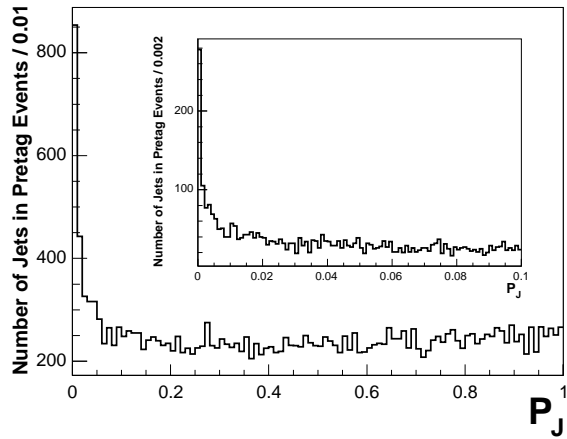}
\caption{\label{f:jp_dist} The $P_J$ distribution for taggable jets in 
the pretag sample (note that the vertical axis does not start at 0). The inset shows a zoom of the $P_J$ distribution from 0.0 to 0.1.} 
\end{center}
\end{figure}
 
\subsection{Optimized Selection}
The variable $H_T$, defined as the scalar sum of all the transverse energy in 
the event, i.e., the sum of the $\met$, the electron $E_T$ or muon $p_T$, 
and the $E_T$ of the jets, is a measure of the energy in the hard scatter, and  is a powerful discriminant between the $t\bar t$ pair production signal 
events ($S$) and background events ($B$).
In order to find the optimal $H_T$ cut, we maximize the statistical 
significance ($S/\sqrt{S+B}$) in the signal region. Figure~\ref{f:ht} (top)
shows the $H_T$ distribution of the $t\bar t$ Monte Carlo simulation, together with the 
various background contributions, properly normalized. Figure~\ref{f:ht} (bottom) shows the
statistical significance as a function of the $H_T$ cut. Details about the 
background estimates and datasets used can be found in Section~\ref{sec:bkg}. 
Optimal statistical significance is reached with a cut of $H_T >$ 200~$\Gev$. 
\begin{figure}[!htb]
  \centering
    \includegraphics[width=0.5\textwidth,clip=]{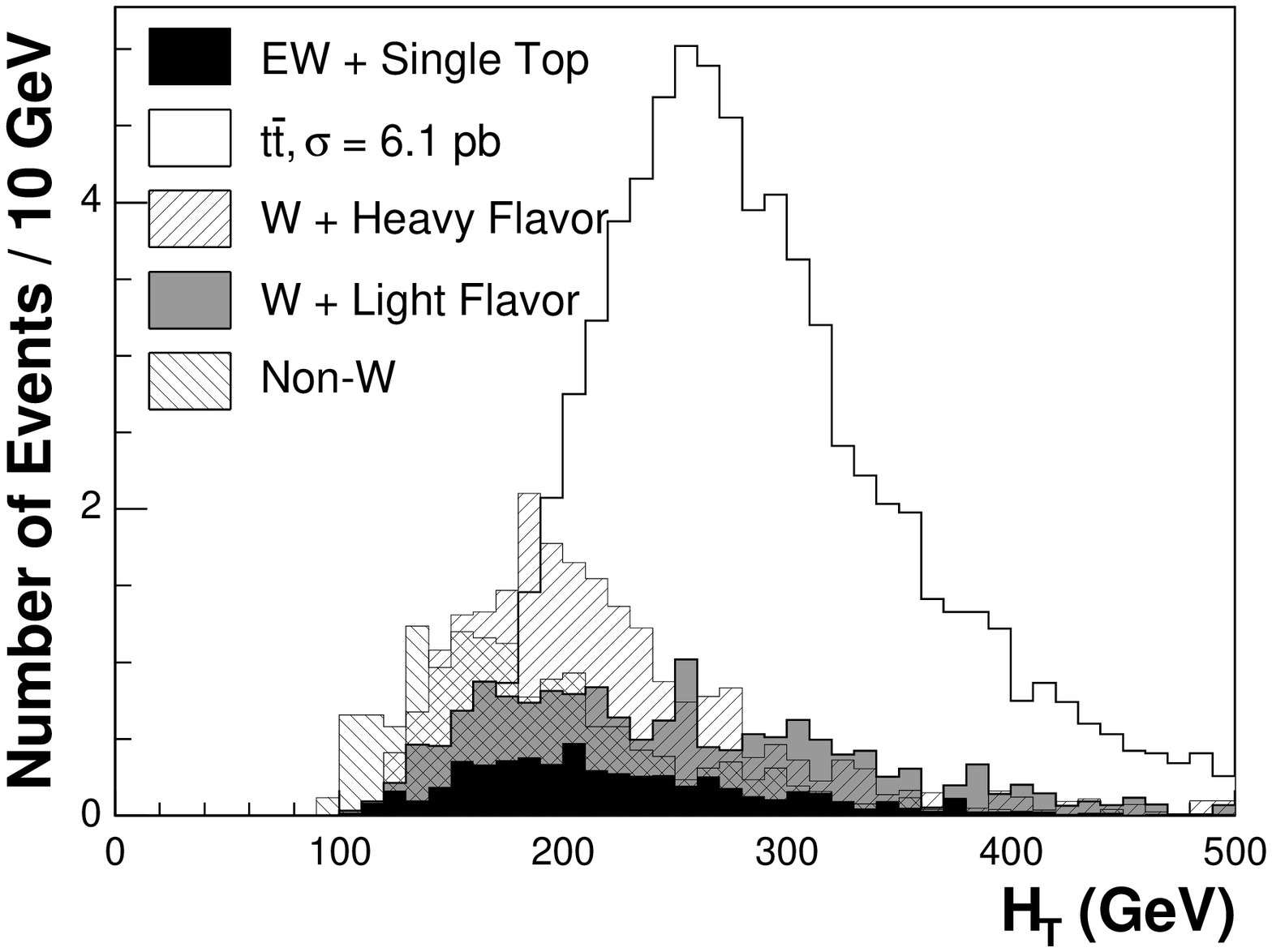}
    \includegraphics[width=0.5\textwidth,clip=]{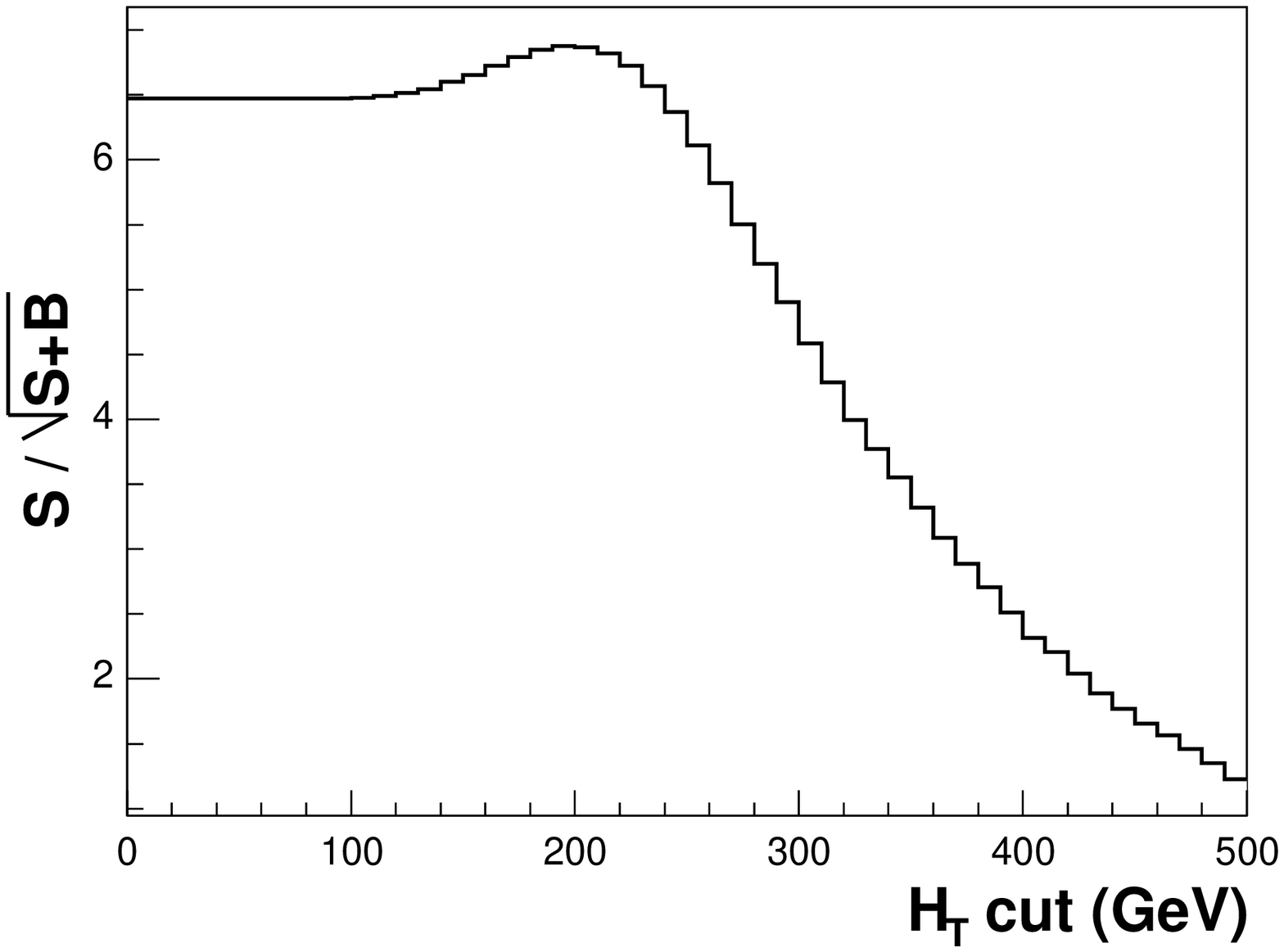}
\caption{Top: $H_T$ distribution for tagged events with 
3 or more tight jets in
$t\bar t$ Monte Carlo simulation (6.1~pb) and main backgrounds, for an integrated 
luminosity of 318~pb$^{-1}$. Bottom: statistical significance as a 
function of the cut applied.} 
\label{f:ht}
\end{figure}

In addition, we enhance the $W$ component of the sample by requiring the 
transverse mass of the lepton and the missing energy,
\mbox{$M_T^W = \sqrt{(E_T(l)+E_T(\nu))^2 - (\vec{P}_T(l)+\vec{P}_T(\nu))^2}$},~be consistent with $W$ boson production.
Figure~\ref{f:mtw} (top) shows the $M_T^W$ distribution for the $t\bar{t}$
Monte Carlo simulation together with the various normalized background contributions. 
Figure~\ref{f:mtw} (bottom) shows
the statistical significance as a function of the $M_T^W$ cut. Note that the
non-$W$ background lies at lower values of $M_T^W$. In the optimization of the 
$M_T^W$ cut, 
we take $S$ to be the number of events from real $W$ bosons and $B$ to be the 
number of events from non-W background. A cut of $M_T^W > 20~\Gevcc$ gives 
optimal statistical significance.

\begin{figure}[!htb]
  \centering
    \includegraphics[width=0.5\textwidth,clip=]{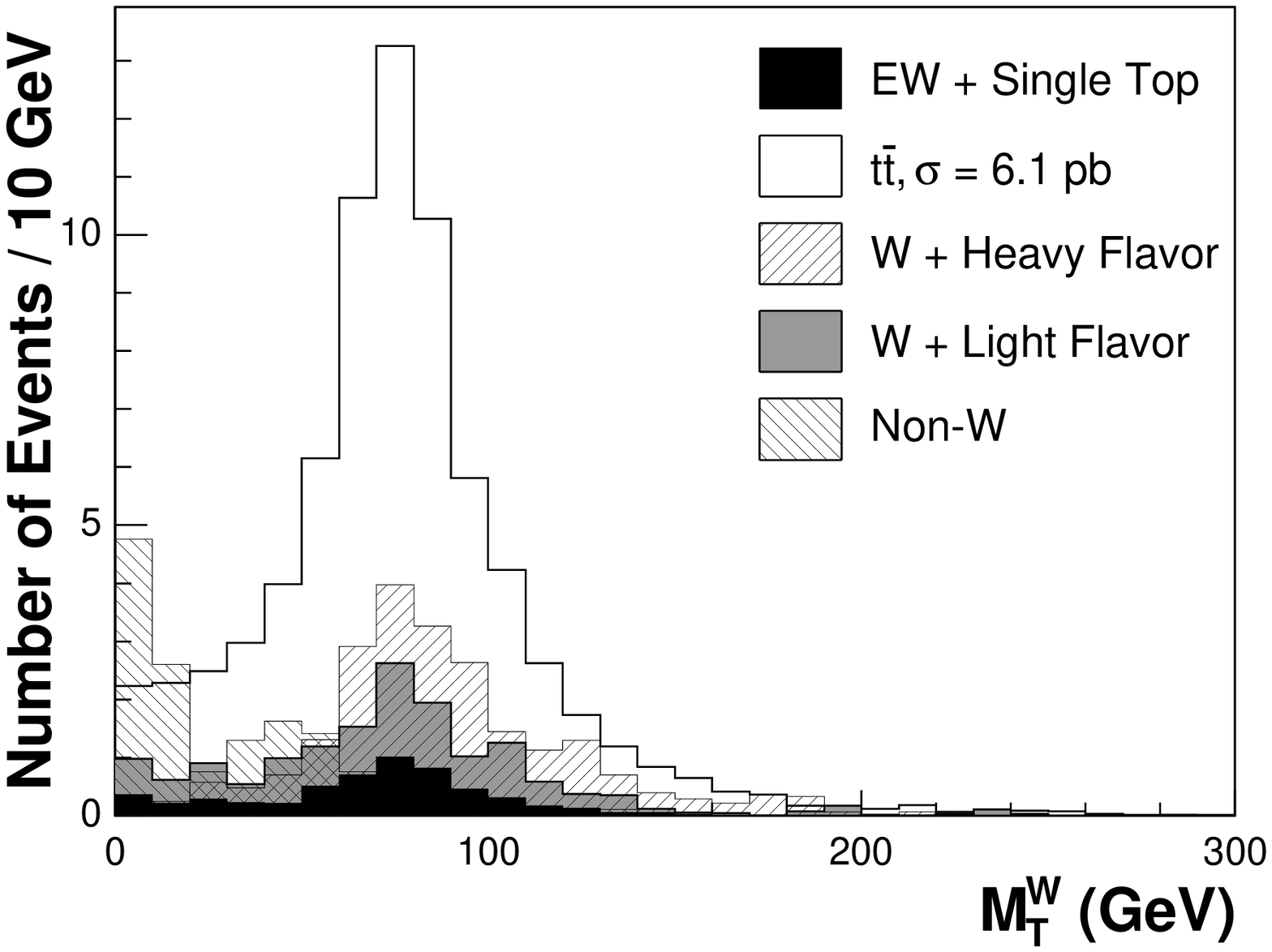}
    \includegraphics[width=0.5\textwidth,clip=]{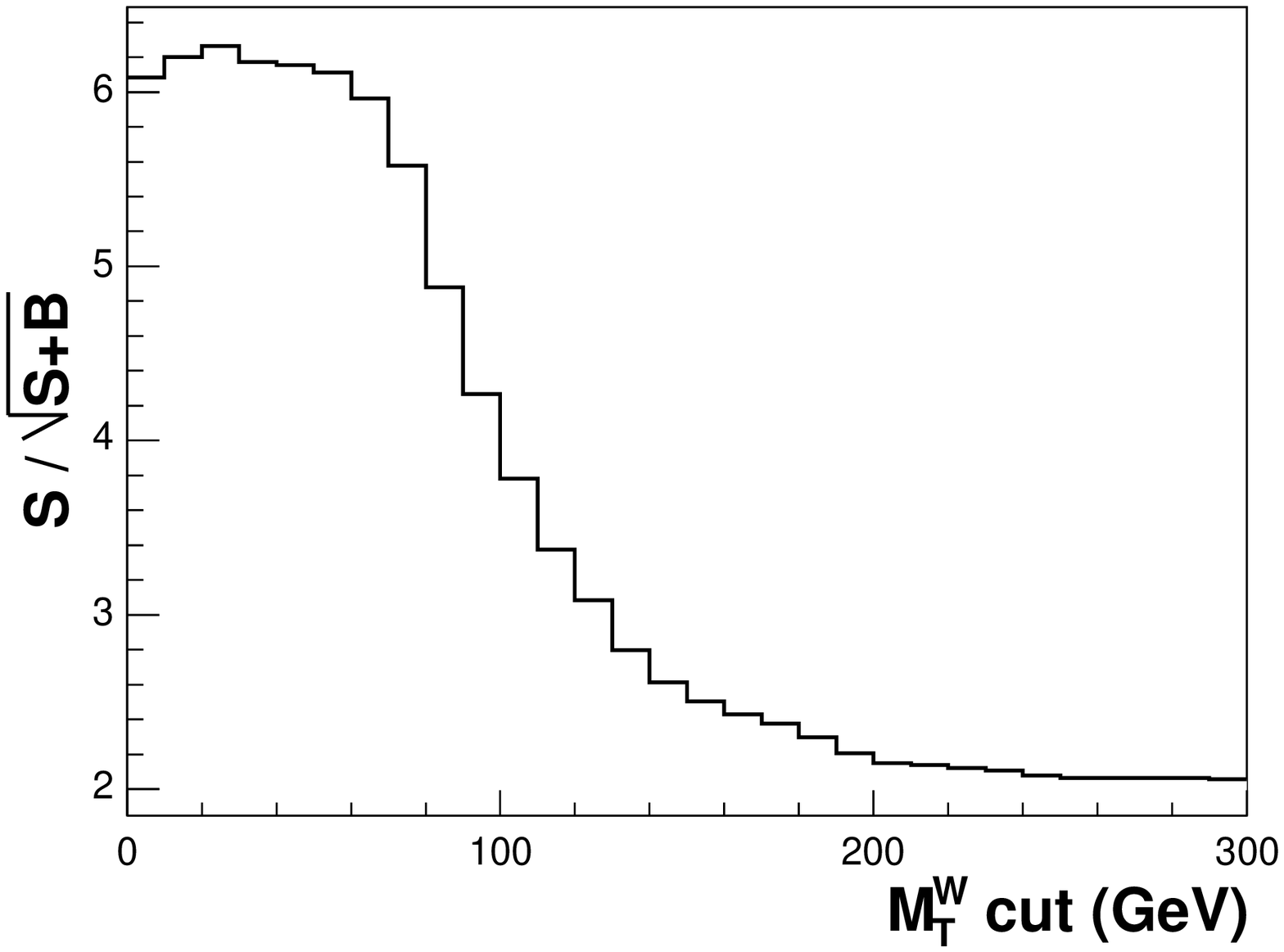}
\caption{Top: $W^M_T$ distribution for tagged events with 
3 or more tight jets in
$t\bar t$ Monte Carlo simulation (6.1~pb) and main backgrounds, for an integrated 
luminosity of 318~pb$^{-1}$. Bottom: statistical significance as a 
function of the cut applied.} 
\label{f:mtw}
\end{figure}

\subsection{Yields of Events}
Events which pass the selection criteria described so far, before applying 
$b$-tagging, form the pretag sample. The number of observed events in both the 
pretag and tagged samples for $P_J<$ 1\% and $P_J<$ 5\% are summarized in 
Table~\ref{t:yields} as a function of the number of tight jets in the event.

\begin{table*}[!htb]
\begin{center}
\caption{\label{t:yields}Yield of events in 318~$\pb^{-1}$ of data for $P_J<$ 
1\% and $P_J<$ 5\%.}
\begin{tabular}{l@{\hspace{0.9cm}}c@{\hspace{1.3cm}}c@{\hspace{1.3cm}}c@{\hspace{1.3cm}}c}
\hline
\hline
Jet Multiplicity & 1 jet & 2 jets & 3 jets & $\ge$ 4 jets \\
\hline
\hline
           \multicolumn{5}{c}{Pretag Events}  \\
\hline
{\CEM} & 16897 & 2657 & 182 & 105 \\
{\CMUP} & 8169 & 1175 & 83 & 44 \\
{\CMX} & 4273 & 610 & 35 & 17 \\
\hline
Total & 29339 & 4442 & 300 & 166 \\
\hline
\hline
           \multicolumn{5}{c}{Single Tagged Events, $P_J<$ 1\% (5\%)} \\
\hline
{\CEM}  & 207 (571) & 106 (230) & 33 (53) & 36 (53) \\
{\CMUP} & 92  (256) & 58  (105) & 13 (24) & 24 (29) \\
{\CMX}  & 51  (148) & 27  (50)  & 6  (10) & 8  (11) \\
\hline
Total & 350 (975) & 191 (385) & 52 (87) & 68 (93) \\
\hline
\hline
          \multicolumn{5}{c}{Double Tagged Events, $P_J<$ 1\% (5\%)} \\
\hline
{\CEM}  & --- & 8 (16) & 7 (15) & 9 (18) \\
{\CMUP} & --- & 3 (9)  & 4 (4)  & 8 (17) \\
{\CMX}  & --- & 2 (3)  & 1 (3)  & 1 (4)  \\
\hline
Total & --- & 13 (28) & 12 (22) & 18 (39) \\
\hline
\hline

\end{tabular}
\end{center}
\end{table*}

\section{\label{sec:bkg} \boldmath Backgrounds: Expected Composition of the $b$-Tagged Lepton+Jets Sample}

Other processes besides $t\bar t$ are expected to contribute to the 
tagged lepton+jets sample. 
The main contribution comes
from heavy flavor production in association with a $W$ boson ($Wb\bar b$,
$Wc\bar c$, $Wc$). $W$+light flavor production also gives a significant
contribution due to mistagged jets. Smaller 
contributions come from electroweak processes (diboson production, 
$Z\rightarrow\tau^+\tau^-$ events or single top)
and generic {\QCD} jet production with misidentified $W$ bosons.
These backgrounds are described in the following subsections.

\subsection{\label{sec:other}Electroweak Processes}

Electroweak processes are studied
using Monte Carlo simulated samples. Diboson events ($WW$, $WZ$ and $ZZ$)
can contribute to the tagged lepton+jets sample when one boson decays 
leptonically and the other decays into heavy quarks. The process 
$Z\rightarrow\tau^+\tau^-$ can also give a contribution due to the leptonic 
decays of the tau. Finally, there is a contribution from
single top quarks produced in association with a $b$ quark through
$q\bar q$ annihilation in $W^*$ (s-channel) or $W$-gluon fusion (t-channel),
in which an initial gluon splits into a $b\bar b$ pair and a b quark 
interacts with a virtual W.

The number of events from these processes are predicted based on their 
theoretical cross sections~\cite{bib:9,bib:9.5,bib:incW} (listed in Table~\ref{t:cs}), 
the measured integrated luminosity, and the acceptances and tagging 
efficiencies derived from Monte Carlo simulations.
The expectations for these backgrounds are corrected for differences 
between Monte Carlo simulations and data, which include the lepton identification scale 
factor, trigger efficiencies, the $z$ vertex cut efficiency
and the tagging scale factor.

The total diboson, $Z\rightarrow\tau^+\tau^-$ and single top predictions 
for $P_J<$ 1\% (5\%) are shown
in Table~\ref{t:backres1tag1jp} (Table~\ref{t:backres_1tag_5jp}) and account 
for 2.5\% (3.0\%) of the number of events in the signal region of 3 and 
$\geq$4 jets.
Following the same procedure, we also compute the electroweak 
background contributions to the pretag sample. The results are shown 
in Table~\ref{t:backres_pretag}.

\begin{table}[!htb] 
\begin{center}
\caption{\label{t:cs} Cross sections used to estimate electroweak
backgrounds. For diboson and single top production, the theoretical values 
are used. For $Z\rightarrow\tau^+\tau^-$, we use the cross section measured by 
{\CDF}.}
\begin{tabular}{lc}
\hline
\hline
 Process & Cross Section ($\pb$) \\
\hline
$WW$ & 13.25 $\pm$ 0.25 \\
$WZ$ & 3.96 $\pm$ 0.06 \\
$ZZ$ & 1.58 $\pm$ 0.02 \\
Single Top $W-g$ (t-channel) & 1.98 $\pm$ 0.08 \\
Single Top $W^*$ (s-channel) & 0.88 $\pm$ 0.05 \\
$Z\rightarrow \tau^+\tau^-$ & 254.3 $\pm$ 5.4 \\
\hline
\hline
\end{tabular}
\end{center}
\end{table}

\subsection{\label{sec:non-w}Non-W Background}

The non-$W$ background consists of events for which the
lepton+$\met$ signature is not due to the decay of a $W$ boson.
The main contribution to this source of background comes from {\QCD} jet
production where a jet provides the signature of a lepton and the missing 
transverse energy
is due to a bad measurement of the jet energies. Semileptonic decays of
$b$ mesons and misidentified photon conversions can also contribute. 
Due to its inherent instrumental nature, this background
is difficult to estimate. In the event selection, its contribution to
the lepton+jets sample is minimized by the requirement on the $W$ boson transverse mass $M_T^W$.
In particular, note that the optimization of this cut has been performed by 
requiring the lepton to be non-isolated in order to have an independent data sample 
to construct the kinematical variables (we use region C of Fig.~\ref{f:isoreg}).

The method used to estimate the non-$W$ background assumes that the isolation
of the high-$p_T$ lepton and the event $\met$ are uncorrelated for {\QCD} 
processes, so that the ratio of non-$W$ events with low lepton isolation to those with high lepton 
isolation in the region at low $\met$ is the same as in the high $\met$ region.
Four regions in the lepton isolation $versus$ missing transverse energy plane are 
defined (see Fig.~\ref{f:isoreg}):
\begin{itemize}
  \item{ Region A: Isolation $>$ 0.2 and $\met$ $<$ 15~$\Gev$}
  \item{ Region B: Isolation $<$ 0.1 and $\met$ $<$ 15~$\Gev$}
  \item{ Region C: Isolation $>$ 0.2 and $\met$ $>$ 20~$\Gev$}
  \item{ Region D: Isolation $<$ 0.1 and $\met$ $>$ 20~$\Gev$.}
\end{itemize}

\begin{figure}[!h]
\centering
    \includegraphics[width=8.7cm,clip=]{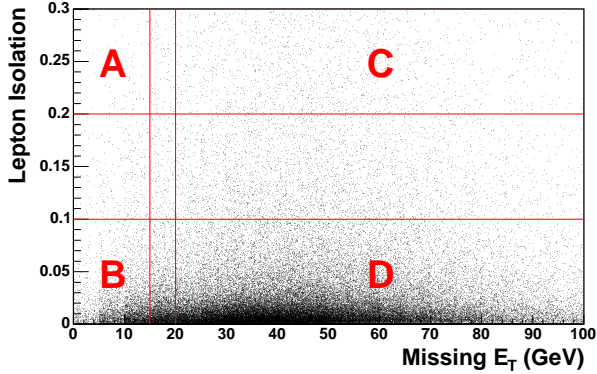}
\caption{Definition of the sideband regions used to estimate the non-$W$ background. Lepton isolation $versus$ missing transverse energy distribution for $t\bar t$ simulated events is also shown.}
\label{f:isoreg}
\end{figure}

The $t\bar t$ signal is expected to populate region D (signal region),
while the non-$W$ events dominate regions A, B and C (sideband regions).
We can therefore estimate the fraction of events in the signal region 
which originate from non-W backgrounds as follows:
  \begin{equation}\label{eq:nonwf}
     F_{non-W} = {N_B \times N_C \over N_A \times N_D},
  \end{equation}
where $N_A$, $N_B$, $N_C$ and $N_D$ are the total numbers of observed events
in the four regions.
We describe next the estimate of the non-$W$ events in both the pretag and 
tagged samples.

\subsubsection{Fraction of non-W Events in the Pretag Sample}\label{s:nonwfr}

An estimate of the contribution of the non-$W$ events to the pretag sample 
is mandatory to correctly normalize most of the backgrounds in the tagged 
sample. Table~\ref{t:frac} summarizes the results for the non-$W$ fractions in 
the pretag sample as a function of the jet multiplicity for electrons and 
muons. Note that we do not apply the $H_T$ and $M_T^W$ cuts in regions A and B 
to preserve statistics. We correct the yields in regions A, B and C by
subtracting the expected contribution from $t\bar t$ events 
assuming $\sigma_{t\bar t}=6.1~\pb$ (this assumption is found to 
have a negligible impact on the final non-$W$ estimate).
Uncertainties in Table~\ref{t:frac} are statistical only. The main source of
systematic uncertainty comes from the lepton isolation and missing 
transverse energy not being fully uncorrelated for {\QCD} events. We
study the effect of this assumption by varying the values of the $\met$ and 
lepton isolation cuts in the definition of the sideband regions. We observe 
a maximum variation of $50\%$ in the resulting non-$W$ fraction, which we
assign as a systematic uncertainty on our estimates.

\begin{table*}[!htb]
\begin{center}
\caption{\label{t:frac} Number of events in the sideband regions and 
                        fraction of non-$W$ events in the signal region
                        before and after correcting for $t\bar t$ 
                        contribution. Quoted errors are statistical only.}
\begin{tabular}{l@{\hspace{1cm}}c@{\hspace{1cm}}c@{\hspace{1cm}}c@{\hspace{1cm}}c}
\hline
\hline
Jet Multiplicity & 1 jet & 2 jets & 3 jets & $\ge$ 4 jets \\
\hline
\hline
 \multicolumn{5}{c}{                  Pretag Electrons    }       \\
\hline
Region A & 100600 & 12756 & 1745 & 216 \\
Region B & 61818 & 5228 & 593 & 98 \\
Region C & 1651 & 428 & 27 & 15 \\
Region D & 16897 & 2657 & 182 & 105 \\
F$_{non-W}^{uncorr.}$ & 0.060 $\pm$ 0.002 & 0.066 $\pm$ 0.004 & 0.05 $\pm$ 0.01 & 0.06 $\pm$ 0.02 \\
F$_{non-W}$           & 0.060 $\pm$ 0.002 & 0.066 $\pm$ 0.004 & 0.05 $\pm$ 0.01 & 0.05 $\pm$ 0.02 \\
\hline
\hline
            \multicolumn{5}{c}{                Pretag Muons         }      \\
\hline
Region A & 36599 & 5248 & 657 & 97 \\
Region B & 11718 & 968 & 114 & 21 \\
Region C & 737 & 181 & 12 & 11 \\
Region D & 12442 & 1785 & 118 & 61 \\
F$_{non-W}^{uncorr.}$ & 0.0190 $\pm$ 0.0007 & 0.019 $\pm$ 0.002 & 0.018 $\pm$ 0.006 & 0.04 $\pm$ 0.02 \\
F$_{non-W}$           & 0.0190 $\pm$ 0.0007 & 0.019 $\pm$ 0.002 & 0.014 $\pm$ 0.005 & 0.03 $\pm$ 0.01 \\
\hline
\hline
\end{tabular}
\end{center}
\end{table*}

To further cross check the accuracy of the predictions, 
we define new intermediate isolation regions B$^{\prime}$ and D$^{\prime}$:
\begin{itemize}
   \item{Region B: $\met< 15$, Isol$< 0.1\rightarrow$ Region B$^{\prime}$: $\met< 15$, 0.1$<$Isol$< 0.2$,}
   \item{Region D: $\met> 20$, Isol$< 0.1\rightarrow$ Region D$^{\prime}$: $\met> 20$, 0.1$<$Isol$< 0.2$.}
\end{itemize}
From the intermediate region B$^{\prime}$, we estimate the number of non-$W$ events
in region D$^{\prime}$. The predicted non-$W$ fractions are shown in Table~\ref{t:nonwcheck}. 
The uncertainties quoted are statistical only. 
In the same table, these fractions are compared with the expected
non-$W$  fractions computed from the difference between the observed
events and the contributions from $t\bar t$ and $W+$jets events.
The expected number of $t\bar t$ events is derived by normalizing the Monte 
Carlo prediction to a cross section of $6.1~\pb$. 
In order to estimate the $W+$jets contribution in region D$^{\prime}$, we 
compute the ratio of $W$+jets events in the regions D$^{\prime}$ and D using 
simulations and normalize the expectations for $W$+jets production in 
D$^{\prime}$ to the number of events in the signal region after removing 
$t\bar t$, electroweak contributions.
We compute the relative differences as the ratio of the difference between 
expected and predicted non-$W$ fractions to the predicted fraction.
For each jet multiplicity bin,
the differences between predicted and expected non-$W$ fractions in region 
D$^{\prime}$ are consistent with the $50\%$ uncertainty we derived varying the 
$\met$ and lepton isolation cuts in the definition of the sideband regions.

\begin{table*}[!htb]
\begin{center}
\caption{\label{t:nonwcheck} Predicted and expected fractions of non-$W$ 
events in the intermediate region D$^{\prime}$ for the electron and muon samples. 
Errors are statistical only.}
\begin{tabular}{l@{\hspace{1cm}}c@{\hspace{1cm}}c@{\hspace{1cm}}c@{\hspace{1cm}}c}
\hline
\hline
Jet Multiplicity & 1 jet & 2 jets & 3 jets & $\ge$ 4 jets \\
\hline
\hline
    \multicolumn{5}{c}{     Electron+Jets Sample    }                    \\
\hline
 Predicted Non-$W$ Fraction& 0.82 $\pm$ 0.03 & 0.60 $\pm$ 0.05 & 0.53 $\pm$ 0.16 & 0.49 $\pm$ 0.21 \\
 Expected Non-$W$ Fraction& 0.78 $\pm$ 0.01 & 0.82 $\pm$ 0.02 & 0.76 $\pm$ 0.10 & 0.59 $\pm$ 0.16 \\
 \hline
 Fractional Relative Difference  & -0.05 & 0.37 & 0.42 & 0.20 \\
\hline
\hline
   \multicolumn{5}{c}{ Muon+Jets Sample } \\
\hline
 Predicted Non-$W$ Fraction& 0.41 $\pm$ 0.03 & 0.41 $\pm$ 0.06 & 0.40 $\pm$ 0.23 & 0.40 $\pm$ 0.33 \\
 Expected Non-$W$ Fraction& 0.70 $\pm$ 0.02 & 0.62 $\pm$ 0.05 & 0.44 $\pm$ 0.22 & 0.26 $\pm$ 0.25 \\
 \hline
 Fractional Relative Difference  & 0.70 & 0.50 & 0.11 & -0.35 \\
 \hline
\hline
\end{tabular}
\end{center}
\end{table*}

\subsubsection{Non-W Events in the Tagged Sample}

The non-$W$ background contributes to the tagged sample through both
real heavy flavor production ($b\bar b$ and $c\bar c$ events) and 
mistags. We compute the number of non-$W$ events with tagged jets in the
signal region using equation~\ref{eq:nonwf} with the numbers of tagged events
in the sideband regions. Yields in regions A, B and C are corrected
for $t\bar t$ contributions.
The results are summarized in Table~\ref{t:restag}: $(N_B/N_A)^{tagged}$
is the ratio of tagged events in regions B and A and it is
used to normalize the number $N_C^{tagged}$ of tagged events in the region C 
to get the expected number $N^{tagged,l}_{non-W}$of non-W events on the signal region. 
The precision of these 
estimates is limited by the number of tagged events in the sideband regions.

\begin{table*}[!htb]
\begin{center}
\caption{\label{t:restag} The number of non-$W$ events in the signal region D 
estimated from the corrected numbers of tagged events in the sideband regions 
with equation~\ref{eq:nonwf}. Uncertainties are statistical only.}
\begin{tabular}{l@{\hspace{1cm}}c@{\hspace{1cm}}c@{\hspace{1cm}}c@{\hspace{1cm}}c}
\hline
\hline
Jet Multiplicity & 1 jet & 2 jets & 3 jets & $\ge$ 4 jets \\
\hline
\hline
               \multicolumn{5}{c}{        Electron+Jets Sample ($P_J<$ 1\%)      }       \\
\hline
$(N_B/N_A)^{tagged}$ & 0.36 $\pm$ 0.01 & 0.26$\pm$ 0.02& 0.26 $\pm$ 0.05 & 0.4 $\pm$ 0.2 \\
$N_C^{tagged}$ & 74.8 & 25.1 & 1.8 & 1.0 \\
$N^{tagged,e}_{non-W}$ & 26.7 $\pm$ 3.3 & 6.6 $\pm$ 1.4 & 0.5 $\pm$ 0.4 & 0.4 $\pm$ 0.4 \\
\hline
               \multicolumn{5}{c}{        Muon+Jets Sample ($P_J<$ 1\%)      }       \\
\hline
$(N_B/N_A)^{tagged}$ & 0.102 $\pm$ 0.008 & 0.10 $\pm$ 0.02 & 0.11 $\pm$ 0.04 & 0.2 $\pm$ 0.1 \\
$N_C^{tagged}$ & 36.9 & 20.3 & 4.0 & 0.81 \\
$N^{tagged,\mu }_{non-W}$ & 3.8 $\pm$ 0.7 & 2.0 $\pm$ 0.6 & 0.5 $\pm$ 0.3 & 0.2 $\pm$ 0.1 \\
\hline
\hline
               \multicolumn{5}{c}{        Electron+Jets Sample ($P_J<$ 5\%)      }       \\
\hline
$(N_B/N_A)^{tagged}$ & 0.42 $\pm$ 0.01 & 0.33 $\pm$ 0.02 & 0.29 $\pm$ 0.04 & 0.5 $\pm$ 0.1 \\
$N_C^{tagged}$ & 142.8 & 52.9 & 3.5 & 1.5 \\
$N^{tagged,e}_{non-W}$ & 59.6 $\pm$ 5.2 & 17.6 $\pm$ 2.6 & 1.0 $\pm$ 0.5 & 0.7 $\pm$ 0.6 \\
\hline
               \multicolumn{5}{c}{        Muon+Jets Sample ($P_J<$ 5\%)      }       \\
\hline
$(N_B/N_A)^{tagged}$ & 0.141 $\pm$ 0.007 & 0.12 $\pm$ 0.01 & 0.09 $\pm$ 0.03 & 0.17 $\pm$ 0.07 \\
$N_C^{tagged}$ & 65.8 & 32.1 & 3.7 & 0.6 \\
$N^{tagged,\mu }_{non-W}$ & 9.3 $\pm$ 1.2 & 3.8 $\pm$ 0.8 & 0.3 $\pm$ 0.2 & 0.1 $\pm$ 0.1 \\
\hline
\hline
\end{tabular}
\end{center}
\end{table*}

We cross check these results by estimating the non-$W$ contribution to
the tagged lepton+jets sample following two alternative methods. In the first
one ($check~1$), we assume the tag rate in region D to be the same as in region B:
\begin{equation}
   N_{non-W}^{tagged} = F_{non-W} \times N_D \times \epsilon_{B},
\end{equation}
where $\epsilon_{B}$ is the event tag rate in region B and $N_D$ is
the number of events in region D. 
This method has a large systematic uncertainty since the tag rate could depend 
on the missing transverse energy due to the contribution of $b\bar b$ events 
with a $b$ quark decaying into leptons. Events with large $\met$ would have a 
larger  heavy flavor contribution due to real neutrino production from semileptonic $b$ decay. 
In the second alternative method ($check~2$), we compute the tagging rates per jet in the
sideband regions, and then we predict the tag rate per jet in the signal region 
D as
\begin{equation}
   Pred ~Tag ~Rate ~D = {Tag ~Rate ~B \times Tag ~Rate ~C \over Tag ~Rate ~A}.
\end{equation}
We compute the jet tagging rate by assuming it to be the same in all the jet multiplicity bins 
and use this estimate to predict the non-$W$ background in the signal
region taking into account the jet multiplicity and the number of non-$W$ events expected in the 
pretag lepton+jets sample. Table~\ref{t:nonwtagsum} compares the non-$W$ 
contributions predicted by the three methods.

\begin{table*}[!htb]
\begin{center}
\caption{\label{t:nonwtagsum} Number of non-$W$ events expected in the tagged 
lepton+jets sample as a function of the jet multiplicity for the three methods 
described. Uncertainties are statistical only.}
\vspace{0.75ex}
\begin{tabular}{l@{\hspace{1cm}}c@{\hspace{1cm}}c@{\hspace{1cm}}c@{\hspace{1cm}}c}
\hline
\hline
Jet Multiplicity & 1 jet & 2 jets & 3 jets & $\ge$ 4 jets \\
\hline
\hline
               \multicolumn{5}{c}{        $P_J<$ 1\%      }       \\
\hline

$N^{tagged}_{non-W}$          & 30.5 $\pm$ 3.3 & 8.6 $\pm$ 1.5 & 0.9 $\pm$ 0.5 & 0.5 $\pm$ 0.4\\
$N^{tagged,~check~1}_{non-W}$ & 19.0 $\pm$ 0.7 & 6.7 $\pm$ 0.6 & 0.6 $\pm$ 0.1 & 0.6 $\pm$ 0.3 \\
$N^{tagged,~check~2}_{non-W}$ & 27.7 $\pm$ 2.6 & 9.3 $\pm$ 0.9 & 0.7 $\pm$ 0.1 & 0.6 $\pm$ 0.2 \\ 
\hline
\hline
               \multicolumn{5}{c}{        $P_J<$ 5\%      }       \\
\hline

$N^{tagged}_{non-W}         $ & 68.8 $\pm$ 5.4 & 21.4 $\pm$ 2.8 & 1.3 $\pm$ 0.6 & 0.8 $\pm$ 0.6 \\
$N^{tagged,~check~1}_{non-W}$ & 43.5 $\pm$ 1.3 & 16.2 $\pm$ 1.0 & 1.1 $\pm$ 0.2 & 1.6 $\pm$ 0.5 \\
$N^{tagged,~check~2}_{non-W}$ & 65.4 $\pm$ 4.4 & 21.9 $\pm$ 1.8 & 1.6 $\pm$ 0.3 & 1.5 $\pm$ 0.4  \\ 
\hline
\hline
\end{tabular}
\end{center}
\end{table*}

Finally, we  use the results of the two alternative estimates to assign a 
systematic uncertainty of 50\% which takes into account the differences with 
the base method.
The total non-$W$ background accounts for 1.2\% of the observed events with tagged
jets in the signal region, both for $P_J<$ 1\% and $P_J<$ 5\%. 

\subsection{\label{sec:whf}W + Heavy Flavor Processes}

$W$+heavy flavor production is the main source of background in the
tagged lepton+jets sample. It is estimated using the heavy flavor fractions in 
$W$ boson production in association with partons and 
the tagging efficiency for these processes. These quantities are derived from 
Monte Carlo simulations. The overall normalization is 
obtained from the number of observed events in the pretag sample.

The estimate of the heavy flavor fraction in $W$+jets events is described
elsewhere~\cite{bib:secvtx}. We use the {\ALPGEN} event generator, which is 
able to compute exact matrix element calculations at leading order for parton 
level {\QCD} and electroweak processes. 
We can therefore compute the ratio between the $W$+heavy flavor production 
cross section and the inclusive $W$+jets cross section since it is expected 
to be stable in the transition from leading-order to next-to-leading-order 
matrix elements.
We generate events where
inclusive $W$, $Wb\bar b$, $Wc\bar c$ and $Wc$ are produced in association 
with $n$ light partons. Parton level events from {\ALPGEN} are fed to the 
{\HERWIG} parton shower program which generates additional jets from
gluon radiation, and a full {\CDF} detector simulation is applied.
Events containing a different number $n$ of light partons are combined 
following a rigorous prescription in order to avoid double counting due to
parton shower radiation, which causes $W+n$ parton events to populate 
part of the phase space described by the $W+$($n+1$) parton sample.
The $Wb\bar b$ and $Wc\bar c$ samples are further divided 
into two classes according to the number of reconstructed heavy flavor jets 
in the event. We refer to these classes as 1B and 2B
(1C and 2C) for $Wb\bar b$ ($Wc\bar c$).
By means of these combined Monte Carlo simulated samples, the heavy 
flavor fractions for $W$+jets events are measured as the ratio between the 
computed $W$+heavy flavor and $W$+jets cross sections. Jet data samples are
used to correct for residual discrepancies between data and Monte Carlo 
simulations: a factor 1.5 $\pm$ 0.4 
is applied to the $Wb\bar b$ and $Wc\bar c$ fractions~\cite{bib:secvtx}, 
where the uncertainty is dominated by the systematic uncertainties associated 
with the {\ALPGEN} heavy flavor calculations.
The final heavy flavor fractions are shown in Table~\ref{t:whffrac}.

\begin{table*}[!htb]
\begin{center}
\caption{\label{t:whffrac} Summary of $Wb\bar b$, $Wc\bar c$ and $Wc$ 
fractions. 1B and 2B (1C and 2C) indicate the $Wb\bar b$ ($Wc\bar c$) events 
with one and two $b$-jets ($c$-jets) reconstructed, respectively. Uncertainties
are statistical only.}
\begin{tabular}{l@{\hspace{1.2cm}}c@{\hspace{1.4cm}}c@{\hspace{1.4cm}}c@{\hspace{1.4cm}}c}
\hline
\hline
Jet Multiplicity & 1 jet & 2 jets & 3 jets & $\geq$ 4 jets\\
\hline
 1B   & 0.010 $\pm$ 0.003 & 0.014 $\pm$ 0.004 & 0.024 $\pm$ 0.006 & 0.022 $\pm$ 0.006 \\ 
 2B   &         ---       & 0.014 $\pm$ 0.004 & 0.023 $\pm$ 0.006 & 0.026 $\pm$ 0.007 \\
 1C   & 0.016 $\pm$ 0.004 & 0.024 $\pm$ 0.006 & 0.038 $\pm$ 0.010 & 0.035 $\pm$ 0.010 \\ 
 2C   &         ---       & 0.018 $\pm$ 0.005 & 0.029 $\pm$ 0.008 & 0.037 $\pm$ 0.010 \\
\hline
 $Wc$ & 0.043 $\pm$ 0.009 & 0.060 $\pm$ 0.013   & 0.060 $\pm$ 0.013 & 0.059 $\pm$ 0.013 \\ 
\hline
\hline
\end{tabular}
\end{center}
\end{table*}

\begin{table*}[!htb]
\begin{center}
\caption{\label{t:whfeff} Jet probability tagging efficiencies for $Wb\bar b$, 
$Wc\bar c$ and $Wc$ events. The first uncertainty is statistical, 
while the second is systematic.}
\begin{tabular}{l@{\hspace{1cm}}c@{\hspace{1cm}}c@{\hspace{1cm}}c@{\hspace{1cm}}c}
\hline
\hline
Jet Multiplicity      &        1 jet    &    2 jet    &    3 jet    &    $\geq$ 4 jets \\
\hline
\hline
    \multicolumn{5}{c}{Event Tagging Efficiencies (\%), $P_J<$ 1\%}                    \\
\hline
1B ($\ge$1 tag) & 29.5 $\pm$ 0.3 $\pm$ 2.5 & 30.7 $\pm$ 0.6 $\pm$ 2.6 & 37.0 $\pm$ 1.5 $\pm$ 3.2 & 33.5 $\pm$ 3.2 $\pm$ 2.9 \\
2B ($\ge$1 tag) & --- & 50.5 $\pm$ 0.7 $\pm$ 4.3 & 56.0 $\pm$ 1.6 $\pm$ 4.8 & 54.6 $\pm$ 2.2 $\pm$ 4.7 \\
1C ($\ge$1 tag) & 6.8 $\pm$ 0.2 $\pm$ 0.9 & 7.8 $\pm$ 0.4 $\pm$ 1.0 & 9.1 $\pm$ 1.0 $\pm$ 1.2 & 8.0 $\pm$ 1.9 $\pm$ 1.0 \\
2C ($\ge$1 tag) & --- & 13.5 $\pm$ 0.6 $\pm$ 1.7 & 16.8 $\pm$ 1.5 $\pm$ 2.2 & 14.3 $\pm$ 1.7 $\pm$ 1.8 \\
Wc ($\ge$1 tag) & 7.2 $\pm$ 0.2 $\pm$ 0.9 & 7.9 $\pm$ 0.3 $\pm$ 1.0 & 8.3 $\pm$ 0.9 $\pm$ 1.1 & 6.9 $\pm$ 1.1 $\pm$ 0.9 \\
\hline
\hline
    \multicolumn{5}{c}{Event Tagging Efficiencies (\%), $P_J<$ 5\%}                    \\
\hline
1B ($\ge$1 tag) & 39.6 $\pm$ 0.3 $\pm$ 3.4 & 40.8 $\pm$ 0.6 $\pm$ 3.4 & 47.1 $\pm$ 1.6 $\pm$ 4.0 & 41.4 $\pm$ 3.4 $\pm$ 3.5 \\
2B ($\ge$1 tag) & --- & 63.5 $\pm$ 0.7 $\pm$ 5.4 & 68.4 $\pm$ 1.5 $\pm$ 5.8 & 66.7 $\pm$ 2.1 $\pm$ 5.6 \\
1C ($\ge$1 tag) & 14.8 $\pm$ 0.3 $\pm$ 1.9 & 17.2 $\pm$ 0.5 $\pm$ 2.2 & 18.7 $\pm$ 1.4 $\pm$ 2.4 & 16.3 $\pm$ 2.6 $\pm$ 2.1 \\
2C ($\ge$1 tag) & --- & 27.1 $\pm$ 0.8 $\pm$ 3.4 & 33.2 $\pm$ 1.9 $\pm$ 4.2 & 32.3 $\pm$ 2.3 $\pm$ 4.1 \\
Wc ($\ge$1 tag) & 15.3 $\pm$ 0.3 $\pm$ 1.9 & 16.4 $\pm$ 0.5 $\pm$ 2.1 & 19.4 $\pm$ 1.3 $\pm$ 2.5 & 18.6 $\pm$ 1.7 $\pm$ 2.4 \\
\hline
\hline
\end{tabular}
\end{center}
\end{table*}

The contribution of $W$+heavy flavor production to the pretag lepton+jets
sample is estimated by multiplying heavy flavor fractions by the observed 
number of events in the pretag sample, corrected for the non-$W$, 
and electroweak background expectations. The results are shown in 
Table~\ref{t:backres_pretag}. 

The above Monte Carlo simulated samples after pretag selection are used to compute 
the tagging efficiencies. In order to avoid double counting of the mistag
background, the {\JP} algorithm is applied only to jets
known to be due  
to a $b$ or $c$ quark. Each tagged jet is weighted according to
the scale factor. Results are summarized in Table~\ref{t:whfeff}. The 
systematic uncertainties are dominated by the uncertainties on the scale factor for $b$
and $c$ jets. The pretag expectations are multiplied by the 
tagging efficiencies to estimate the contributions of these processes to the tagged sample.
The numbers of $Wb\bar b$, $Wc\bar c$ and $Wc$ events expected in the tagged lepton+jets sample 
for $P_J<$ 1\% (5\%) are shown, along with the rest of the backgrounds, 
in Table~\ref{t:backres1tag1jp} (Table~\ref{t:backres_1tag_5jp}) and 
account for 12.3\% (13.2\%) of the observed number of events in the signal region.

\subsection{\label{sec:mistag}Mistag Background}

Events in which jets from light partons are tagged as heavy flavor jets can
contribute to the tagged sample.
The number of events with negative tags in the pretag sample would be a simple 
estimate of this background, but this method has the problem of a large statistical uncertainty. 
Instead, we count the events in the pretag sample weighted by their probability to have at
least one mistagged jet. This probability is computed by applying the negative tag 
rate matrix to all the taggable jets in the event. 

This estimate is corrected for the mistag asymmetry derived in Section~\ref{sec:misasy}. 
In order to take into account the dependence of the mistag asymmetry on the 
jet $E_T$, we convolute it with the jet $E_T$ spectra in events with $W$+three or more jets
for data and $t\bar t$ Monte Carlo simulation, 
as shown in Fig.~\ref{f:etdepjp1}. 
The observed difference between the means of the distributions and the mistag asymmetry values
measured in Section~\ref{sec:misasy} are negligible within the uncertainties. We therefore
decide to use the former in our analysis.
The RMS of the distributions gives an estimate of how much the asymmetry changes over the jets
in our samples,
and it is taken as an additional uncertainty on the mistag asymmetry. 
The final mistag asymmetry scale factors are 1.56 $\pm$ 0.17 for $P_J<$ 1\% and 1.27 $\pm$ 0.20 for $P_J<$ 5\%.

\begin{figure}[!htb]
   \centering 
    \includegraphics[width=8.5cm,clip=]{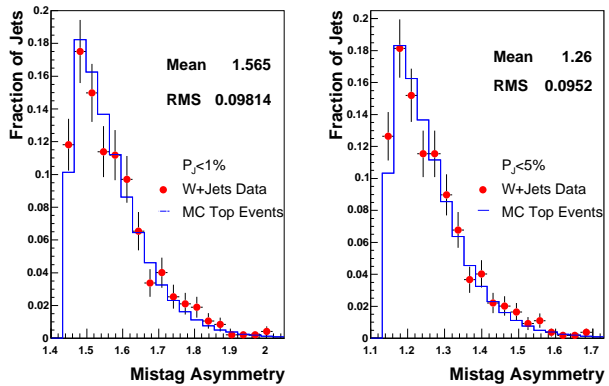}
\caption{Mistag asymmetry distribution for jets in data (dots) and 
         $t\bar t$ Monte Carlo simulated events (histogram) in the signal region for 
         $P_J<$ 1\% (5\%) in the left (right) side.}
\label{f:etdepjp1}
\end{figure}

The estimate of the mistag background is also scaled down 
by one minus the fraction of pretag events which are due to non-$W$, 
and electroweak backgrounds. The contribution of the 
mistag background to the lepton+jets sample when a jet with 
$P_J<$ 1\% (5\%) is required is shown in Table~\ref{t:backres1tag1jp} 
(Table~\ref{t:backres_1tag_5jp}) and accounts for 12.8\% (27.9\%) of the 
observed number of events in the signal region. 

\subsubsection{Mistag Cross Check}\label{s:mistag_crosscheck}

The negative tag rate matrix has been extensively tested on inclusive jet
samples. Results are discussed in Section~\ref{sec:jp_mistag}. The
mistag matrix is found to correctly predict the number of
events with negatively tagged jets observed in independent samples to 
within a few percent.
To further test the mistag matrix reliability on lepton+jets data,
we select a subsample of events by requiring $\met <20~\Gev$. This
sample is expected to be dominated by {\QCD} jet production with 
the high-$p_T$ lepton signature provided by a jet. Figure~\ref{f:qcdlike111}
compares the observed number of events with negative tags and the matrix 
prediction as a function of the jet multiplicity. Good agreement is
observed confirming the
reliability of the mistag matrix to predict the negative tag rate in 
events dominated by prompt jets.  
\begin{figure}[!htb]
    \includegraphics[width=8.5cm,clip=]{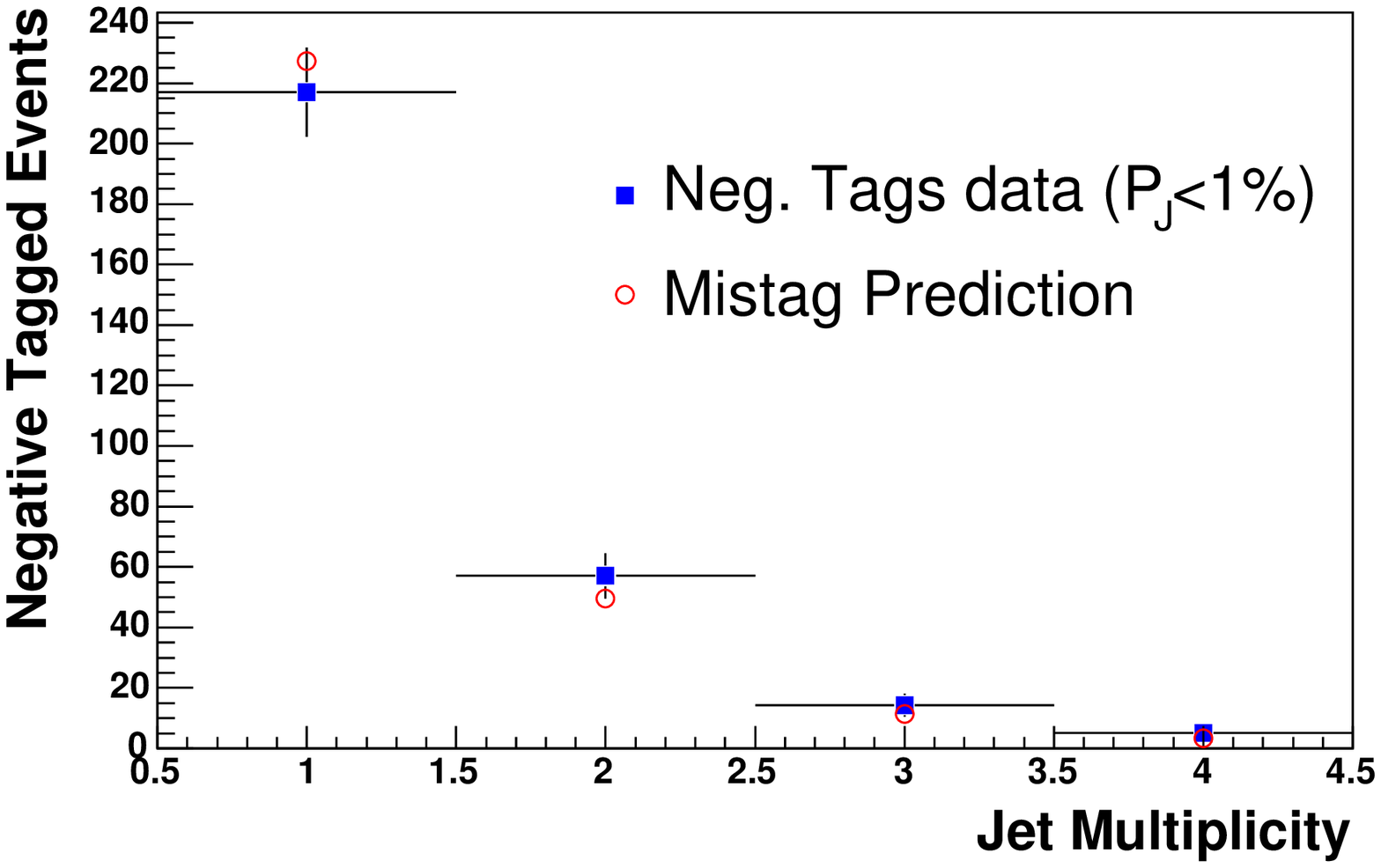}
    \includegraphics[width=8.5cm,clip=]{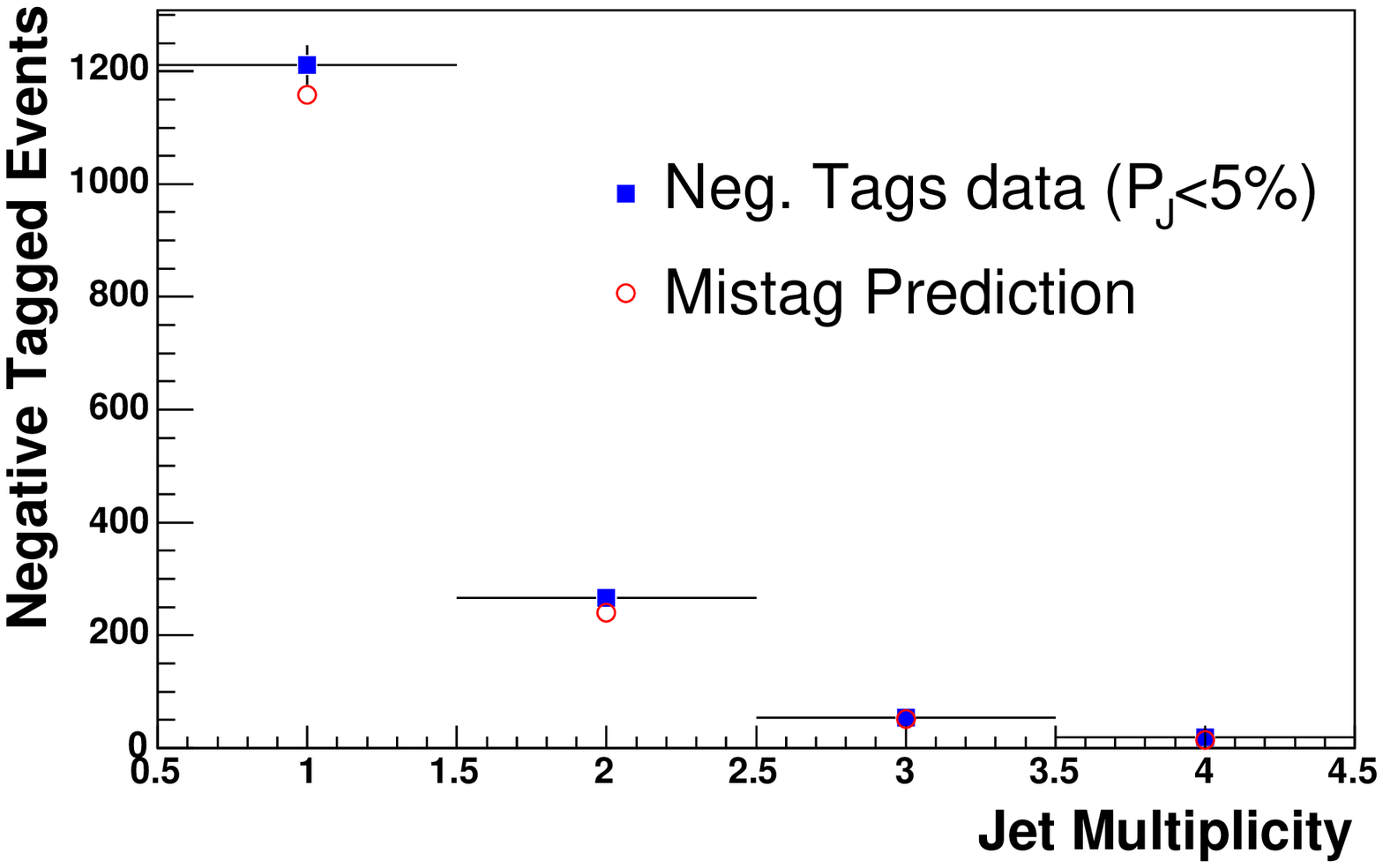}
\caption{Events with negative tagged jets compared to the
         prediction using the mistag matrix. We select events in the 
         high-$p_T$ lepton sample with $\met <20~\Gev$. 
         $P_J$ cuts of 1\% and 5\% are used on the top and bottom plots respectively.} 
\label{f:qcdlike111}
\end{figure}
We repeat the test on the pretag lepton+jets sample, where a $\met >20~\Gev$  
requirement is applied. Results are shown in Fig.~\ref{f:tagneg}.
We observe a discrepancy between observed and predicted negative tags 
which we attribute to the higher fraction of heavy flavor in lepton+jets
events with high value of $\met$ with respect to the inclusive jet samples 
where the matrix has been computed. 
To corroborate this hypothesis, we make a first-order correction to
the mistag prediction by using the heavy flavor fractions ($f_T$) in $W$+jets 
events
 (see Table~\ref{t:whffrac}). We compute the negative tag rates 
for light ($M_l$) and heavy ($M_h$) flavor jets in $t\bar t$ Monte Carlo 
simulation. 
For each jet multiplicity bin, a scale factor $R$ is then determined as:
\begin{eqnarray}
R&=&(1-\sum f_T) + \sum f_TC_T, \\T&=&1B,~2B,~1C,~2C,~Wc,
\end{eqnarray}
where 
\begin{equation}
C_T=\frac{1-(1-M_l)^{j-k}(1-M_h)^k}{1-(1-M_l)^{j}}.
\end{equation}
The numbers $j$ and $k$ in the formula are the jet multiplicity and the number 
of heavy flavor jets ($k=1$ for $T=1B$, $1C$ or $Wc$ and $k=2$ for $T=2B$ or 
$2C$), respectively. 
The corrected distributions, also shown in Fig.~\ref{f:tagneg}, show a much better agreement
with the observed rates of negative tagged events.
\begin{figure}[!htb]
    \includegraphics[width=8.5cm,clip=]{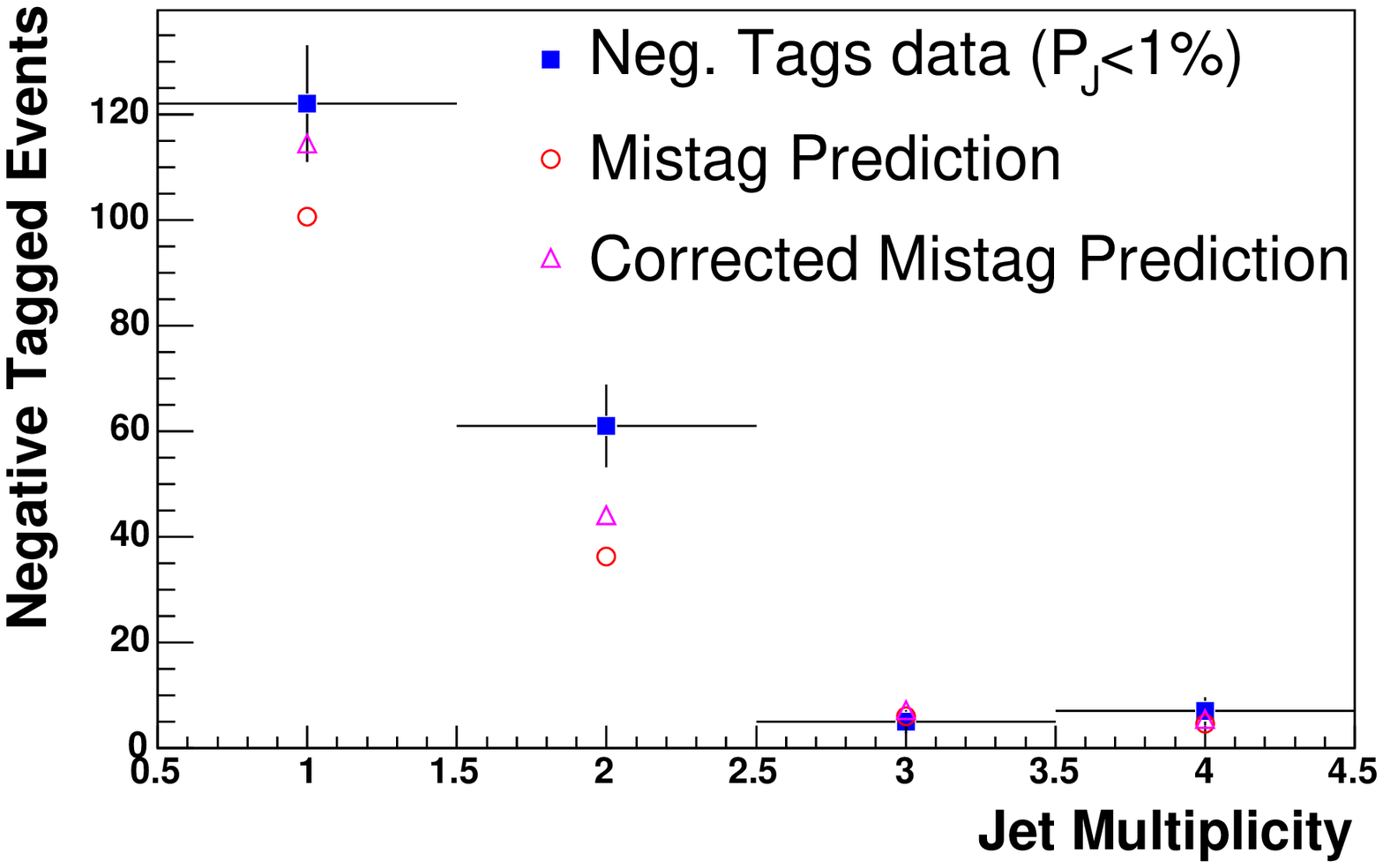}
    \includegraphics[width=8.5cm,clip=]{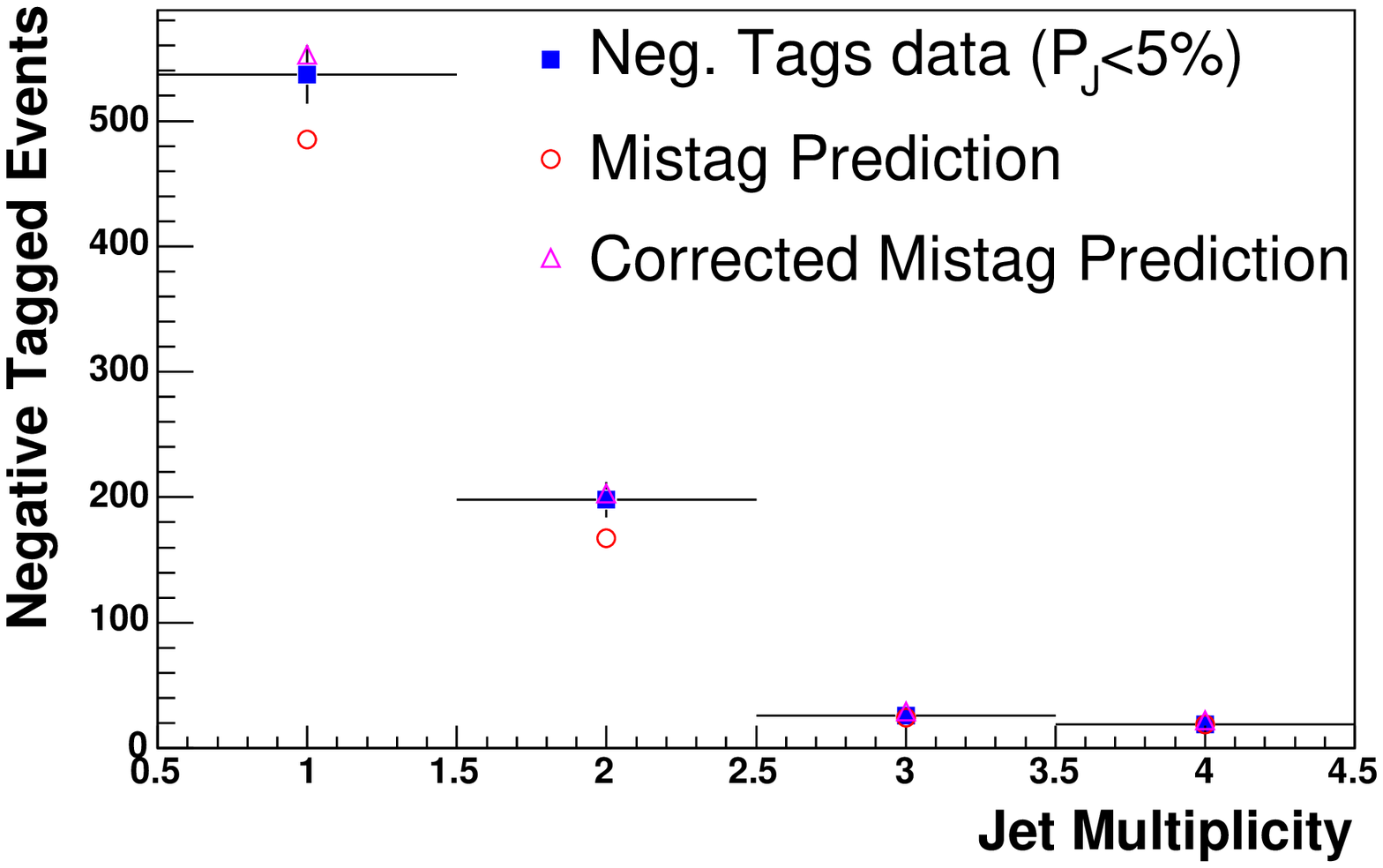}
\caption{Events with negative tagged jets (squares) in the pretag lepton+jets sample 
         compared to the prediction using the mistag matrix before 
         (empty circles) and after (empty triangles) heavy flavor corrections 
         for $P_J<$ 1\% (5\%) in the top (bottom) plot.} 
\label{f:tagneg}
\end{figure}
We therefore use the mistag matrix prediction corrected by the mistag 
asymmetry as an estimate of the number of events with a tagged light jet.

\subsection{\label{sec:summ}Background Summary}

Table~\ref{t:backres_pretag} summarizes the contributions of the different 
background estimates in the pretag sample. The difference between the observed 
number of events and the total background estimate is due to $W$+light flavor 
and $t\bar t$ contributions.

\begin{table*}[!htb]
\begin{center}
\caption{\label{t:backres_pretag}Summary of the background estimate in the 
pretag sample. The difference between the total background estimate and the 
observed number of events is due to $W$+light flavor and $t\bar t$ 
contributions.}
\begin{tabular}{l@{\hspace{1cm}}c@{\hspace{1cm}}c@{\hspace{1cm}}c@{\hspace{1cm}}c}
\hline
\hline
Jet Multiplicity & 1 jet & 2 jets & 3 jets & $\ge$ 4 jets \\
\hline
\hline
                 \multicolumn{5}{c}{    Electroweak}              \\
\hline
$WW$                             & 127 $\pm$ 9 & 123 $\pm$ 9 & 10.0 $\pm$ 0.8 & 3.6 $\pm$ 0.3 \\
$WZ$                             & 16.8 $\pm$ 1.2 & 18.8 $\pm$ 1.4 & 1.7 $\pm$ 0.1 & 0.61 $\pm$ 0.06\\
$ZZ$                             & 0.67 $\pm$ 0.05 & 0.68 $\pm$ 0.05 & 0.14 $\pm$ 0.02 & 0.052 $\pm$ 0.008 \\
Single Top $W-g$               & 13.4 $\pm$ 1.1 & 13.6 $\pm$ 1.1 & 1.44 $\pm$ 0.12 & 0.38 $\pm$ 0.04 \\
Single Top $W^*$               & 4.0 $\pm$ 0.4 & 7.9 $\pm$ 0.7 & 1.02 $\pm$ 0.10 & 0.24 $\pm$ 0.02 \\
$Z\rightarrow \tau^+\tau^-$    & 87 $\pm$ 7 & 16.5 $\pm$ 1.7 & 1.0 $\pm$ 0.3 & 0 $\pm$ 0 \\
\hline
Total                          & 249 $\pm$ 18 & 180 $\pm$ 13 & 15.3 $\pm$ 1.2 & 4.9 $\pm$ 0.4 \\
\hline
\hline
            \multicolumn{5}{c}{          W + Heavy Flavor }                   \\
\hline
$Wb\bar b$         & 281 $\pm$ 75 & 116 $\pm$ 31 & 12.9 $\pm$ 3.3 & 7.4 $\pm$ 2.0 \\
$Wc\bar c$         & 459 $\pm$ 123 & 170 $\pm$ 46 & 18.4 $\pm$ 5.0 & 11.1 $\pm$ 3.1 \\
$Wc$               & 1197 $\pm$ 252 & 243 $\pm$ 53 & 16.9 $\pm$ 3.6 & 9.1 $\pm$ 2.0 \\
\hline
Total              & 1938 $\pm$ 322 & 530 $\pm$ 94 & 47.8 $\pm$ 9.0 & 27.5 $\pm$ 5.5 \\
\hline
\hline
                  \multicolumn{5}{c}{    Others     }                        \\
\hline
Non-$W$                          & 1250 $\pm$ 626 & 208 $\pm$ 104 & 10.0 $\pm$ 5.3 & 7.3 $\pm$ 4.1 \\
\hline
\hline
Total Background               & 3436 $\pm$ 741 & 917 $\pm$ 150 & 73 $\pm$ 11 & 39.7 $\pm$ 7.2 \\
Data                           &                29339 &                4442 &                300 &        166 \\
\hline
\hline

\end{tabular}
\end{center}
\end{table*}

Table~\ref{t:backres1tag1jp} and Table~\ref{t:backres_1tag_5jp} summarize the contributions 
of the different background sources in the tagged lepton+jets sample for $P_J<$ 1\% and
$P_J<$ 5\% respectively. 

\begin{table*}[!htb]
\begin{center}
\caption{\label{t:backres1tag1jp}Summary of the background estimate in the lepton+jets sample when a jet with $P_J<$ 1\% is required.}
\begin{tabular}{l@{\hspace{1cm}}c@{\hspace{1cm}}c@{\hspace{1cm}}c@{\hspace{1cm}}c}
\hline
\hline
Jet Multiplicity & 1 jet & 2 jets & 3 jets & $\ge$ 4 jets \\
\hline
\hline
                 \multicolumn{5}{c}{    Electroweak}              \\
\hline
$WW$                          & 2.2 $\pm$ 0.3 & 5.0 $\pm$ 0.6 & 0.7 $\pm$ 0.1 & 0.28 $\pm$ 0.06 \\
$WZ$                          & 1.0 $\pm$ 0.1 & 2.0 $\pm$ 0.2 & 0.23 $\pm$ 0.03 & 0.09 $\pm$ 0.02 \\
$ZZ$                 & 0.027 $\pm$ 0.006 & 0.09 $\pm$ 0.01 & 0.012 $\pm$ 0.004 & 0.007 $\pm$ 0.002 \\
Single Top $W-g$               & 4.1 $\pm$ 0.5 & 4.9 $\pm$ 0.6 & 0.73$\pm$ 0.08 & 0.20 $\pm$ 0.02 \\
Single Top $W^*$               & 1.3 $\pm$ 0.2 & 4.2 $\pm$ 0.5 & 0.60 $\pm$ 0.07 & 0.14 $\pm$ 0.02 \\
$Z\rightarrow \tau^+\tau^-$    & 0.7 $\pm$ 0.3 & 0.4 $\pm$ 0.2 & 0.04 $\pm$ 0.04 & 0 $\pm$ 0 \\
\hline
Total                          & 9.3 $\pm$ 1.1 & 16.6 $\pm$ 1.8 & 2.3 $\pm$ 0.3 & 0.71 $\pm$ 0.09 \\
\hline
\hline
            \multicolumn{5}{c}{          W + Heavy Flavor }                   \\
\hline
$Wb\bar b$                     & 83 $\pm$ 23 & 47 $\pm$ 13 & 6.0 $\pm$ 1.6 & 3.3 $\pm$ 0.9 \\
$Wc\bar c$                     & 31 $\pm$ 9 & 17.5 $\pm$ 5.2 & 2.3 $\pm$ 0.7 & 1.2 $\pm$ 0.4 \\
$Wc$                           & 86 $\pm$ 21 & 19.2 $\pm$ 5.0 & 1.4 $\pm$ 0.4 & 0.6 $\pm$ 0.2 \\
\hline
Total                          & 200 $\pm$ 42 & 84 $\pm$ 20 & 9.6 $\pm$ 2.4 & 5.2 $\pm$ 1.4 \\
\hline
\hline
                  \multicolumn{5}{c}{    Others     }                        \\
\hline
Mistag                         & 149 $\pm$ 17  & 51.8 $\pm$ 5.9  & 8.5 $\pm$ 1.0  & 6.7 $\pm$ 0.8  \\
Non-$W$                          & 31 $\pm$ 16 & 8.6 $\pm$ 4.6 & 0.9 $\pm$ 0.6 & 0.5 $\pm$ 0.5 \\
\hline
\hline
Total Background               & 389 $\pm$ 49  & 161 $\pm$ 22  & 21.4 $\pm$ 2.7  & 13.1 $\pm$ 1.7  \\
Data                            &                350 &                191 &                52 &                68 \\
\hline
\hline
\end{tabular}
\end{center}
\end{table*}

\begin{table*}[!htb]
\begin{center}
\caption{\label{t:backres_1tag_5jp}Summary of the background estimate in the lepton+jets sample when a jet with $P_J<$ 5\% is required.}
\begin{tabular}{l@{\hspace{1cm}}c@{\hspace{1cm}}c@{\hspace{1cm}}c@{\hspace{1cm}}c}
\hline
\hline
Jet Multiplicity & 1 jet & 2 jets & 3 jets & $\ge$ 4 jets \\
\hline
\hline
                 \multicolumn{5}{c}{    Electroweak}              \\
\hline
$WW$                          & 5.5 $\pm$ 0.6 & 12.5 $\pm$ 1.4 & 1.81 $\pm$ 0.21 & 0.74 $\pm$ 0.10 \\
$WZ$                          & 1.6 $\pm$ 0.2 & 3.3 $\pm$ 0.3 & 0.40 $\pm$ 0.05 & 0.16 $\pm$ 0.02 \\
$ZZ$                 & 0.049 $\pm$ 0.009 & 0.14 $\pm$ 0.02 & 0.027 $\pm$ 0.006 & 0.014 $\pm$ 0.004 \\
Single Top $W-g$               & 5.4 $\pm$ 0.6 & 6.5 $\pm$ 0.7 & 0.92 $\pm$ 0.10 & 0.26 $\pm$ 0.03 \\
Single Top $W^*$               & 1.7 $\pm$ 0.2 & 5.2 $\pm$ 0.6 & 0.74 $\pm$ 0.08 & 0.17 $\pm$ 0.02 \\
$Z\rightarrow \tau^+\tau^-$ & 2.1 $\pm$ 0.5 & 1.1 $\pm$ 0.3 & 0.13 $\pm$ 0.10 & 0 $\pm$ 0 \\
\hline
Total                          & 16.3 $\pm$ 1.8 & 28.8 $\pm$ 3.0 & 4.0 $\pm$ 0.4 & 1.4 $\pm$ 0.1 \\
\hline
\hline
            \multicolumn{5}{c}{          W + Heavy Flavor }                   \\
\hline
$Wb\bar b$                     & 111 $\pm$ 31 & 61 $\pm$ 17 & 7.4 $\pm$ 2.0 & 4.1 $\pm$ 1.2 \\
$Wc\bar c$                     & 68 $\pm$ 20 & 36 $\pm$ 11 & 4.6 $\pm$ 1.4 & 2.7 $\pm$ 0.8 \\
$Wc$                           & 184 $\pm$ 45 & 40 $\pm$ 10 & 3.2 $\pm$ 0.8 & 1.7 $\pm$ 0.5 \\
\hline
Total                          & 363 $\pm$ 75 & 137 $\pm$ 31 & 15.2 $\pm$ 3.6 & 8.5 $\pm$ 2.1 \\
\hline
\hline
                  \multicolumn{5}{c}{    Others     }                        \\
\hline
Mistag                         & 585 $\pm$ 92  & 194 $\pm$ 30  & 28.2 $\pm$ 4.4  & 22.1 $\pm$ 3.5  \\
Non-$W$                          & 69 $\pm$ 35 & 21 $\pm$ 11 & 1.3 $\pm$ 0.9 & 0.79 $\pm$ 0.74 \\
\hline
\hline
Total Background               & 1033 $\pm$ 125 & 381 $\pm$ 46  & 48.8 $\pm$ 5.9  & 32.7 $\pm$ 4.2  \\
Data                            &                975 &                385 &                87 &            93 \\
\hline
\hline
\end{tabular}
\end{center}
\end{table*}
 
We observe good agreement between data and background predictions in events 
with one and two jets, which supports the validity
of our background estimates. In events with three or more jets, we observe
an excess of tagged events in data which we attribute to $t\bar t$ events.
The estimates of the $W$+heavy flavor and mistag background contributions
have been normalized to the data in the pretag sample assuming no signal. 
Having actually observed a significant number of $t\bar t$ events in the tagged sample,
we need to correct those estimates by the number of signal events in the
pretag sample. We make this correction through an iterative procedure 
which is described in Section~\ref{sec:single_jp}.

\section{\label{sec:acc}Signal Acceptance}

The signal acceptance, or $t\bar t$ event detection efficiency, is defined as 
the fraction of $t\bar t$ events that satisfy all selection 
requirements, and includes trigger and reconstruction efficiencies as well as 
the efficiencies of the kinematic selection and of the $b$-tagging algorithm. 
We measure it using a {\PYTHIA} $t\bar t$ Monte Carlo sample 
generated with a top quark mass $m_t = 178~\Gevcc$ and simulated as discussed in
Section~\ref{sec:data}. Wherever possible, effects which are not sufficiently well 
modeled in the simulation are measured using data.
The acceptance is defined as
\begin{align}
  \epsilon_{t\bar t} &= (A_{t\bar t} \times K_{lep}\times \epsilon_{trig}
  \times \epsilon_{z_0} \times \epsilon_{zvtx}) 
  \times \epsilon_{b-tag} \nonumber\\&= \epsilon_{t\bar t}^{\rm pretag} \times \epsilon_{b-tag} 
\end{align}
where $A_{t\bar t}$ is the fraction of Monte Carlo simulated $t\bar t$ events which pass 
the kinematic requirements (except $b$-tagging) and includes the branching 
fraction for $t\bar t\rightarrow e/\mu +jets$, the lepton identification 
efficiency (including isolation and cosmic/conversion veto efficiency, as described in 
Section~\ref{sec:data}), 
the dilepton and $Z^0\rightarrow l^+l^-$ veto efficiencies, 
and the kinematic and geometric acceptances. $A_{t\bar t}$ is measured 
separately for electron and muon events. 
$K_{lep}$ is a scale factor which takes into account the difference in lepton 
identification efficiency between data and Monte Carlo simulations estimated using 
$Z \rightarrow l^+l^-$ events;
$\epsilon_{trig}$ is the trigger efficiency for identifying high $p_{T}$ 
leptons and is measured using data from independent triggers. Both $K_{lep}$ and
$\epsilon_{trig}$ are discussed in Section~\ref{sec:data}.
$\epsilon_{z_0}$ and
$\epsilon_{zvtx}$ are the efficiencies for the $z$ vertex cuts described in 
Section~\ref{sec:evt_sel} and
$\epsilon_{b-tag}$ is the efficiency to tag at least one tight jet in a 
$t\bar t$ event and includes a tagging scale factor to account for differences between Monte Carlo simulations and data. 
\begin{table*}[!htb]
\begin{center}
\caption{\label{t:acceff} Summary of acceptances for $t\bar t$ events. 
Efficiencies are expressed as percentages. The average $\epsilon_{b-tag}$ is the
luminosity-weighted {\CEM}/{\CMUP}/{\CMX} tagging efficiency. First uncertainty is statistical and the second one corresponds to systematics.}
\begin{tabular}{l@{\hspace{1.4cm}}c@{\hspace{1.4cm}}c@{\hspace{1.4cm}}c}
\hline
\hline
Quantity    &   {\CEM}         &    {\CMUP}   &    {\CMX}   \\
\hline
\hline
$\epsilon_{t\bar t}^{\rm pretag}$ & 3.67 $\pm$ 0.02 $\pm$ 0.22 & 1.92 $\pm$ 0.01 $\pm$ 0.12 & 0.751 $\pm$ 0.008 $\pm$ 0.046 \\
$\int L dt$ ($\pb^{-1}$) & 318 $\pm$ 19 & 318 $\pm$ 19 & 305 $\pm$ 18 \\
\hline
\hline
     \multicolumn{4}{c}{Single Tag, $P_J<$ 1\%, $SF$ = 0.817 $\pm$ 0.070}   \\
\hline
$\epsilon_{b-tag}$ & 54.7 $\pm$ 0.2 $\pm$ 3.6 & 54.1 $\pm$ 0.3 $\pm$ 3.5 & 55.2 $\pm$ 0.5 $\pm$ 3.6 \\
\hline
Average $\epsilon_{b-tag}$ &  \multicolumn{3}{c}{  54.5 $\pm$ 0.2 $\pm$ 3.6} \\
\hline
$\epsilon_{t\bar t}$ & 2.00 $\pm$ 0.01 $\pm$ 0.18 & 1.04 $\pm$ 0.01 $\pm$ 0.09 & 0.41 $\pm$ 0.01 $\pm$ 0.04 \\
$\epsilon_{t\bar t}\int L dt$ ($\pb^{-1}$) & 6.38 $\pm$ 0.04 $\pm$ 0.68 & 3.30 $\pm$ 0.03 $\pm$ 0.36 & 1.32 $\pm$ 0.02 $\pm$ 0.14 \\
\hline
Total $\epsilon_{t\bar t}\int L dt$  &  \multicolumn{3}{c}{ $11.00 \pm 0.05 ({\rm stat}) \pm 1.17 ({\rm syst})$~pb$^{-1}$} \\
\hline
\hline
     \multicolumn{4}{c}{Single Tag, $P_J<$ 5\%, $SF$ = 0.852 $\pm$ 0.072}   \\
\hline
$\epsilon_{b-tag}$ & 68.8 $\pm$ 0.2 $\pm$ 3.7 & 68.6 $\pm$ 0.3 $\pm$ 3.7 & 69.6 $\pm$ 0.5 $\pm$ 3.7 \\
\hline
Average $\epsilon_{b-tag}$ &  \multicolumn{3}{c}{  68.8 $\pm$ 0.2 $\pm$ 3.7} \\
\hline
$\epsilon_{t\bar t}$ & 2.52 $\pm$ 0.01 $\pm$ 0.20 & 1.315 $\pm$ 0.009 $\pm$ 0.108 & 0.523 $\pm$ 0.006 $\pm$ 0.042 \\
$\epsilon_{t\bar t}\int L dt$ ($\pb^{-1}$) & 8.03 $\pm$ 0.05 $\pm$ 0.80 & 4.19 $\pm$ 0.03 $\pm$ 0.42 & 1.67 $\pm$ 0.02 $\pm$ 0.17 \\
\hline
Total $\epsilon_{t\bar t}\int L dt$  &  \multicolumn{3}{c}{ $13.89 \pm 0.06 ({\rm stat}) \pm 1.38 ({\rm syst})$~pb$^{-1}$} \\
\hline
\hline
\end{tabular}
\end{center}
\end{table*}

The event tagging efficiency $\epsilon_{b-tag}$ is obtained from the same
$t\bar t$
Monte Carlo simulated sample. We compute, for each $t\bar t$ event, the probability of
having $n$ tagged jets in the event by assigning to each jet a probability to
be tagged. The sum of these probabilities over all the events 
returns the number of expected events with at least $n$ tags, from which we 
calculate the tagging efficiency.
For light flavor jets, this probability is computed using the mistag matrix,
while for heavy flavor jets the probability is the value
of the tagging scale factor (see Section~\ref{sec:jp}) if the jet is tagged 
and zero otherwise. 
We estimate the systematic uncertainty on the event tagging efficiency by 
varying the tagging scale factor and mistag prediction by $\pm 1 \sigma$.

Table~\ref{t:acceff} summarizes the acceptance for $t\bar t$ events. 
For $P_J$ cuts of 1\% and 5\%, 
the combined acceptance times integrated luminosity are, respectively, 
$11.00 \pm 0.05 ({\rm stat}) \pm 1.17 ({\rm syst})$~pb$^{-1}$ and
$13.89 \pm 0.06 ({\rm stat}) \pm 1.38 ({\rm syst})$~pb$^{-1}$, where the
statistical uncertainty is uncorrelated between the lepton types, and the 
systematic uncertainty is assumed to be 100\% correlated since it is dominated
by the luminosity and tagging scale factor uncertainties.

Table~\ref{t:accsyshp} summarizes the contributions to the systematic 
uncertainty
on the signal acceptance. Trigger, lepton identification and $z$ vertex cuts
have already been discussed in Sections~\ref{sec:data} and~\ref{sec:evt_sel}.
The observed difference in the conversion veto efficiency between $t\bar t$ 
events and the $Z\rightarrow e^+e^-$ sample used to measure the electron 
identification scale factor is added as an uncertainty 
on the tight electron identification efficiency.
The efficiency of the cosmic ray veto is measured from data and accounts for 
a 1\% uncertainty on the tight muon identification efficiency.
Additional uncertainties in the electron (muon) acceptance are due to $E_T$ 
($p_T$) scale, $E_T$ ($p_T$) resolution and material (geometrical) effects, 
and are found to be 0.3\% (1.2\%) in inclusive $W$ events. The lepton isolation
uncertainty accounts for differences in the modeling of the lepton 
identification in events with different jet multiplicity. It has been evaluated
by comparing data to Monte Carlo simulations for $W$+jets and $t\bar t$ events.
The uncertainty due to the jet energy scale is estimated by the shift in signal
acceptance observed by changing the jet energy corrections within their 
uncertainties.
The uncertainty due to parton distribution functions ({\PDF}) is estimated 
by re-weighting the $t\bar t$ events generated with {\CTEQcL} for different 
sets of {\PDF}s~\cite{bib:cteq5l}. In particular, we consider the difference in signal acceptance
between {\NLOCTEQsM} and {\CTEQcL}, between {\MRST} for two different values of
$\alpha_S$, and between {\NLOCTEQsM} and the 20 {\CTEQ} eigenvectors, and we 
add in quadrature all the contributions.
Differences in the modeling of $t\bar t$ production and decay are evaluated 
as the difference in acceptance between samples of signal events generated with {\HERWIG}
and {\PYTHIA}. Samples of $t\bar t$ 
events with different levels of initial and final state radiation ({\ISR/\FSR}) 
are used to evaluate the effect of this source of uncertainty on the signal 
acceptance. 
The systematic uncertainty on the event tagging efficiency is estimated by 
varying the tagging scale factor and the mistag prediction by $\pm$ 1$\sigma$.
The total systematic uncertainty on the signal acceptance is 8.9\% (8.0\%) for
$P_J<$ 1\% (5\%), and is dominated by the tagging scale factor and the jet 
energy scale uncertainties.
\begin{table*}[!htb]
\begin{center}
\caption{\label{t:accsyshp} Summary of the systematic uncertainties on the 
signal acceptance. The second column quotes the relative uncertainty on the 
indicated quantities, while the third column shows the effect on the overall 
$t\bar t$ acceptance.}
\begin{tabular}{l@{\hspace{1cm}}c@{\hspace{1cm}}c}
\hline
\hline
Source   & Relative Uncertainty (\%) & Uncertainty on the Acceptance (\%) \\
\hline 
Central Electron Trigger     & 0.6 & 0.3 \\
Central Electron ID $SF$       & 0.5 & 0.3 \\
Conversion Veto Eff.         & 1.4 & 0.8 \\
$E_T$ Scale of Electron      & 0.3 & 0.2 \\
Central Muon Trigger         & 0.5 & 0.2 \\
Central Muon ID $SF$           & 1.0 & 0.3 \\
{\CMX} Muon Trigger          & 0.4 & 0.05 \\
{\CMX} Muon ID $SF$            & 0.6 & 0.07 \\
Cosmic Veto Eff.             & 1.0 & 0.4 \\
$p_T$ Scale of Muon          & 1.2 & 0.5 \\
Lepton Isolation             & 2.0 & 2.0 \\
$|Z_{vtx}|$ Cut Eff.         & 0.3 & 0.3 \\
$Z_{vtx}^{JetProb}$ Cut Eff. & 2 & 2 \\
Jet Energy Scale             & --- & 4.2 \\
\PDF                         & --- & 2   \\
MC Modeling                  & --- & 1.6 \\
\ISR/\FSR                      & --- & 1.3 \\
Tagging $SF$ $P_J<$ 1\% (b's/c's) & 8.6/12.9 & 6.5 \\
Mistag Asymmetry $P_J<$ 1\%     & 11.0     & 0.2 \\ 
Tagging $SF$ $P_J<$ 5\% (b's/c's) & 8.5/12.7 & 5.4 \\
Mistag Asymmetry $P_J<$ 5\%     & 15.5     & 0.4 \\ 
\hline
Total Uncertainty  ($P_J<$ 1\%)           & ---  & 8.9 \\ 
Total Uncertainty  ($P_J<$ 5\%)           & ---  & 8.0 \\ 
\hline
\hline
\end{tabular}
\end{center}
\end{table*}

\section{\label{sec:single_jp}Cross Section for Single Tagged Events}

We measure the cross section as
\begin{equation}
  \sigma_{t\bar t} = {N_{obs} - N_{bck} \over \epsilon_{t\bar t} \times \int L~dt},
\end{equation}
where $N_{obs}$ is the observed number of events with at least one jet tagged, $N_{bck}$ is the
background estimate in the signal region, $\epsilon_{t\bar t}$ is the signal
acceptance including the tagging efficiency and $\int L~dt$ is the integrated 
luminosity.
The estimated number of background events must be corrected for the $t\bar t$ 
contribution, since we normalize mistag and $W$+heavy flavor backgrounds  
assuming no $t\bar t$ signal events in the pretag sample. We apply an iterative
procedure in which we first estimate the number of tagged top candidates in the
sample as the number of tagged signal events minus the total background in the
$\geq$ 3 jet bins. Successively, the obtained signal cross section is used to 
estimate the number of $t\bar t$ events before the $b$-tagging requirement, and 
this contribution is subtracted from the total number of events to which we 
normalize the mistag, $Wb\bar b$, $Wc\bar c$ and $Wc$ backgrounds.  The 
expectations for single top, diboson and 
Z$\rightarrow$$\tau^+$$\tau^-$ do not change with the number of 
$t\bar t$ events in the signal region. The change for non-$W$ background is 
found to be negligible compared to its uncertainty. Therefore, this background 
is also kept fixed.
Having obtained a new estimate for the tagged background, we re-evaluate the 
number of $t\bar t$ candidates. The procedure is repeated until the cross 
section $\sigma_{t\bar t}$ changes by less than 0.1\%. 

Starting with the backgrounds shown in Tables~\ref{t:backres1tag1jp} and~\ref{t:backres_1tag_5jp},
we apply the iterative procedure and measure
\begin{displaymath}
  \sigma_{t\bar t} = 8.9^{+1.0}_{-1.0} \rm{(stat.)}^{+1.1}_{-1.0}\rm{(syst.)}~\rm{pb}
\end{displaymath}
for $P_J<$ 1\%. As a cross check, we apply the iterative procedure for $P_J<$ 5\% and measure
\begin{displaymath}
  \sigma_{t\bar t} = 9.6^{+1.0}_{-0.9} \rm{(stat.)}^{+1.2}_{-1.1}\rm{(syst.)}~\rm{pb}.
\end{displaymath}
The final signal and background estimates are shown in Table~\ref{t:backres1tag1jp_after}, together
with the observed number of events. 

\begin{table*}[!htb]
\begin{center}
\caption{\label{t:backres1tag1jp_after}Summary of the final signal and 
background estimates and observed data in the single tag sample.}
\begin{tabular}{l@{\hspace{1.6cm}}c@{\hspace{1.6cm}}c@{\hspace{1.6cm}}c@{\hspace{1.6cm}}c}
\hline
\hline
Jet Multiplicity & 1 jet & 2 jets & 3 jets & $\ge$ 4 jets \\
\hline
\hline
Pretag Data      &                29339 &                4442 &                300 &        166 \\
\hline
\hline
                 \multicolumn{5}{c}{$P_J<$ 1\%}              \\
\hline
Electroweak                     & 9.3 $\pm$ 1.1 & 16.6 $\pm$ 1.8 & 2.3 $\pm$ 0.3 & 0.71 $\pm$ 0.09 \\
$Wb\bar b$                     & 83 $\pm$ 23 & 47 $\pm$ 13 & 4.3 $\pm$ 1.2 & 1.1 $\pm$ 0.3 \\
$Wc\bar c$                     & 31 $\pm$ 9 & 17.3 $\pm$ 5.2 & 1.6 $\pm$ 0.5 & 0.4 $\pm$ 0.1 \\
$Wc$                           & 86 $\pm$ 21 & 19.0 $\pm$ 4.9 & 1.0 $\pm$ 0.3 & 0.21 $\pm$ 0.06 \\
Mistag                         & 149 $\pm$ 17  & 51 $\pm$ 6  & 6.1 $\pm$ 0.7  & 2.2 $\pm$ 0.3  \\
Non-$W$                          & 31 $\pm$ 16 & 8.6 $\pm$ 4.6 & 0.9 $\pm$ 0.6 & 0.5 $\pm$ 0.5 \\
\hline
Total Background               & 389 $\pm$ 49 & 159 $\pm$ 22 & 16.3 $\pm$ 2.0  & 5.1 $\pm$ 0.7  \\
$t\bar t$ (8.9 pb)       & 2.5 $\pm$ 0.5 & 20.6 $\pm$ 2.4 & 40.4 $\pm$ 4.5 & 58.1 $\pm$ 6.2 \\
Data                            &                350 &                191 &                52 &    68 \\
\hline
\hline
                 \multicolumn{5}{c}{$P_J<$ 5\%}              \\
 
\hline
Electroweak                          & 16.3 $\pm$ 1.8 & 28.8 $\pm$ 3.0 & 4.0 $\pm$ 0.4 & 1.4 $\pm$ 0.1 \\
$Wb\bar b$                     & 111 $\pm$ 31 & 60 $\pm$ 17 & 5.2 $\pm$ 1.4 & 1.1 $\pm$ 0.3 \\
$Wc\bar c$                     & 68 $\pm$ 20 & 36 $\pm$ 11 & 3.2 $\pm$ 1.0 & 0.76 $\pm$ 0.24 \\
$Wc$                           & 184 $\pm$ 45 & 40 $\pm$ 10 & 2.2 $\pm$ 0.6 & 0.5 $\pm$ 0.13 \\
Mistag                         & 585 $\pm$ 92  & 191 $\pm$ 30  & 19.6 $\pm$ 3.1  & 6.1 $\pm$ 1.0  \\
Non-$W$                          & 69 $\pm$ 35 & 21 $\pm$ 11 & 1.3 $\pm$ 0.9 & 0.8 $\pm$ 0.7 \\
\hline
Total Background               & 1033 $\pm$ 125  & 377 $\pm$ 46  & 35.5 $\pm$ 4.2  & 10.6 $\pm$ 1.4  \\
$t\bar t$ (9.6 pb)       & 3.6 $\pm$ 0.6 & 28.4 $\pm$ 3.1 & 55.1 $\pm$ 5.7 & 78.6 $\pm$ 7.8 \\
Data                            &                975 &                385 &                87 &    93 \\
\hline
\hline
\end{tabular}
\end{center}
\end{table*}

Figure~\ref{f:backplot} compares the numbers of observed data to background and signal expectations, 
for $P_J<$ 1\% and 5\%, for the measured $t\bar t$ production cross sections. 
\begin{figure}[!htb]
    \includegraphics[width=0.49\textwidth,clip=]{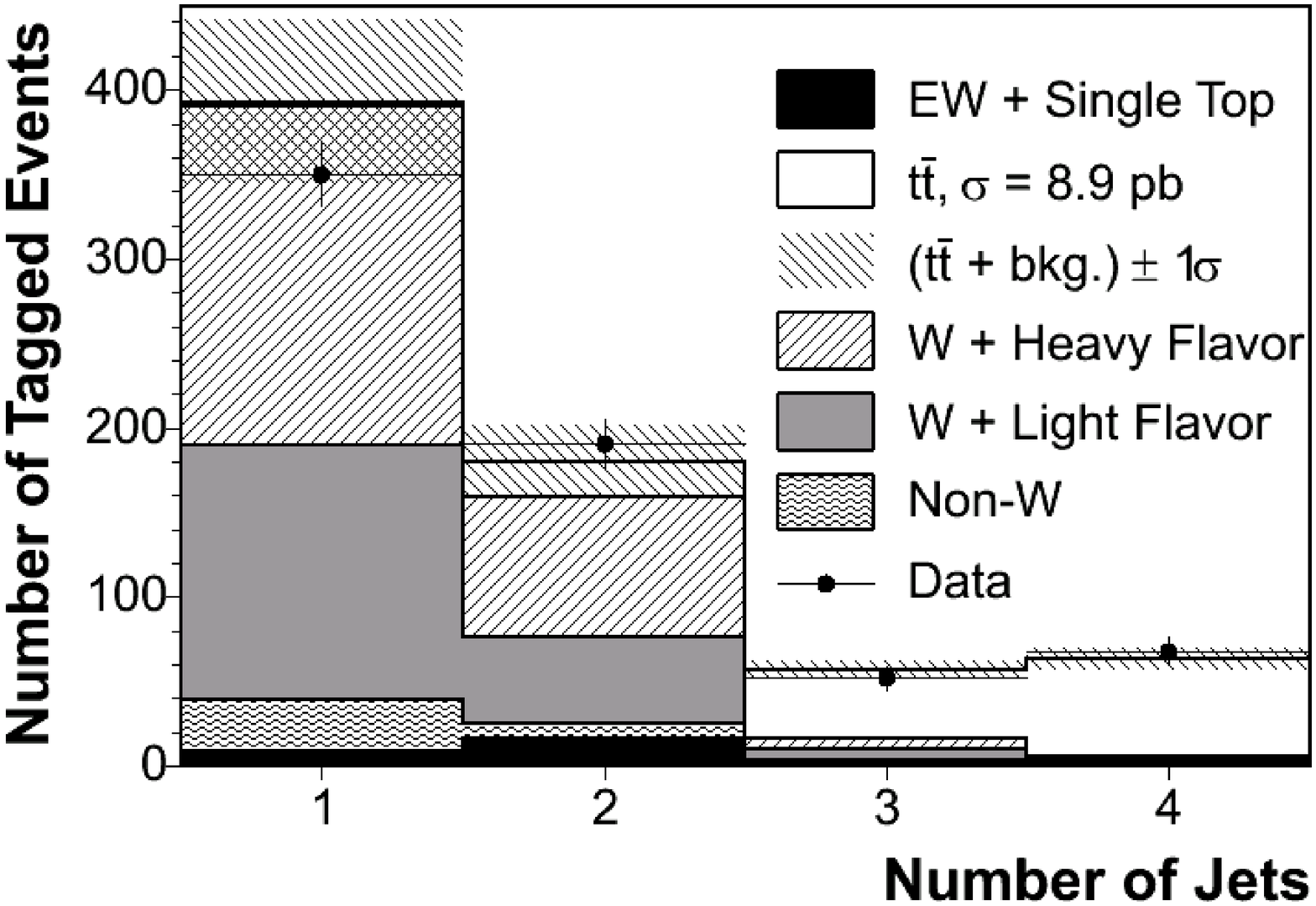}
    \includegraphics[width=0.49\textwidth,clip=]{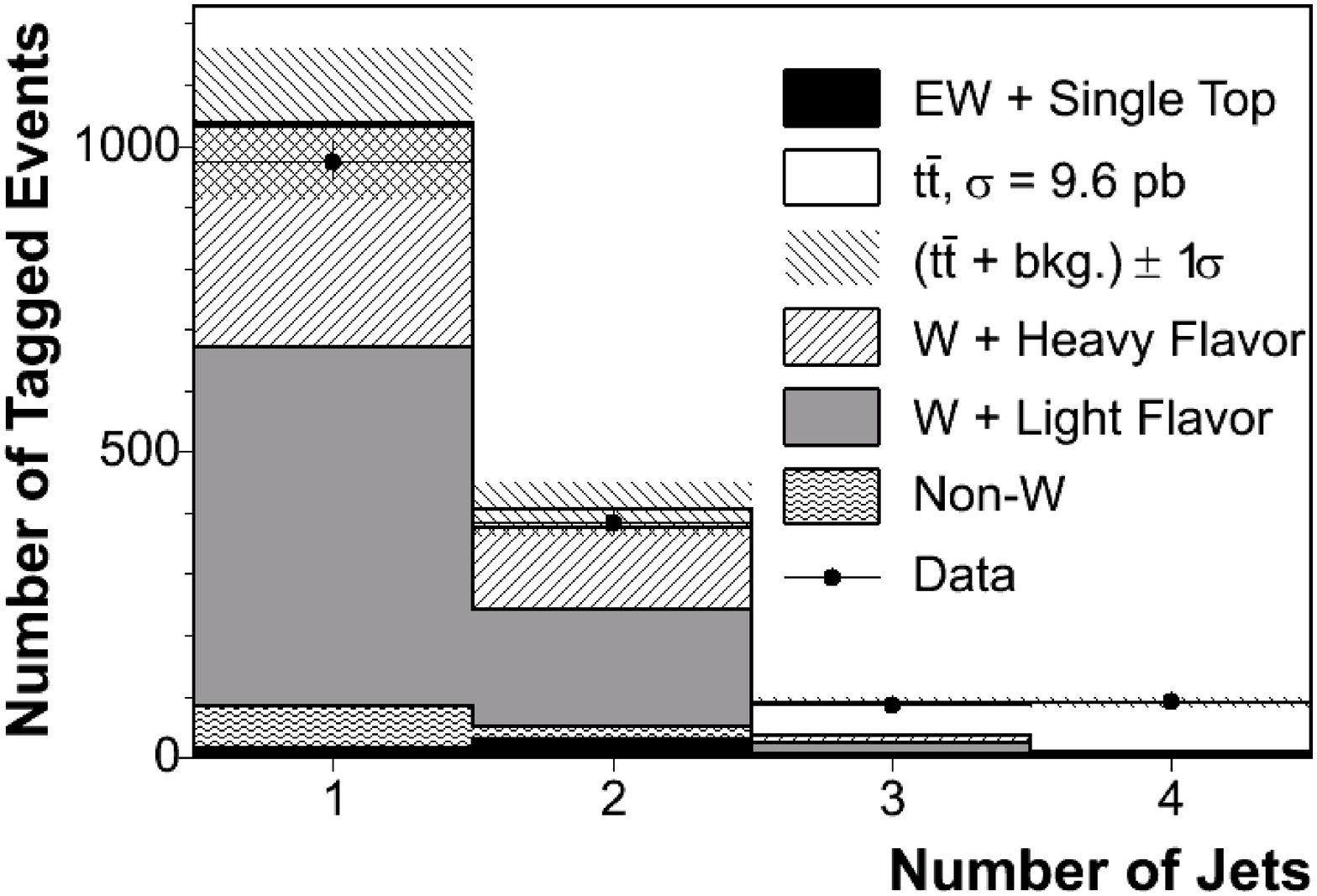}
\caption{Single tag data and background contributions (for an integrated luminosity of 318~pb$^{-1}$) as a function of the
event jet multiplicity for $P_J<$ 1\% (top) and $P_J<$ 5\% (bottom). A top mass of $m_{t}=178~\Gevcc$ is assumed.}
\label{f:backplot}
\end{figure}

The statistical uncertainty on the measured cross section is dominated by the data sample size. 
Table~\ref{t:usyst} summarizes the systematic
contributions to the cross section uncertainty. The correlations in acceptances,
tagging scale factor and luminosity uncertainty are taken into account. 
$Wb\bar b$ and $Wc\bar c$ systematics are considered correlated across all the
bins. All the other uncertainties are treated as uncorrelated. 
\begin{table*}[!htb]
\begin{center}
\caption{\label{t:usyst}Summary of the systematic uncertainties in the single tag analysis.}
\begin{tabular}{l@{\hspace{1cm}}c@{\hspace{1cm}}c@{\hspace{1.2cm}}c}
\hline
\hline
Source                & Fractional Syst. Uncert. (\%) & \multicolumn{2}{c}{Contribution to $\sigma_{t\bar t}$ (\%)} \\
\hline
                 \multicolumn{2}{c}{} & $P_J<$ 1\% &    $P_J<$ 5\%           \\
\hline  
Central Electron ID   & 1.6 & +0.99/-0.97 &  +1.00/-0.98 \\
Central Muon ID       & 1.9 & +0.61/-0.61 &  +0.62/-0.61 \\
{\CMX} Muon ID        & 1.8 & +0.22/-0.22 &  +0.22/-0.22 \\
\PDF                  &   2 &  +2.1/-2.0 &   +2.1/-2.0 \\
Jet Energy Scale      & 4.2 &  +4.5/-4.2 &   +4.6/-4.2 \\
Lepton Isolation      &   2 &  +2.1/-2.0 &   +2.1/-2.0 \\
\ISR/\FSR             & 1.3 &  +1.4/-1.3 &   +1.4/-1.3 \\
MC Modeling           & 1.6 &  +1.7/-1.6 &   +1.7/-1.6 \\
Z Vertex              & 2.0 &  +2.1/-2.1 &   +2.2/-2.1 \\
Tagging $SF$ $P_J<$ 1\% (b's/c's) & 8.6/12.9 & +8.2/-7.2 &   --- \\
Tagging $SF$ $P_J<$ 5\% (b's/c's) & 8.5/12.7 & --- & +7.0/-6.3 \\
Mistag Asymmetry $P_J<$ 1\% & 11.0 & +0.93/-0.93 & ---  \\
Mistag Asymmetry $P_J<$ 5\% & 15.5 & --- & +3.0/-3.0 \\
Non-$W$ Fraction           & 50 & 0.33 &   0.56 \\
Non-$W$ Prediction         & 50 & 0.71 &   0.79 \\
$W$+HF Prediction          & 30 & 2.6 &   2.9 \\
Cross Sections Bck.     & 1.8 & 0.056 &   0.072 \\
Luminosity & 5.9 & +6.5-5.7 &   +6.5-5.8 \\
\hline
Total Systematic Uncertainty &   & +12.5/-11.3  &   +12.3/-11.3 \\
\hline
\hline
\end{tabular}
\end{center}
\end{table*}

\subsection{\label{sec:massdep1}\boldmath $t\bar t$ Cross Section Dependence on the Top Quark Mass}

The signal acceptance used in this analysis has been computed using a sample of
$t\bar t$ events gene\-ra\-ted with {\PYTHIA} for $m_{t}=178~\Gevcc$, which 
co\-rres\-ponds  to the combined Run I top mass measurement at the Tevatron 
Collider~\cite{bib:runimass}. 
We study the dependence of the $t\bar t$ cross section on the top quark mass 
by reevaluating the signal acceptance through a set of Monte Carlo 
simulated samples 
generated by {\HERWIG} for different values of the top mass. 
Results are shown in Fig.~\ref{fig:massdep1}.
A linear fit to the measured cross sections as a function of the top mass 
returns a slope of -0.052 $\pm$ 0.008~pb/($\Gevcc$) and -0.066 $\pm$ 0.008~pb/($\Gevcc$) for $P_J<$1\% and 5\% respectively, where the uncertainties are
due to Monte Carlo simulation statistics.
Note that the fit results for $m_{t}=178~\Gevcc$ agree with the measured 
cross section within the 1.6\% uncertainty estimated in Section~\ref{sec:acc}
due to different modeling in {\PYTHIA} and {\HERWIG}.
\begin{figure}[!h]
    \includegraphics[width=0.5\textwidth,clip=]{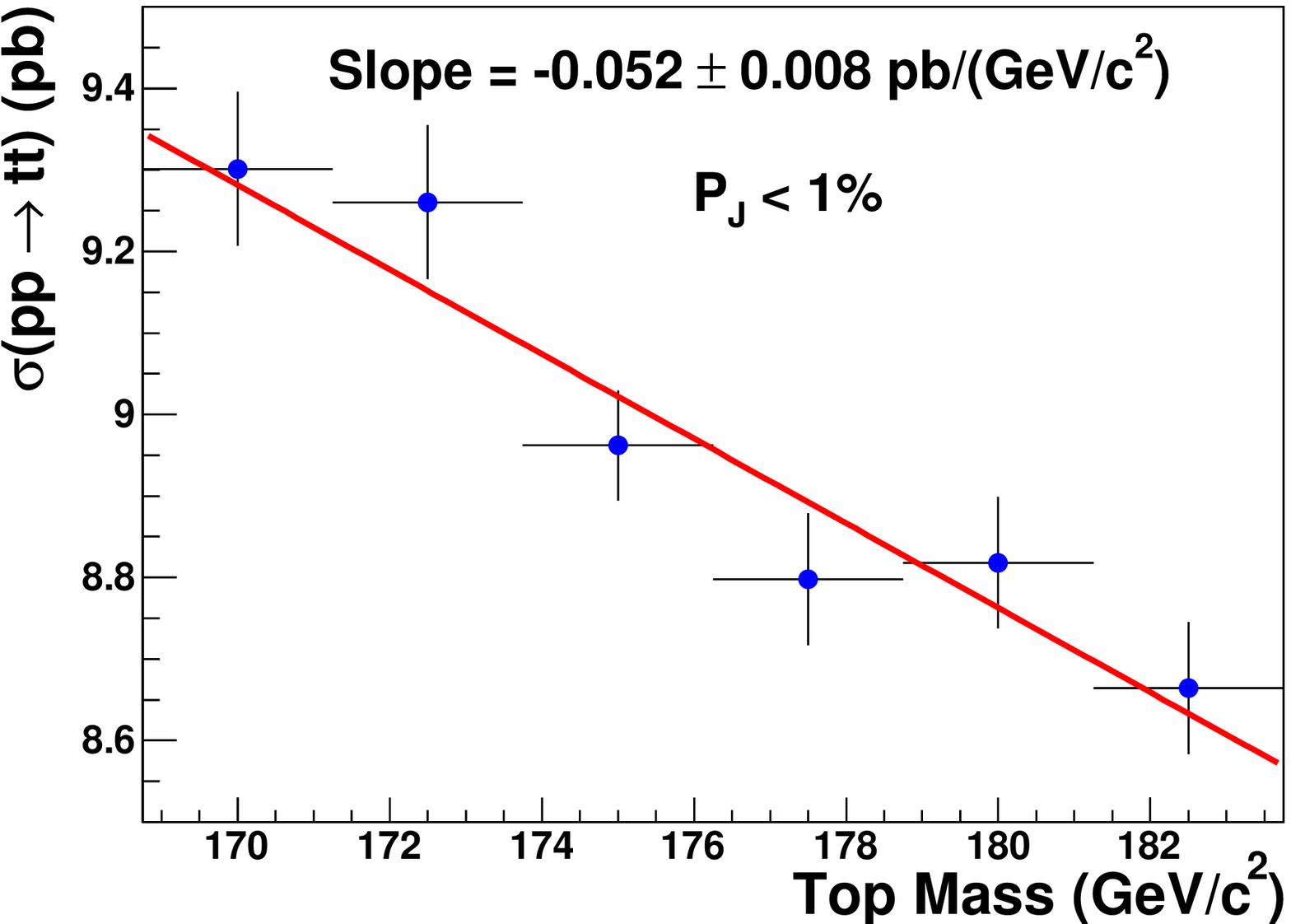}
    \includegraphics[width=0.5\textwidth,clip=]{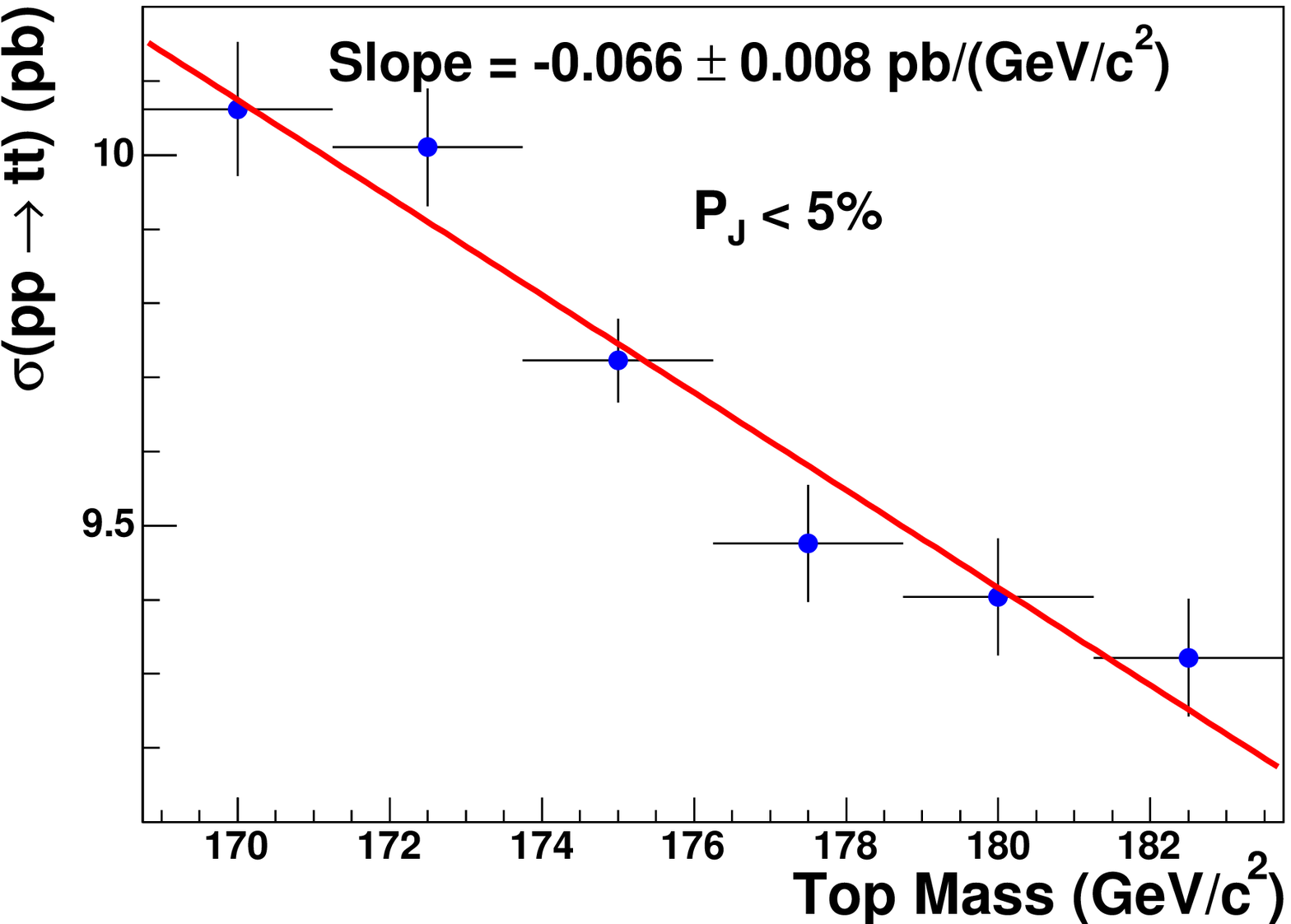}
\caption{Top pair production cross sections as a function of the top quark mass
 for $P_J<$ 1\% (top) and $P_J<$ 5\% (bottom). The uncertainties shown are the
statistical uncertainties on the acceptances for each mass.}
\label{fig:massdep1}
\end{figure}

\subsection{\boldmath Electron versus Muon $t\bar t$ Cross Section Measurements}

As an additional cross check, we measure the cross section separately for events with tight electrons and muons.
Table~\ref{t:ele_muo_summ} summarizes the cross sections for the two analyses
with $P_J<$ 1\% and $P_J<$ 5\%. 
The cross section measurements in the electron and muon+jets samples agree within
their statistical uncertainty. 
%
\begin{table*}[!htb]
\begin{center}
\caption{\label{t:ele_muo_summ}Summary of the cross sections for  $P_J<$ 1\% and $P_J<$ 5\% 
and for each lepton type. Results are expressed in pb.}
\begin{tabular}{l@{\hspace{1cm}}c@{\hspace{1cm}}c@{\hspace{1cm}}c}
\hline
\hline
  & Total & Electrons & Muons \\
\hline
$P_J<$ 1\%         & 8.9$^{+1.0}_{-1.0}$(stat.)$^{+1.1}_{-1.0}$(syst.)  
                   & 8.6$^{+1.4}_{-1.2}$(stat.)$^{+1.1}_{-1.0}$(syst.)  
                   & 9.4$^{+1.7}_{-1.4}$(stat.)$^{+1.2}_{-1.0}$(syst.)  \\
$P_J<$ 5\%           & 9.6$^{+1.0}_{-0.9}$(stat.)$^{+1.2}_{-1.1}$(syst.) 
                   & 9.4$^{+1.3}_{-1.2}$(stat.)$^{+1.2}_{-1.1}$(syst.) 
                   & 9.9$^{+1.6}_{-1.4}$(stat.)$^{+1.2}_{-1.1}$(syst.) \\
\hline
\hline
\end{tabular}
\end{center}
\end{table*}

\section{\label{sec:double_jp}\boldmath Cross Section for Double Tagged Events}

The measurement of the ttbar cross 
section in a sample with at least two b tags follows the same procedure as the single 
tag analysis, with a much purer sample of $t\bar t$ events.
As shown in Table~\ref{t:yields}, after requiring the event selection 
described in Section~\ref{sec:evt_sel}, we observe 30 (61) events with two $b$-tagged jets out of the 120 (180) 
events with  at least one $b$-tagged jet for 
$P_J<$ 1\% ($P_J<$ 5\%).
Table~\ref{t:acceff_2tag_jp1} shows the signal acceptances and the
efficiencies to tag two jets in signal events passing the pretag selection. 
The total acceptance times luminosity for $P_J<$ 1\% and $P_J<$ 5\% is  
$2.57 \pm 0.02 ({\rm stat}) \pm 0.49 ({\rm syst})$~pb$^{-1}$ and
$4.92 \pm 0.04 ({\rm stat}) \pm 0.87 ({\rm syst})$~pb$^{-1}$ respectively.

\begin{table*}[!htb]
\begin{center}
\caption{\label{t:acceff_2tag_jp1}Summary of acceptances for $t\bar t$ events. 
Efficiencies are expressed as percentages. The first uncertainty quoted is 
statistical and the second is systematic. The average $\epsilon_{b-tag}$ is the
luminosity-weighted {\CEM}/{\CMUP}/{\CMX} tagging efficiency. }
\begin{tabular}{l@{\hspace{1.3cm}}c@{\hspace{1.3cm}}c@{\hspace{1.3cm}}c}
\hline
\hline
Quantity    &   {\CEM}         &    {\CMUP}   &    {\CMX}   \\
\hline
$\epsilon_{t\bar t}^{\rm pretag}$ & 3.67 $\pm$ 0.02 $\pm$ 0.22 & 1.92 $\pm$ 0.01 $\pm$ 0.12 & 0.751 $\pm$ 0.008 $\pm$ 0.046 \\
$\int L dt$ ($pb^{-1}$) & 318 $\pm$ 19 & 318 $\pm$ 19 & 305 $\pm$ 18 \\
\hline
\hline
   \multicolumn{4}{c}{Double Tag, $P_J<$ 1\%, $SF$ = 0.817 $\pm$ 0.070}   \\
\hline
 $\epsilon_{b-tag}$ & 12.7 $\pm$ 0.2 $\pm$ 2.1 & 12.6 $\pm$ 0.2 $\pm$ 2.0 & 13.4 $\pm$ 0.4 $\pm$ 2.2 \\
\hline
Average $\epsilon_{b-tag}$ &  \multicolumn{3}{c}{  12.7 $\pm$ 0.1 $\pm$ 2.1} \\
\hline
$\epsilon_{t\bar t}$ & 0.465 $\pm$ 0.006 $\pm$ 0.081 & 0.241 $\pm$ 0.004 $\pm$ 0.042 & 0.101 $\pm$ 0.003 $\pm$ 0.018 \\
$\epsilon_{t\bar t}\int L dt$ ($\pb^{-1}$) & 1.48 $\pm$ 0.02 $\pm$ 0.27 & 0.77 $\pm$ 0.01 $\pm$ 0.14 & 0.32 $\pm$ 0.01 $\pm$ 0.06 \\
\hline
Total $\epsilon_{t\bar t}\int L dt$ &  \multicolumn{3}{c}{$2.57 \pm 0.02 ({\rm stat}) \pm 0.49 ({\rm syst})$ pb$^{-1}$ } \\
\hline
\hline
     \multicolumn{4}{c}{Double Tag, $P_J<$ 5\%, $SF$ = 0.852 $\pm$ 0.072}   \\
\hline
$\epsilon_{b-tag}$ & 24.4 $\pm$ 0.2 $\pm$ 3.6 & 24.1 $\pm$ 0.3 $\pm$ 3.6 & 25.2 $\pm$ 0.5 $\pm$ 3.7 \\
\hline
Average $\epsilon_{b-tag}$  &  \multicolumn{3}{c}{  24.4 $\pm$ 0.2 $\pm$ 3.6} \\
\hline
$\epsilon_{t\bar t}$  & 0.895 $\pm$ 0.009 $\pm$ 0.142 & 0.462 $\pm$ 0.006 $\pm$ 0.074 & 0.189 $\pm$ 0.004 $\pm$ 0.030 \\
$\epsilon_{t\bar t}\int L dt$ ($\pb^{-1}$) & 2.85 $\pm$ 0.03 $\pm$ 0.48 & 1.47 $\pm$ 0.02 $\pm$ 0.25 & 0.60 $\pm$ 0.01 $\pm$ 0.10 \\
\hline
Total $\epsilon_{t\bar t}\int L dt$ &  \multicolumn{3}{c}{$4.92 \pm 0.04 ({\rm stat}) \pm 0.87 ({\rm syst})$ pb$^{-1}$ } \\
\hline
\hline
\end{tabular}
\end{center}
\end{table*}

\subsection{\boldmath Backgrounds in the Double $b$-Tag Sample}\label{s:2t_bkg}

A few differences with respect to the single tag analysis must be taken into 
account in order to estimate the backgrounds.
We define the mistag background as the events with at least two mistagged jets.
The negative tag rate matrix is applied to negatively taggable jets in the 
event and the probability to have at least two mistagged jets is summed over 
all events. The mistag prediction is scaled by the fraction of non-$W$, 
electroweak backgrounds and by the mistag asymmetry as is done for the 
single tag analysis.

Events with one real heavy flavor tag plus a mistag are included in the other 
background sources.
The contribution of mistags to the $W$+heavy flavor background is taken into 
account by applying the mistag rate matrix to light flavor jets in events with 
an extra real tag when computing the tagging efficiency. 
Results are summarized in Table~\ref{t:whfeff_2tag_jp1}.
\begin{table*}[!htb]
\begin{center}
\caption{\label{t:whfeff_2tag_jp1} Jet probability tagging efficiencies for $Wb\bar b$, 
$Wc\bar c$ and $Wc$ events for double tagged events. Values are expressed as 
percentages. The first uncertainty quoted is statistical and the second is 
systematic. 1B and 2B (1C and 2C) refer to $Wb\bar b$ ($Wc\bar c$) events 
with one and two reconstructed heavy flavor jets respectively.}
\begin{tabular}{l@{\hspace{1.3cm}}c@{\hspace{1.3cm}}c@{\hspace{1.3cm}}c}
\hline
\hline
jet multiplicity      &    2 jets    &    3 jets    &    $\geq$ 4 jets \\
\hline
\hline
    \multicolumn{4}{c}{Double Tag Tagging Efficiencies, $P_J<$ 1\%}                    \\
\hline
1B ($\ge$2 tags) & 0.27 $\pm$ 0.06 $\pm$ 0.05 & 0.90 $\pm$ 0.30 $\pm$ 0.15 & 1.3 $\pm$ 0.8 $\pm$ 0.2 \\
2B ($\ge$2 tags) & 10.3 $\pm$ 0.4 $\pm$ 1.8 & 13. 1$\pm$ 1.1 $\pm$ 2.2 & 14.1 $\pm$ 1.5 $\pm$ 2.4 \\
1C ($\ge$2 tags) & 0.067 $\pm$ 0.037 $\pm$ 0.017 & 0.23 $\pm$ 0.17 $\pm$ 0.06 & 0.29 $\pm$ 0.38 $\pm$ 0.08 \\
2C ($\ge$2 tags) & 0.43 $\pm$ 0.12 $\pm$ 0.11 & 1.3 $\pm$ 0.5 $\pm$ 0.3 & 1.1 $\pm$ 0.5 $\pm$ 0.3 \\
Wc ($\ge$2 tags) & 0.05 $\pm$ 0.03$\pm$ 0.01 & 0.20 $\pm$ 0.14 $\pm$ 0.05 & 0.22 $\pm$ 0.21 $\pm$ 0.06 \\
\hline
\hline
    \multicolumn{4}{c}{Double Tag Tagging Efficiencies, $P_J<$ 5\%}                    \\
\hline
1B ($\ge$2 tags) & 1.3 $\pm$ 0.1 $\pm$ 0.2 & 3.7 $\pm$ 0.6 $\pm$ 0.6 & 5.0 $\pm$ 1.5 $\pm$ 0.8 \\
2B ($\ge$2 tags) & 18.6 $\pm$ 0.6 $\pm$ 3.1 & 23.9 $\pm$ 1.4 $\pm$ 4.0 & 26.0 $\pm$ 1.9 $\pm$ 4.4 \\
1C ($\ge$2 tags) & 0.54 $\pm$ 0.11 $\pm$ 0.14 & 1.6 $\pm$ 0.4 $\pm$ 0.4 & 1.8 $\pm$ 0.9 $\pm$ 0.5 \\
2C ($\ge$2 tags) & 2.5 $\pm$ 0.3 $\pm$ 0.6 & 5.6 $\pm$ 0.9 $\pm$ 1.4 & 6.3 $\pm$ 1.2 $\pm$ 1.6 \\
Wc ($\ge$2 tags) & 0.40 $\pm$ 0.08 $\pm$ 0.10 & 1.5 $\pm$ 0.4 $\pm$ 0.4 & 2.1 $\pm$ 0.6 $\pm$ 0.5 \\
\hline
\hline
\end{tabular}
\end{center}
\end{table*}

The strategy to estimate the non-W background is changed, compared to that used for the single tag sample, due to low statistics in the double tagged event sample in the sideband regions 
(see Section~\ref{sec:non-w}). 
We compute a common tag rate for all the jet multiplicity bins by using data 
in region B (isolation $<$ 0.1 and $\met$ $<$ 15~$\Gev$). 
We divide the total number of double tagged events by the sum of the number of 
pretag events scaled by the jet pair multiplicity. Finally, we apply this tag 
rate to the pretag expectation in the signal region derived in 
Section~\ref{s:nonwfr}.

Background predictions for $P_J<$ 1\% and $P_J<$ 5\% are compared to the data in 
Tables~\ref{t:backres_2tag_1jp} and~\ref{t:backres_2tag_5jp}, 
respectively. 
\begin{table*}[!htb]
\begin{center}
\caption{\label{t:backres_2tag_1jp} Summary of the background estimate in the 
double tag sample for $P_J<$ 1\%.}
\begin{tabular}{l@{\hspace{1.3cm}}c@{\hspace{1.3cm}}c@{\hspace{1.3cm}}c}
\hline
\hline
Jet Multiplicity & 2 jets & 3 jets & $\ge$ 4 jets \\
\hline
\hline
                 \multicolumn{4}{c}{ Electroweak}            \\
\hline
$WW$                             & 0.05 $\pm$ 0.02 & 0.03 $\pm$ 0.02 & 0.006 $\pm$ 0.006 \\
$WZ$                             & 0.25 $\pm$ 0.05 & 0.03 $\pm$ 0.01 & 0.013 $\pm$ 0.006 \\
$ZZ$                             & 0.014 $\pm$ 0.005 & 0.001 $\pm$ 0.001 & 0.001 $\pm$ 0.001 \\
Single Top $W-g$               & 0.17 $\pm$ 0.03 & 0.12 $\pm$ 0.03 & 0.05 $\pm$ 0.01 \\
Single Top $W^*$               & 0.88 $\pm$ 0.17 & 0.14 $\pm$ 0.03 & 0.035 $\pm$ 0.007 \\
$Z\rightarrow \tau^+\tau^-$ & 0.06 $\pm$ 0.06 & 0 $\pm$ 0 & 0 $\pm$ 0 \\
\hline
Total                          & 1.4 $\pm$ 0.3 & 0.33 $\pm$ 0.06 & 0.10 $\pm$ 0.02 \\
\hline
\hline
            \multicolumn{4}{c}{          W + Heavy Flavour }                   \\
\hline
$Wb\bar b$                     & 6.2 $\pm$ 2.0 & 0.89 $\pm$ 0.29 & 0.61 $\pm$ 0.21 \\
$Wc\bar c$                     & 0.38 $\pm$ 0.17 & 0.13 $\pm$ 0.06 & 0.077 $\pm$ 0.046 \\
$Wc$                           & 0.13 $\pm$ 0.08 & 0.03 $\pm$ 0.03 & 0.02 $\pm$ 0.02 \\
\hline
Total                          & 6.7 $\pm$ 2.1 & 1.1 $\pm$ 0.3 & 0.71 $\pm$ 0.24 \\
\hline
\hline
                  \multicolumn{4}{c}{    Others     }                        \\
\hline
Mistag                         & 0.21 $\pm$ 0.05  & 0.10 $\pm$ 0.02  & 0.12 $\pm$ 0.03  \\
Non-$W$                          & 0.19 $\pm$ 0.12 & 0.03 $\pm$ 0.02 & 0.05 $\pm$ 0.03 \\
\hline
\hline
Total Background               & 8.5 $\pm$ 2.3  & 1.5 $\pm$ 0.4  & 0.97 $\pm$ 0.25  \\
Data                            &                13 &                12 &                18 \\
\hline
\hline
\end{tabular}
\end{center}
\end{table*}
\begin{table*}[!htb]
\begin{center}
\caption{\label{t:backres_2tag_5jp}Summary of the background estimate in the 
double tag sample for $P_J<$ 5\%.}
\begin{tabular}{l@{\hspace{1.3cm}}c@{\hspace{1.3cm}}c@{\hspace{1.3cm}}c}
\hline
\hline
Jet Multiplicity & 2 jets & 3 jets & $\ge$ 4 jets \\
\hline
\hline
                 \multicolumn{4}{c}{ Electroweak}        \\
\hline
$WW$                             &  0.29 $\pm$ 0.06 & 0.13 $\pm$ 0.04 & 0.07 $\pm$ 0.03 \\
$WZ$                             &  0.51 $\pm$ 0.10 & 0.06 $\pm$ 0.02 & 0.03 $\pm$ 0.01 \\
$ZZ$                             &  0.026 $\pm$ 0.007 & 0.004 $\pm$ 0.002 & 0.002 $\pm$ 0.001 \\
Single Top $W-g$               &  0.39 $\pm$ 0.07 & 0.23 $\pm$ 0.04 & 0.09 $\pm$ 0.02 \\
Single Top $W^*$               &  1.5 $\pm$ 0.3 & 0.26 $\pm$ 0.05 & 0.06 $\pm$ 0.01 \\
$Z\rightarrow \tau^+\tau^-$ &  0.07 $\pm$ 0.07 & 0 $\pm$ 0 & 0 $\pm$ 0 \\
\hline
Total                          &  2.83 $\pm$ 0.51 & 0.70 $\pm$ 0.12 & 0.25 $\pm$ 0.05 \\
\hline
\hline
            \multicolumn{4}{c}{          W + Heavy Flavour }                   \\
\hline
$Wb\bar b$                     &  11.5 $\pm$ 3.7 & 1.8 $\pm$ 0.6 & 1.2 $\pm$ 0.4 \\
$Wc\bar c$                     &  2.4 $\pm$ 0.9 & 0.61 $\pm$ 0.24 & 0.45 $\pm$ 0.19 \\
$Wc$                           &  0.98 $\pm$ 0.38 & 0.25 $\pm$ 0.11 & 0.19 $\pm$ 0.09 \\
\hline
Total                          &  14.9 $\pm$ 4.7 & 2.6 $\pm$ 0.8 & 1.9 $\pm$ 0.6 \\
\hline
\hline
                  \multicolumn{4}{c}{    Others     }                        \\
\hline
Mistag                         & 2.7 $\pm$ 0.9  & 1.0 $\pm$ 0.3  & 1.3 $\pm$ 0.4  \\
Non-$W$                          &  0.63 $\pm$ 0.34 & 0.09 $\pm$ 0.05 & 0.14 $\pm$ 0.09 \\
\hline
\hline
Total Background               & 21.1 $\pm$ 5.1  & 4.4 $\pm$ 0.9  & 3.5 $\pm$ 0.7  \\
Data                            &                        28 &                22 &                39 \\
\hline
\hline
\end{tabular}
\end{center}
\end{table*}

The iterative procedure described in Section~\ref{sec:single_jp} is applied, 
and we obtain a cross section of
\begin{displaymath}
  \sigma_{t\bar t} = 11.1^{+2.3}_{-1.9} {\rm (stat.)}^{+2.5}_{-1.9}{\rm (syst.)~pb}
\end{displaymath}
for $P_J<$ 1\% and
\begin{displaymath}
  \sigma_{t\bar t} = 11.6^{+1.7}_{-1.5} {\rm (stat.)}^{+2.4}_{-1.8}{\rm (syst.)~pb}
\end{displaymath}
for $P_J<$ 5\%. Signal and background estimates after the iterative procedure 
are shown in Table~\ref{t:backres2tag1jp_after}, together with the observed 
number of events. 
\begin{table*}[!htb]
\begin{center}
\caption{\label{t:backres2tag1jp_after} Summary of the final signal and 
background estimates and observed data in the double tag sample. MC derived
refers to electroweak processes.}
\begin{tabular}{l@{\hspace{1.3cm}}c@{\hspace{1.3cm}}c@{\hspace{1.3cm}}c}
\hline
\hline
Jet Multiplicity & 2 jets & 3 jets & $\ge$ 4 jets \\
\hline
\hline
Pretag Data      &            4442 &                300 &        166 \\
\hline
\hline
                  \multicolumn{4}{c}{$P_J<$ 1\%}                        \\
\hline
MC Derived                     & 1.4 $\pm$ 0.3 & 0.33 $\pm$ 0.06 & 0.10 $\pm$ 0.02 \\
$Wb\bar b$                     & 6.1 $\pm$ 1.9 & 0.57 $\pm$ 0.19 & 0.10 $\pm$ 0.03 \\
$Wc\bar c$                     & 0.38 $\pm$ 0.17 & 0.09 $\pm$ 0.04 & 0.013 $\pm$ 0.008 \\
$Wc$                           & 0.12 $\pm$ 0.08 & 0.02 $\pm$ 0.02 & 0.003 $\pm$ 0.003 \\
Mistag                         & 0.21 $\pm$ 0.05  & 0.06 $\pm$ 0.01  & 0.019 $\pm$ 0.004  \\
Non-$W$                          & 0.19 $\pm$ 0.12 & 0.03 $\pm$ 0.02 & 0.05 $\pm$ 0.03 \\
\hline
Total Background               & 8.4 $\pm$ 2.2  & 1.1 $\pm$ 0.3  & 0.28 $\pm$ 0.06  \\
$t\bar t$ (11.1 pb)       & 3.9 $\pm$ 0.9 & 10.2 $\pm$ 2.0 & 18.4 $\pm$ 3.4 \\
Data                            &                13 &                12 &                18 \\
\hline
\hline
                  \multicolumn{4}{c}{$P_J<$ 5\%}                        \\
\hline
MC Derived                          &  2.83 $\pm$ 0.51 & 0.70 $\pm$ 0.12 & 0.25 $\pm$ 0.05 \\
$Wb\bar b$                     &  11.4 $\pm$ 3.6 & 1.1 $\pm$ 0.3 & 0.16 $\pm$ 0.05 \\
$Wc\bar c$                     &  2.3 $\pm$ 0.9 & 0.38 $\pm$ 0.15 & 0.06 $\pm$ 0.03 \\
$Wc$                           &  0.97 $\pm$ 0.37 & 0.16 $\pm$ 0.07 & 0.03 $\pm$ 0.01 \\
Mistag                          & 2.7 $\pm$ 0.8  & 0.65 $\pm$ 0.20  & 0.15 $\pm$ 0.05  \\
Non-$W$                          &  0.63 $\pm$ 0.34 & 0.09 $\pm$ 0.05 & 0.14 $\pm$ 0.09 \\
\hline
Total Background               & 20.9 $\pm$ 5.0  & 3.1 $\pm$ 0.6  & 0.80 $\pm$ 0.15  \\
$t\bar t$ (11.6 pb)       &  7.5 $\pm$ 1.5 & 20.5 $\pm$ 3.7 & 36.6 $\pm$ 6.1 \\
Data                            &                        28 &                22 &                39 \\
\hline
\hline
\end{tabular}
\end{center}
\end{table*}
Figure~\ref{f:backplot_2tag_1jp} compares the numbers of observed data to 
background and signal expectations
for $P_J<$ 1\% and 5\% for the measured $t\bar t$ production cross sections. 
%
\begin{figure}[!htb]
    \includegraphics[width=0.49\textwidth,clip=]{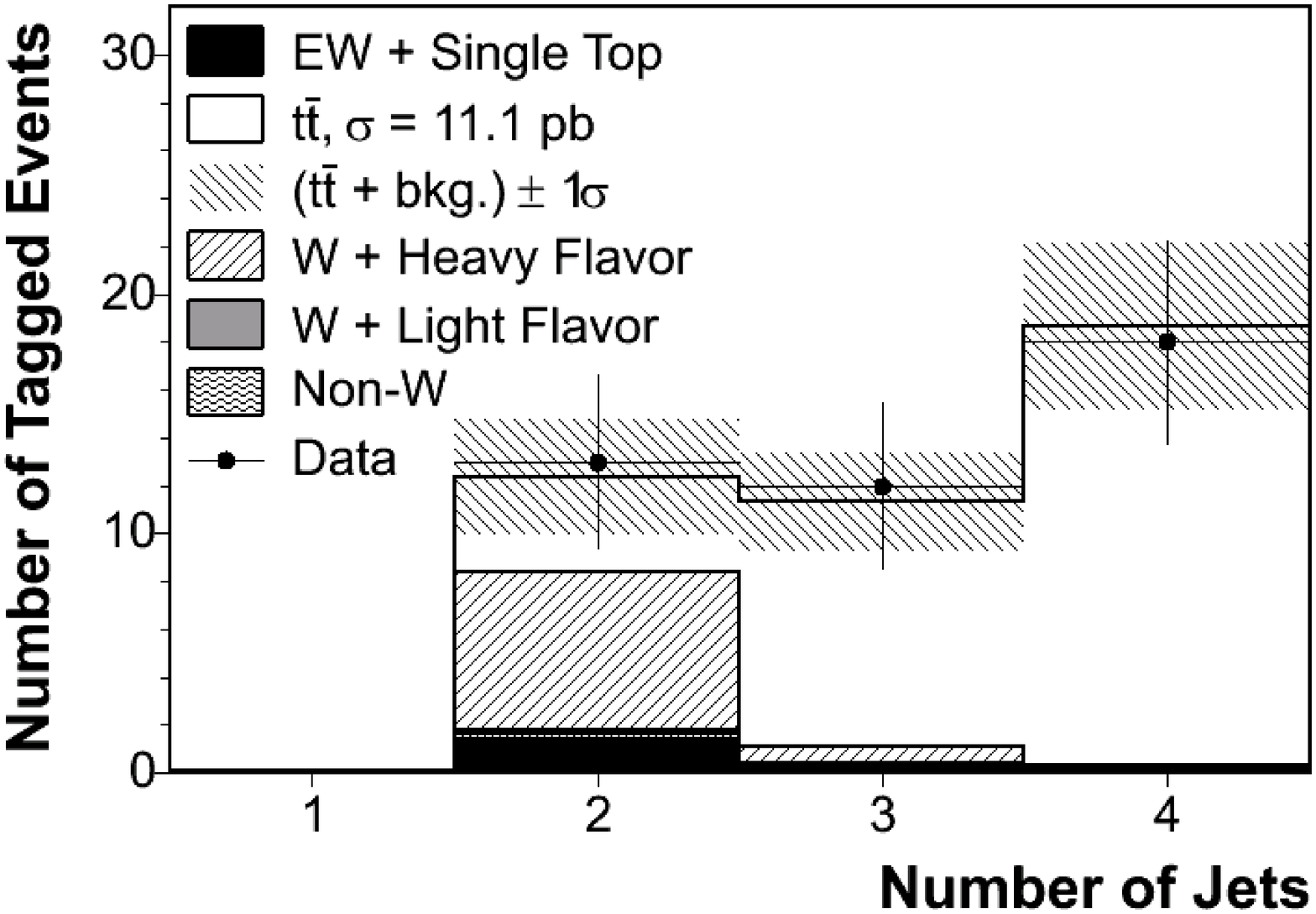}
    \includegraphics[width=0.49\textwidth,clip=]{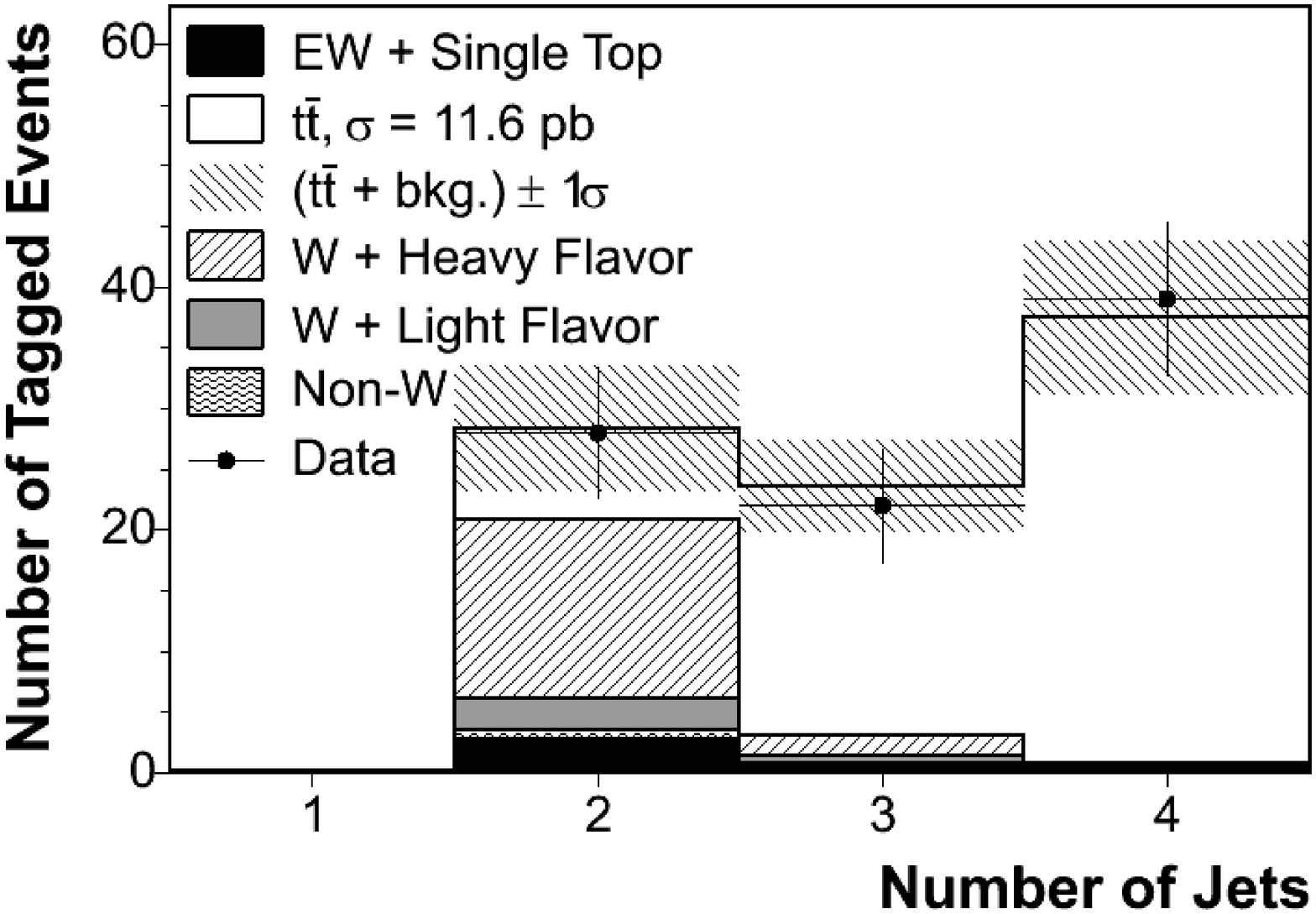}
\caption{Double tag data and background contributions (for an integrated luminosity of 318~pb$^{-1}$) as a function of the
event jet multiplicity for $P_J<$ 1\% (top) and $P_J<$ 5\% (bottom). A top mass of $m_{t}=178~\Gevcc$ is assumed.}
\label{f:backplot_2tag_1jp}
\end{figure}

The statistical uncertainty on the measured cross section is dominated by the data sample size. 
Table~\ref{t:usyst2} summarizes the systematic
contributions to the cross section uncertainty.
%
%
\begin{table*}[!htb]
\begin{center}
\caption{\label{t:usyst2}Summary of the systematical uncertainties in the 
double tag analysis.}
\begin{tabular}{l@{\hspace{1cm}}c@{\hspace{1cm}}c@{\hspace{1.2cm}}c}
\hline
\hline
Source     & Fractional Syst. Uncert. (\%)& \multicolumn{2}{c}{Contribution to $\sigma_{t\bar t}$ (\%)} \\
\hline
                  \multicolumn{2}{c}{} & $P_J<$ 1\% & $P_J<$ 5\%                        \\
\hline
Central Electron ID   & 1.6 & +0.98/-0.96 &  +0.98/-0.96 \\
Central Muon ID       & 1.9 & +0.60/-0.60 &  +0.61/-0.60 \\
{\CMX} Muon ID        & 1.8 & +0.21/-0.21 &  +0.21/-0.21 \\
\PDF                  &   2 &  +2.1/-2.0 &  +2.1/-2.0 \\
Jet Energy Scale      & 4.2 &  +4.5/-4.1 &  +4.5/-4.1 \\
Lepton Isolation      &   2 &  +2.1/-2.0 &  +2.1/-2.0 \\
\ISR/\FSR             & 1.3 &  +1.3/-1.3 &  +1.3/-1.3 \\
MC Modeling           & 1.6 &  +1.7/-1.6 &  +1.7/-1.6 \\
Z Vertex              & 2.0 &  +2.1/-2.0 &  +2.1/-2.0 \\
Tagging $SF$ $P_J<$ 1\% (b's/c's) & 8.6/12.9 & +20.3/-14.7 &  --- \\
Tagging $SF$ $P_J<$ 5\% (b's/c's) & 8.5/12.7 & --- &  +18.3/-13.6 \\
Mistag Asymmetry $P_J<$ 1\% & 11.0 & +0.063/-0.063 &  --- \\
Mistag Asymmetry $P_J<$ 5\% & 15.5 & --- &  +0.44/-0.44 \\
Non-W Fraction           & 50 & 0.060 &  0.092 \\
Non-W Prediction         & 50 &  0.13 &  0.21 \\
W+HF Prediction          & 30 &  0.84 &  1.0 \\
Cross Sections Bkg.     & 1.8 & 0.027 &  0.030 \\
Luminosity & 5.9 & +6.4/-5.7 &  +6.4/-5.7 \\
\hline
Total Systematic Uncertainty &   & +22.2/-16.8  &  +20.4/-15.9\\
\hline
\hline
\end{tabular}
\end{center}
\end{table*}

\subsection{\label{sec:massdep2}Cross Section Dependence on the Top Quark Mass}

We study the dependence of the $t\bar t$ cross section using the double tag
sample on the top quark mass in an analogous way to Section~\ref{sec:massdep1}.
Results are shown in Fig.~\ref{fig:massdep2}.
A linear fit to the measured cross sections as a function of the top mass 
returns a slope of -0.096 $\pm$ 0.022~pb/($\Gevcc$) and 
-0.082 $\pm$ 0.019~pb/($\Gevcc$) for $P_J<$1\% and $P_J<$5\%, respectively, 
where the uncertainties are due to Monte Carlo simulation statistics. 
As before, note that the fit results for $m_{t}=178~\Gevcc$ agree with the 
measured cross section within the 1.6\% uncertainty estimated in 
Section~\ref{sec:acc} due to different modeling in {\PYTHIA} and {\HERWIG}. 
\begin{figure}[!h]
    \includegraphics[width=0.5\textwidth,clip=]{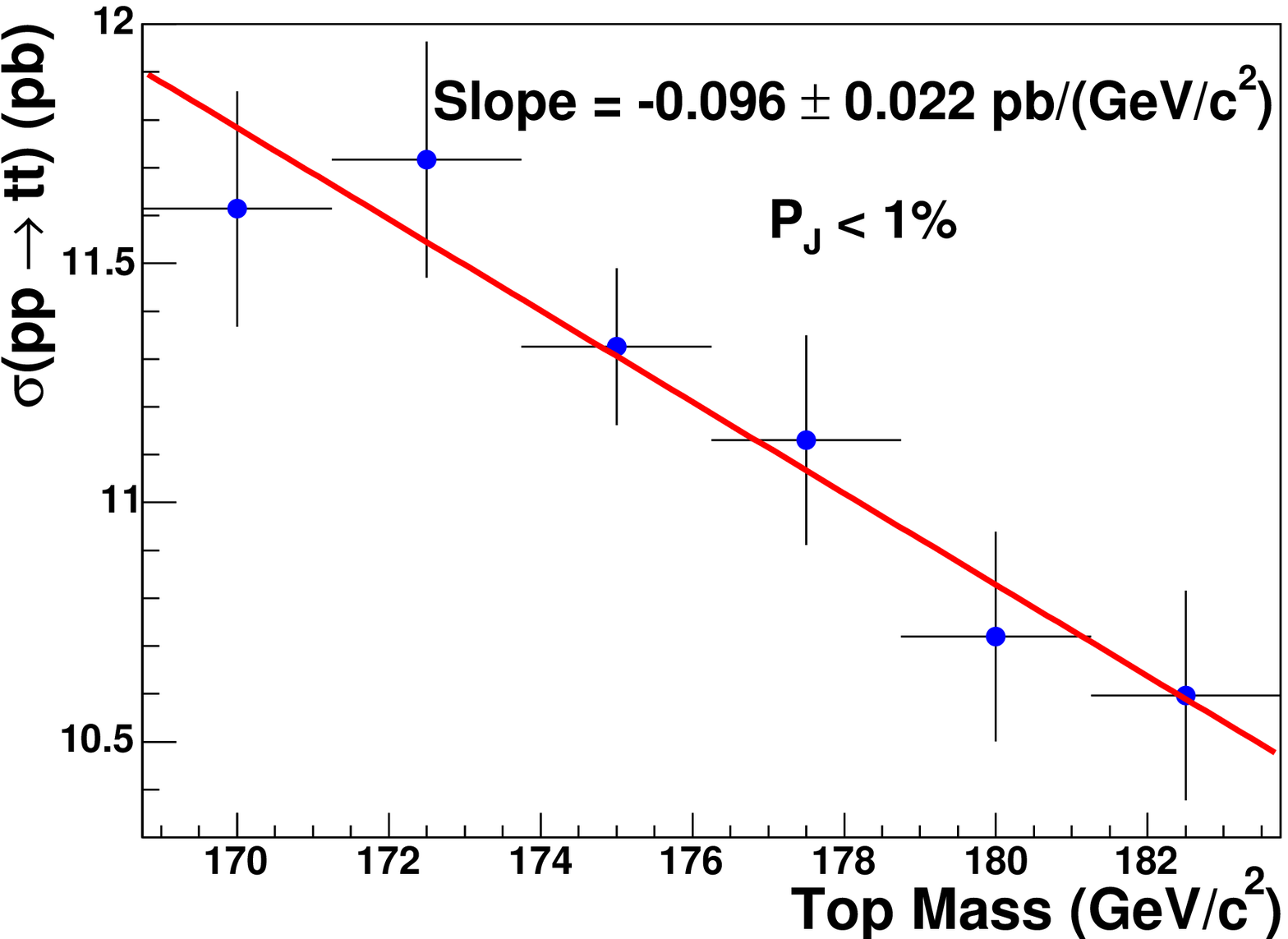}
    \includegraphics[width=0.5\textwidth,clip=]{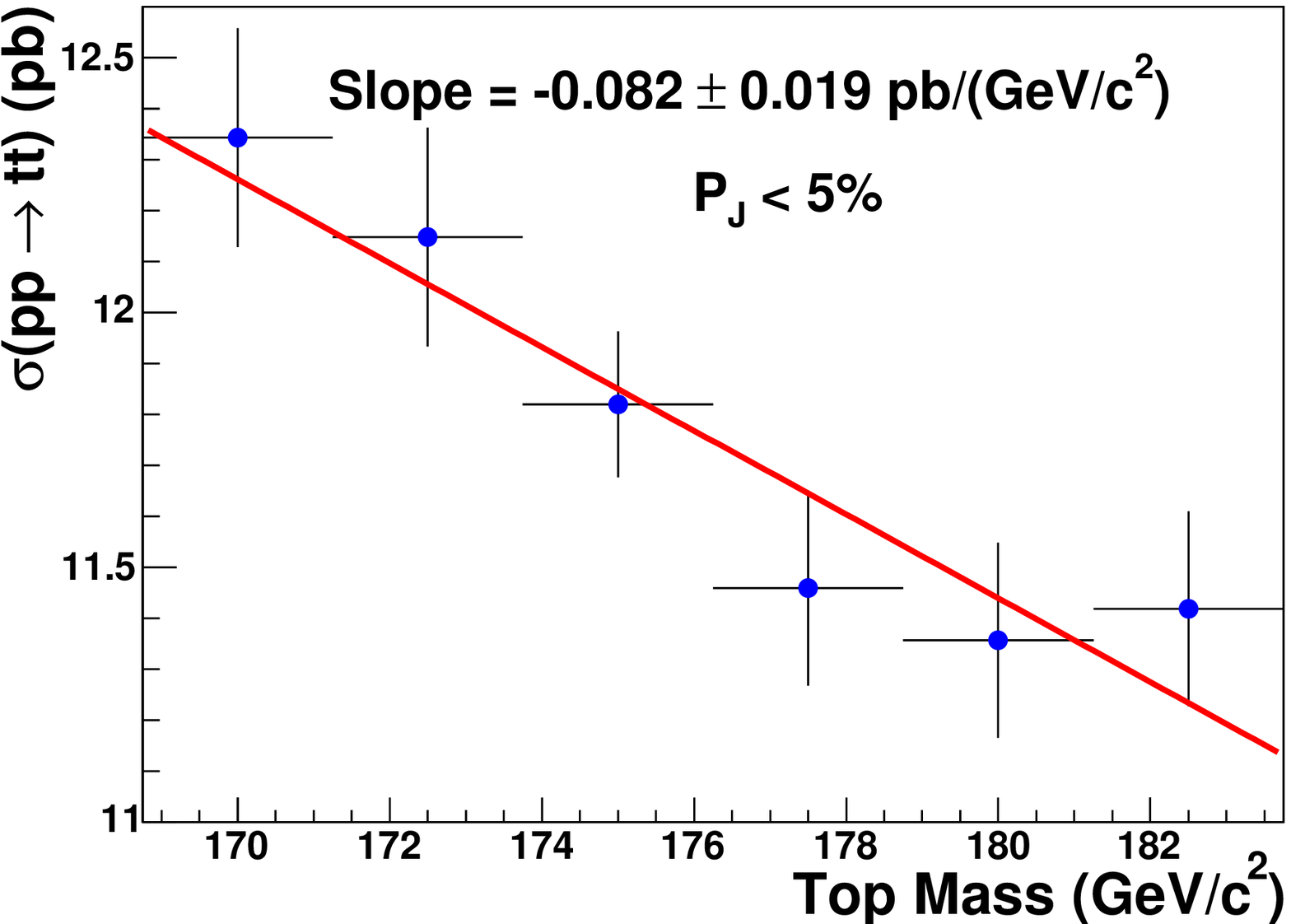}
\caption{Top pair production cross sections as a function of the top quark mass
 for $P_J<$1\% (top) and $P_J<$5\% (bottom) in the double tag analysis. The 
uncertainties shown are the statistical uncertainties on the acceptances for 
each mass.}
\label{fig:massdep2}
\end{figure}

\subsection{\label{sec:double_single}Comparison Between Single and Double Tag Cross Sections}

Although the measurements of the single and double tag cross sections
are statistically compatible, we
observe a ratio of about 1.2 between the measured cross sections 
in the double and single tag samples. 
We use pseudo-experiments to estimate the probability to obtain  a cross 
section greater than the measured double tag cross section
when we assume that the measured single tag cross section is correct. 
For each pseudo-experiment, we vary the
total double tag background estimate 
according to a Gaussian distribution
with a width equal to its uncertainty.
Successively, we add the background to the expected signal
by assuming the single tag $t\bar t$ cross section and
we vary the total number of events according to a Poisson distribution. 
We repeat this procedure 10,000 times and count the number
of pseudo-experiments in which we have a result greater than
the one observed in data. We find a probability of 13.2\% for 
$P_J<$ 1\% and 15.6\% for $P_J<$ 5\%.

The systematic uncertainty in the double
tag measurement is dominated by the uncertainties on the acceptance,
luminosity and tagging scale factor. The systematic uncertainties on the 
background prediction are negligible. A bias on the
values of acceptance and luminosity would affect the cross section
measurement in the single and double tag samples in the same way.
However, a bias on the tagging scale factor would have
a greater effect in the double tag analysis than in the single 
tag one. To study this, we vary the tagging scale factor by
$\pm 1\sigma$ and we repeat the cross section measurements and
the pseudo-experiments. Results are summarized in Tables~\ref{t:pseudo1}
and~\ref{t:pseudo2}. As expected, the cross sections measured in the
double tag sample are more sensitive to a change in the scale factor, 
resulting in a better agreement between the single and double tag cross 
sections when a larger value for the scale factor is used.
\begin{table*}[!htp]
\begin{center}
\caption{\label{t:pseudo1}Cross section for $t\bar t$ event production in single and double tag analysis 
         for $P_J<$ 1\% and $<$ 5\% and different values of the tagging scale factors ($SF$). Results are expressed in pb.}
\begin{tabular}{@{\extracolsep{\fill}}l@{\hspace{0.9cm}}c@{\hspace{0.9cm}}c@{\hspace{0.9cm}}c}
\hline
\hline
   & $SF$ - $1~\sigma$ & $SF$ & $SF$ + $1~\sigma$ \\
\hline
 $P_J<$ 1\%, $\le$1 tag  & $ 9.8^{+1.1}_{-1.0}{\rm (stat.)}^{+1.3}_{-1.1}{\rm (syst.)}$ & $ 8.9^{+1.0}_{-1.0}{\rm (stat.)}^{+1.1}_{-1.0}{\rm (syst.)}$ & $ 8.3^{+1.0}_{-0.9}{\rm (stat.)}^{+1.0}_{-0.9}{\rm (syst.)}$ \\
 $P_J<$ 1\%, $\le$2 tags & $13.3^{+2.8}_{-2.3}{\rm (stat.)}^{+3.3}_{-2.4}{\rm (syst.)}$ & $11.1^{+2.3}_{-1.9}{\rm (stat.)}^{+2.5}_{-1.9}{\rm (syst.)}$ & $ 9.4^{+2.0}_{-1.7}{\rm (stat.)}^{+2.0}_{-1.4}{\rm (syst.)}$ \\ 
 $P_J<$ 5\%, $\le$1 tag  & $10.5^{+1.1}_{-1.0}{\rm (stat.)}^{+1.3}_{-1.2}{\rm (syst.)}$ & $ 9.6^{+1.0}_{-0.9}{\rm (stat.)}^{+1.2}_{-1.1}{\rm (syst.)}$ & $ 9.0^{+1.0}_{-0.9}{\rm (stat.)}^{+1.1}_{-1.0}{\rm (syst.)}$ \\
 $P_J<$ 5\%, $\le$2 tags & $13.7^{+2.0}_{-1.7}{\rm (stat.)}^{+3.3}_{-2.4}{\rm (syst.)}$ & $11.6^{+1.7}_{-1.5}{\rm (stat.)}^{+2.4}_{-1.8}{\rm (syst.)}$ & $ 9.9^{+1.5}_{-1.3}{\rm (stat.)}^{+2.0}_{-1.5}{\rm (syst.)}$ \\
\hline
\hline
\end{tabular}
\end{center}
\end{table*}
\begin{table}[!htp]
\begin{center}
\caption{\label{t:pseudo2} Probability to measure a cross section greater than the 
         one obtained in the double tag analysis when the $t\bar t$ cross 
         section measured in the single tag analysis is assumed.}
\begin{tabular}{l@{\hspace{0.9cm}}c@{\hspace{0.9cm}}c@{\hspace{0.9cm}}c}
\hline
\hline
   & $SF$ - $1~\sigma$ & $SF$ & $SF$ + $1~\sigma$ \\
\hline
 $P_J<$ 1\%    & 4.5\% & 13.2\% & 30\% \\
 $P_J<$ 5\%    & 2.8\% & 15.6\% & 35\% \\
\hline
\hline
\end{tabular}
\end{center}
\end{table}

\section{\label{sec:concl}Conclusions}

We present a measurement of the $t\bar t$ production cross section in $p\bar p$
collisions at $\sqrt{s}=1.96$ TeV with an integrated luminosity of 
318 $\pm$ 18~pb$^{-1}$ at the {\CDF} detector. We select events 
compatible 
with the $t\bar t\rightarrow l\nu q\bar qb\bar b$ 
decay mode by requiring one isolated electron (muon) with transverse energy $E_T~(p_T)>20~\Gev$
and missing transverse energy $\met >20~\Gev$ and at least three jets with 
transverse energy $E_T>15~\Gev$. 
We further require at least one jet tagged by the {\JP} algorithm. 
This selection accepts an estimated (3.5 $\pm$ 0.3)\% of all $t\bar t$ events 
when a $P_J<$ 1\% cut is applied, and an estimated (4.4 $\pm$ 0.4)\% with a 
looser $P_J$ cut at 5\%.
Backgrounds are estimated using data and Monte Carlo simulations. 
We find good agreement with the observed data in a control region defined by 
events with W+one or two jets.
Using the excess of events with three or more jets and at least one 
$b$ tag with $P_J<$ 1\%, we measure a top pair production cross section of
\begin{displaymath}
  \sigma_{t\bar t} = 8.9^{+1.0}_{-1.0} {\rm (stat.)}^{+1.1}_{-1.0}{\rm (syst.)}~{\rm pb}.
\end{displaymath}

As cross checks, we measure the cross section using samples with 
different $b$-tagging requirements. Using events with at least one  
$b$ tag with $P_J<$ 5\% we obtain 
\begin{displaymath}
  \sigma_{t\bar t} = 9.6^{+1.0}_{-0.9} {\rm (stat.)}^{+1.2}_{-1.1}{\rm (syst.)}~{\rm pb}.
\end{displaymath}
We also measure the $t\bar t$ production cross section
in events with at least two tagged jets. The acceptance for signal events is 
estimated to be
(0.8 $\pm$ 0.1)\% for $P_J<$ 1\% and (1.5 $\pm$ 0.3)\% for $P_J<$ 5\%. 
We measure a cross section of
\begin{displaymath}
  \sigma_{t\bar t} = 11.1^{+2.3}_{-1.9} {\rm (stat.)}^{+2.5}_{-1.9}{\rm (syst.)}~{\rm pb}
\end{displaymath}
for $P_J<$ 1\% and 
\begin{displaymath}
  \sigma_{t\bar t} = 11.6^{+1.7}_{-1.5} {\rm (stat.)}^{+2.4}_{-1.8}{\rm (syst.)}~{\rm pb}
\end{displaymath}
for $P_J<$ 5\%.

Figure~\ref{fig:theorycomp} shows our main result together with other CDF 
$t\bar t$ cross section measurements and theoretical predictions. 
Our result is above the central theoretical value by $\sim$1.9 $\sigma$.
It should be noted that our result is highly correlated with the lepton+jets 
measurement using secondary vertex $b$-tagging, described in~\cite{bib:secvtx_prl}, 
where a comparison between the jet probability and secondary vertex $b$-taggers is
given.
\begin{figure}[!htb]
\centering
    \includegraphics[width=0.5\textwidth,clip=]{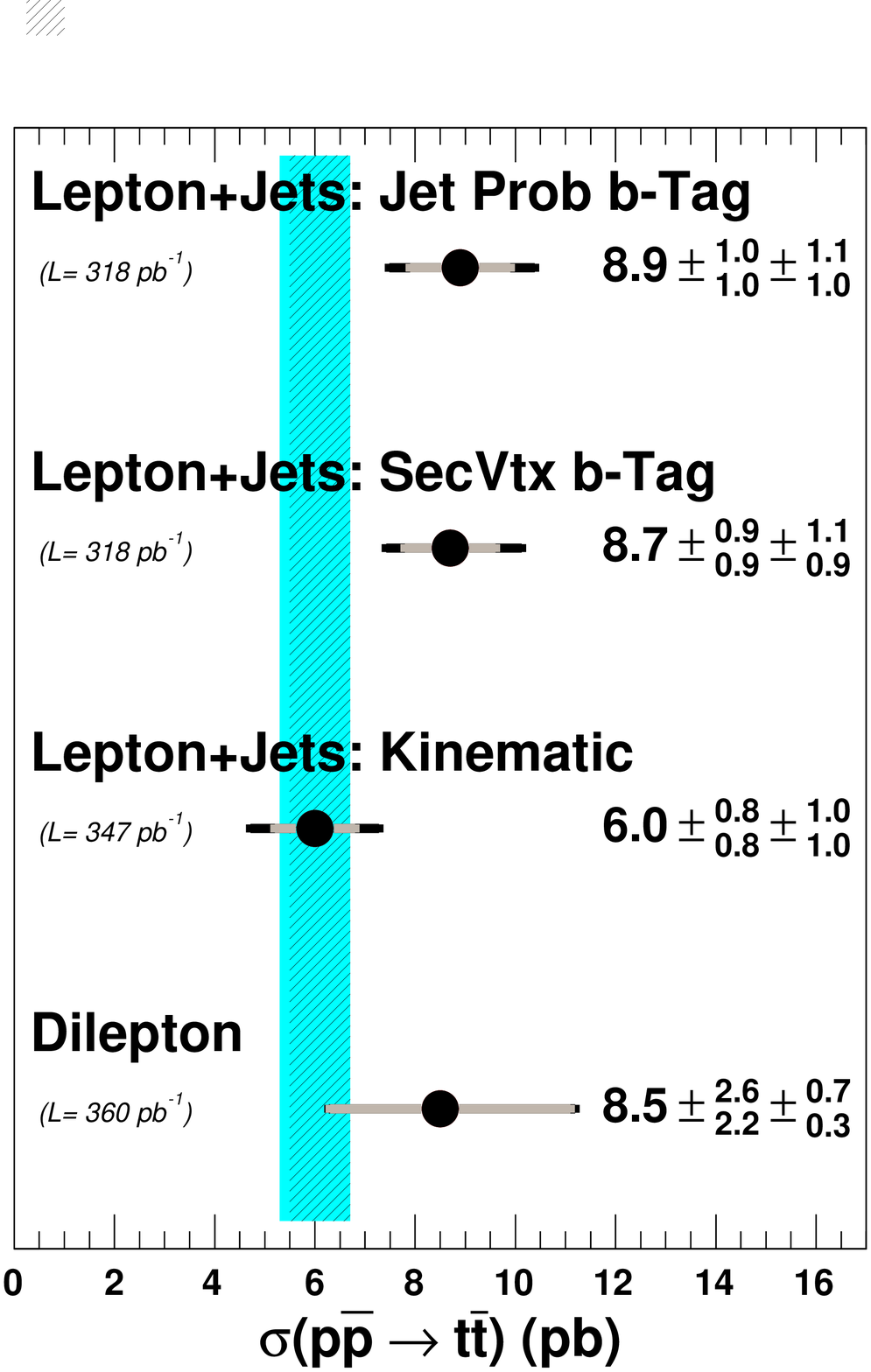}
\caption{Comparison of the $t\bar t$ production cross section measurement 
presented in this paper with theoretical 
predictions (solid band~\cite{bib:ref1}, dashed band~\cite{bib:ref2}). Also shown are the {\CDF} results obtained in the lepton+jets 
channel when using the secondary vertex tagger~\cite{bib:secvtx_prl} and when no $b$-tagging is applied (only kinematic information)~\cite{bib:kin_gen5}, and the result obtained 
in the dilepton channel~\cite{bib:dil_gen5}. 
All the measured cross sections are evaluated at $m_t=178~\Gevcc$. The grey (black) lines represent the statistical (total) uncertainties. For the dilepton analysis, the grey (black) lines represent the uncertainties coming from the fit (shape).}
\label{fig:theorycomp}
\end{figure}

\begin{acknowledgments}
We thank the Fermilab staff and the technical staffs of the participating institutions for their vital contributions. This work was supported by the U.S. Department of Energy and National Science Foundation; the Italian Istituto Nazionale di Fisica Nucleare; the Ministry of Education, Culture, Sports, Science and Technology of Japan; the Natural Sciences and Engineering Research Council of Canada; the National Science Council of the Republic of China; the Swiss National Science Foundation; the A.P. Sloan Foundation; the Bundesministerium f\"ur Bildung und Forschung, Germany; the Korean Science and Engineering Foundation and the Korean Research Foundation; the Particle Physics and Astronomy Research Council and the Royal Society, UK; the Russian Foundation for Basic Research; the Comisi\'on Interministerial de Ciencia y Tecnolog\'{\i}a, Spain; the European Community's Human Potential Programme under contract HPRN-CT-2002-00292; and the Academy of Finland. 
\end{acknowledgments}

\appendix*
\section{Kinematic Distributions}

We compare the distributions for different kinematic variables observed in 
data to the expectations for signal and backgrounds derived 
from a combination of simulation and $t\bar t$ 
cross section measurements. 
Figures~\ref{fig:jp1_1t} to~\ref{fig:jp5_2t} show the results for the
four samples of events passing the selection criteria with at least three jets 
and one or two tags for $P_J<$ 1\% or $P_J<$ 5\%. The considered kinematic 
variables are the sum of the transverse energies of each object in the final 
state ($H_T$), the reconstructed transverse mass of the $W$ boson, the missing 
transverse energy ($\met$) of the event, the $E_T$ of the lepton, the 
transverse energy of the tagged jets, and the pseudo-rapidity of the tagged 
jets with respect to the center of the detector. 
Kolmogorov-Smirnov (KS) probabilities are computed to test the agreement
between observed and expected distributions. The distributions observed
in the data are statistically consistent with the expected 
signal-plus-background distributions.

\begin{figure*}[!htb]
  \includegraphics[width=8.0cm,clip=]{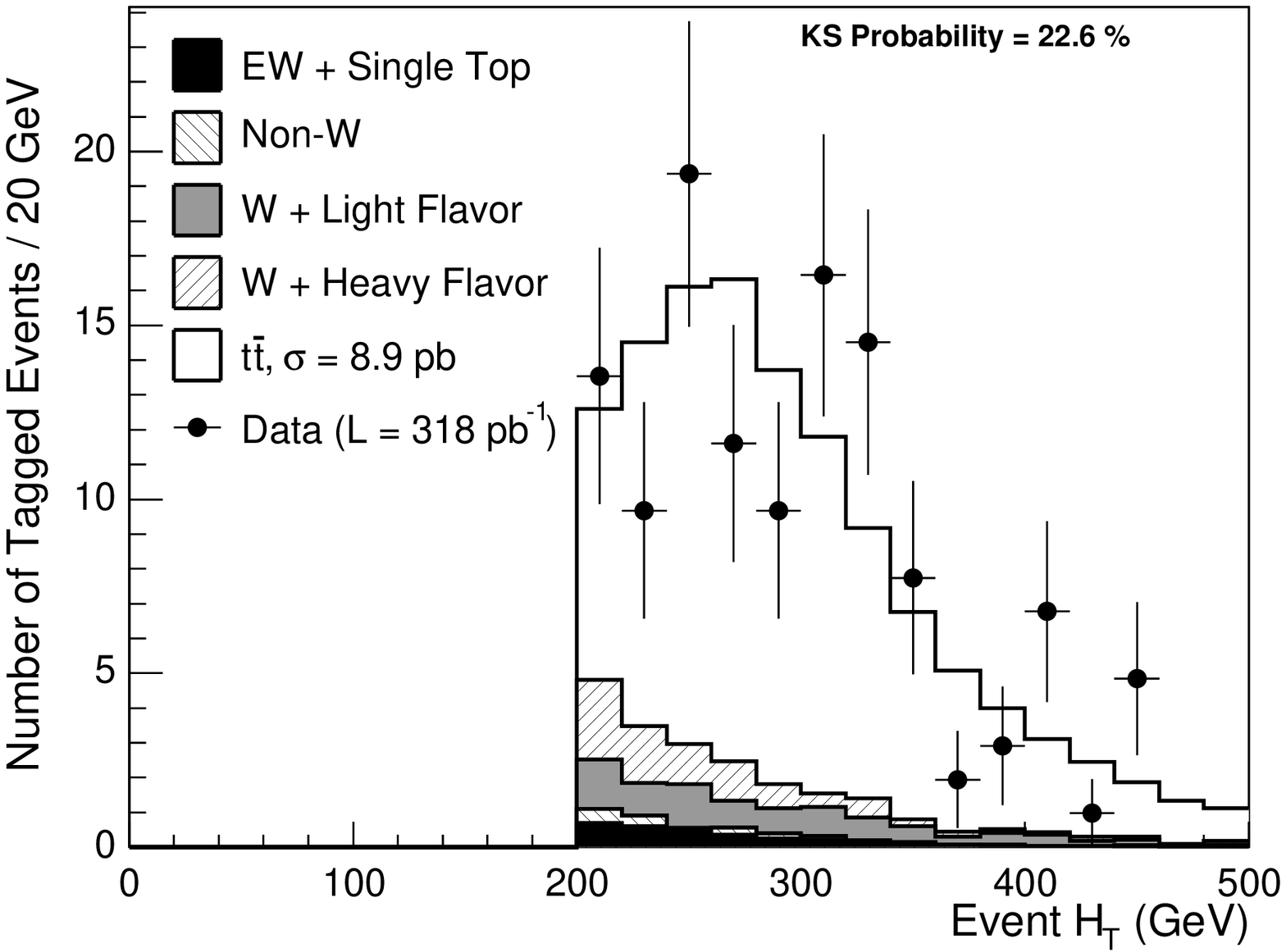}
  \includegraphics[width=8.0cm,clip=]{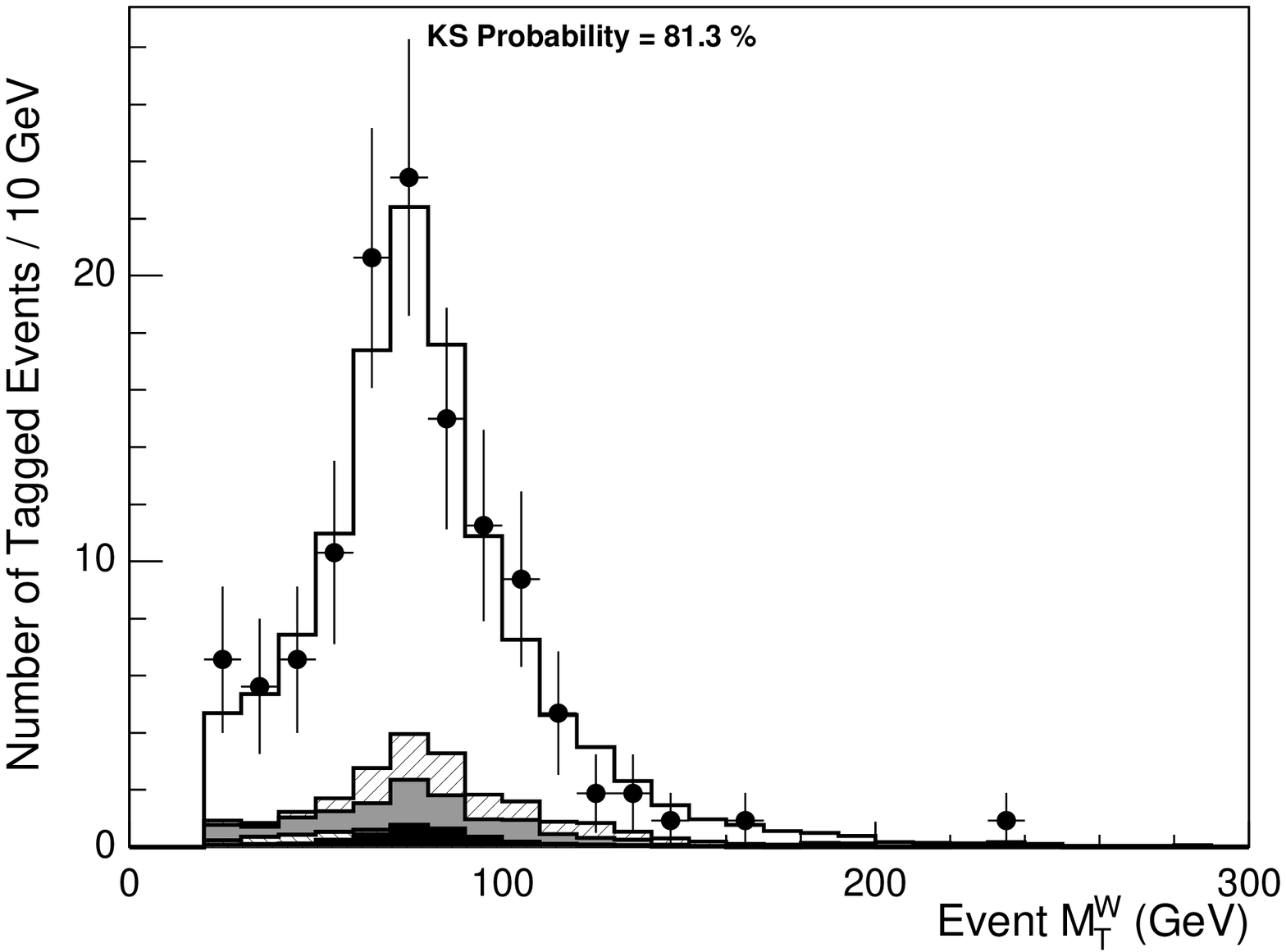}
  \includegraphics[width=8.0cm,clip=]{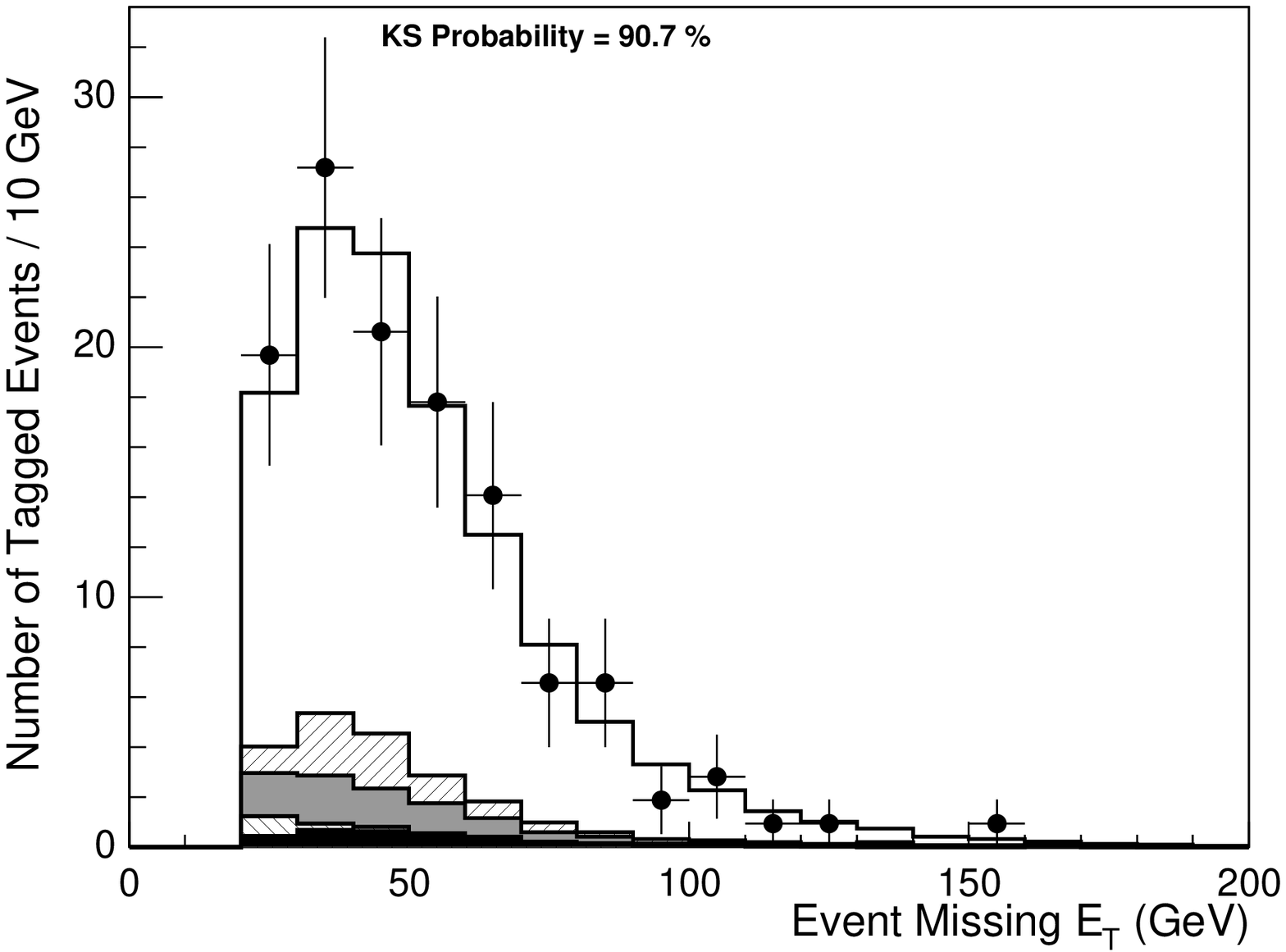}
  \includegraphics[width=8.0cm,clip=]{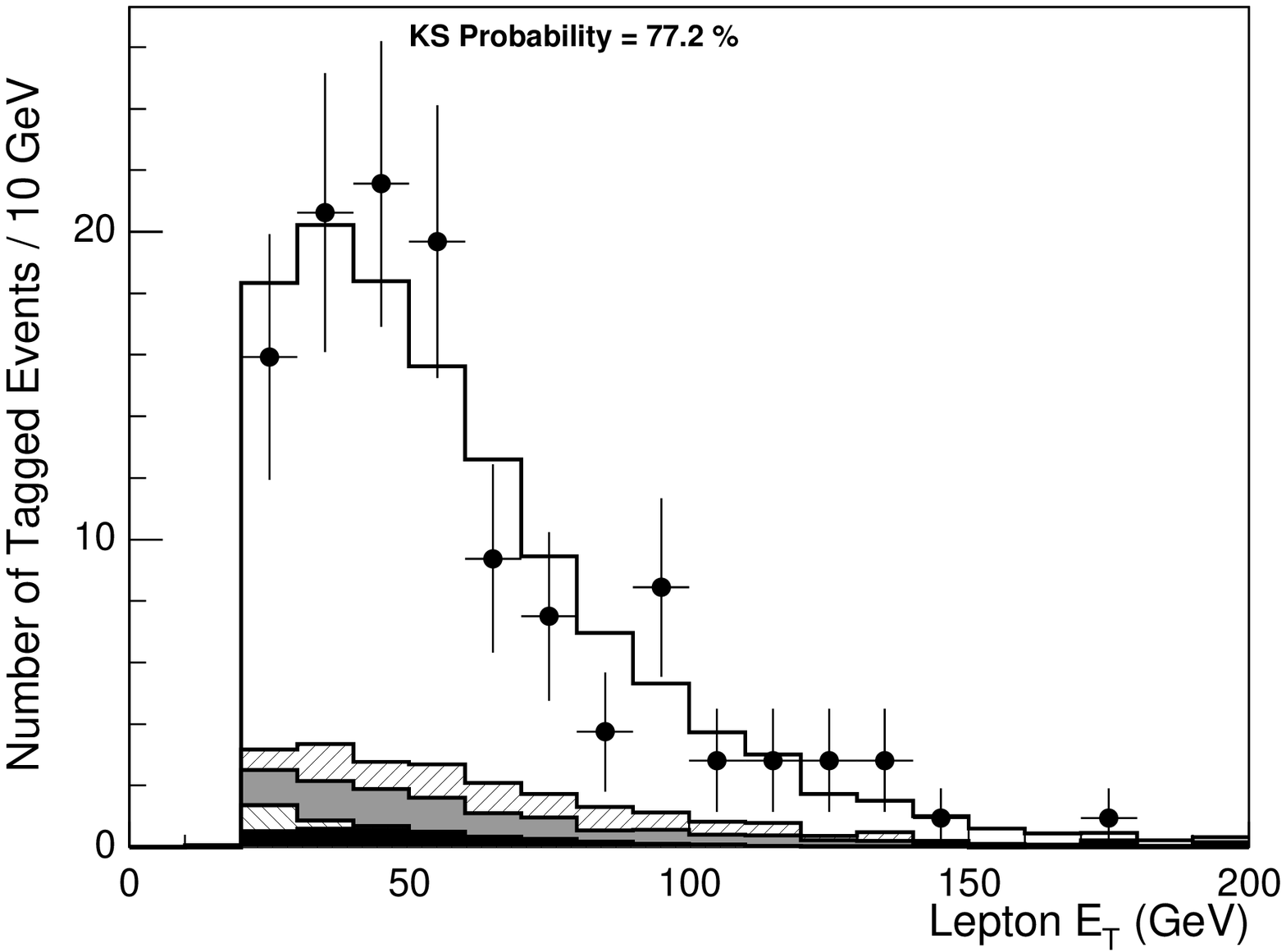}
  \includegraphics[width=8.0cm,clip=]{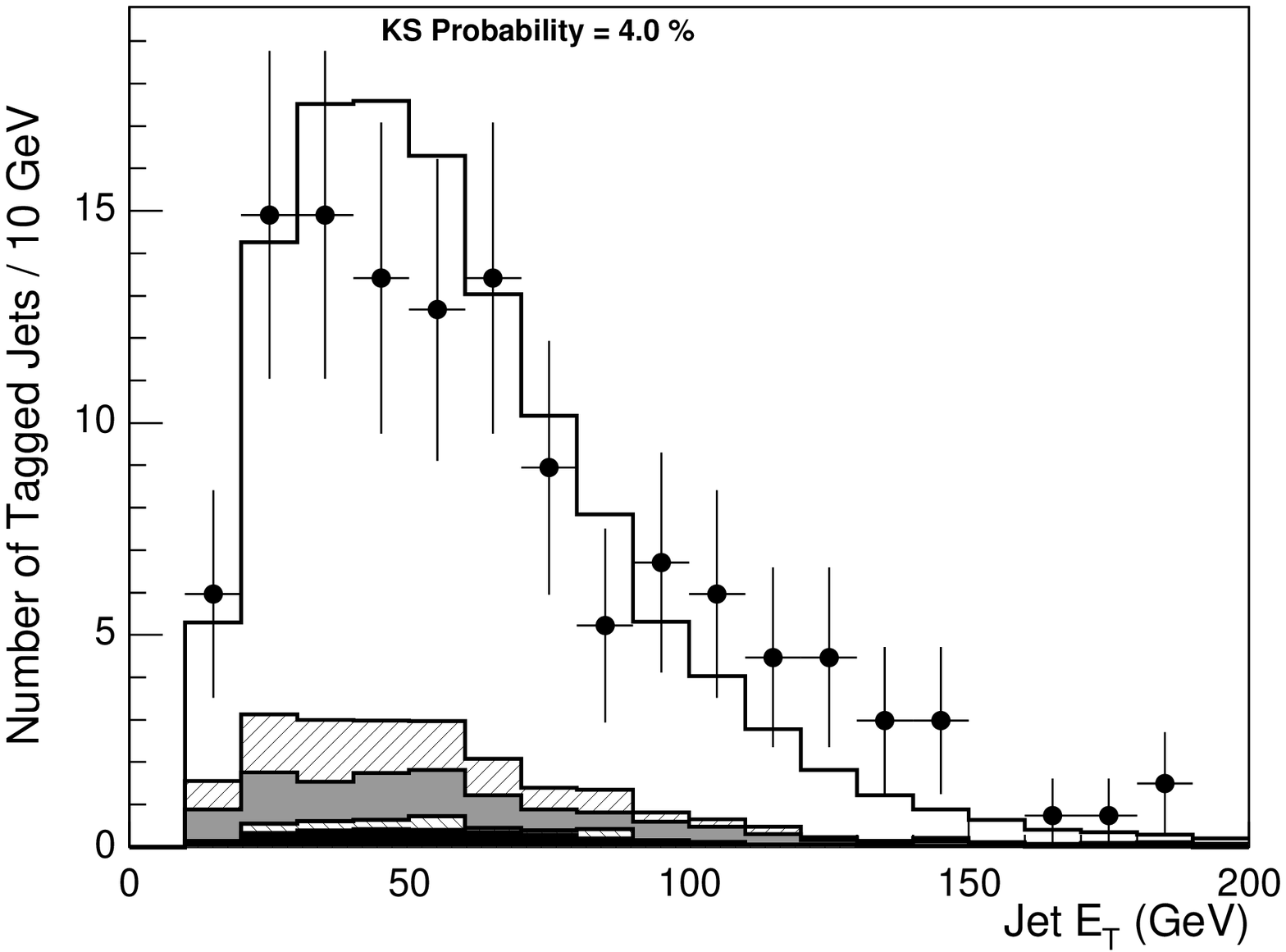}
  \includegraphics[width=8.0cm,clip=]{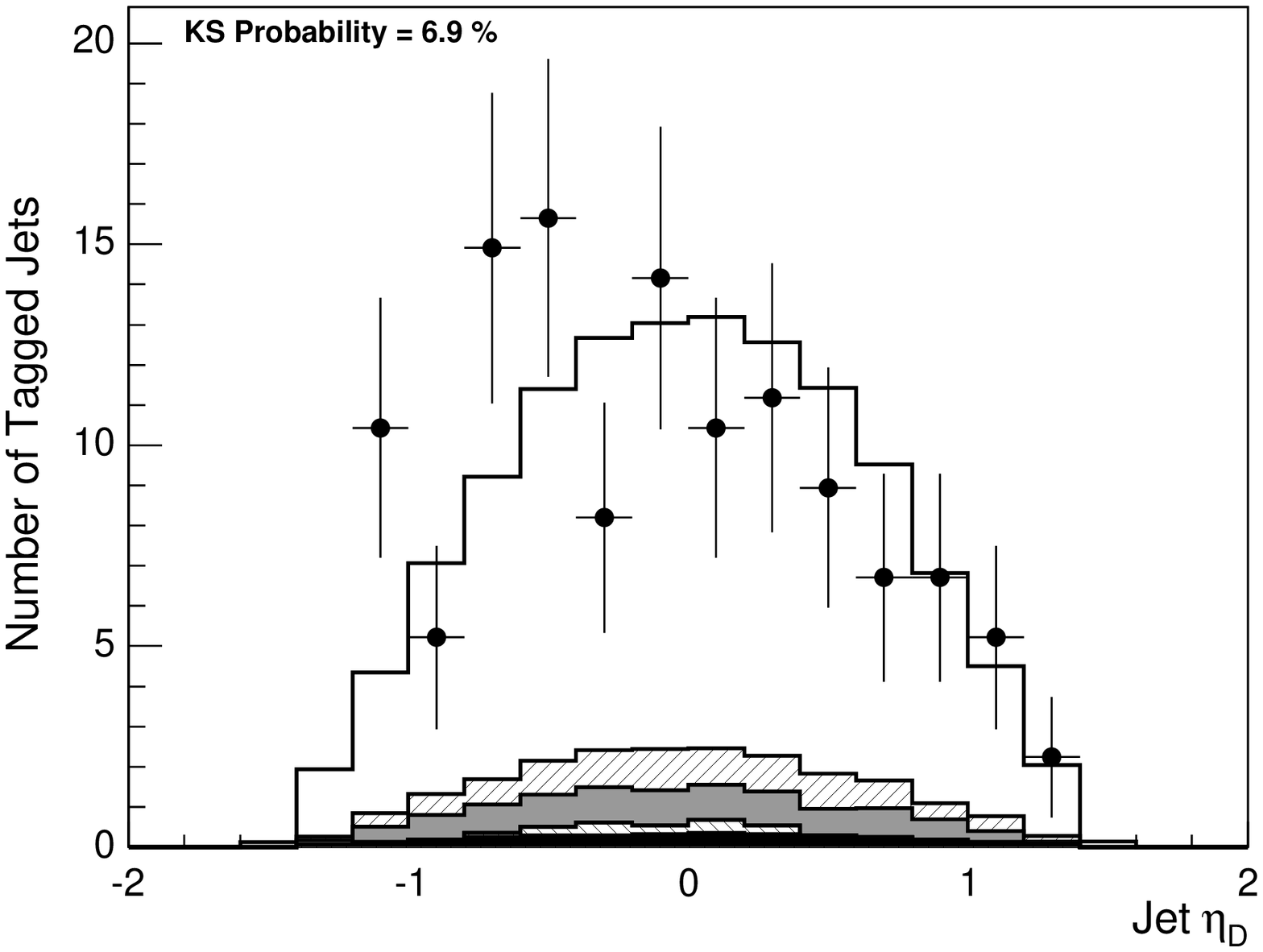}
  \caption{Comparison of kinematic distributions in data to signal and 
background expectations for events passing the selection criteria with at 
least three jets and at least one tag for $P_J<$ 1\%.
From the top-left corner: sum of the transverse energies of each object in the 
final state ($H_T$), reconstructed transverse mass of the $W$ boson, missing 
transverse energy ($\met$), lepton $E_T$, transverse energy of the tagged jets,
and the pseudo-rapidity of the tagged jets with respect to the center of the 
detector.}
  \label{fig:jp1_1t}
\end{figure*}

\begin{figure*}[!htb]
  \includegraphics[width=8.0cm,clip=]{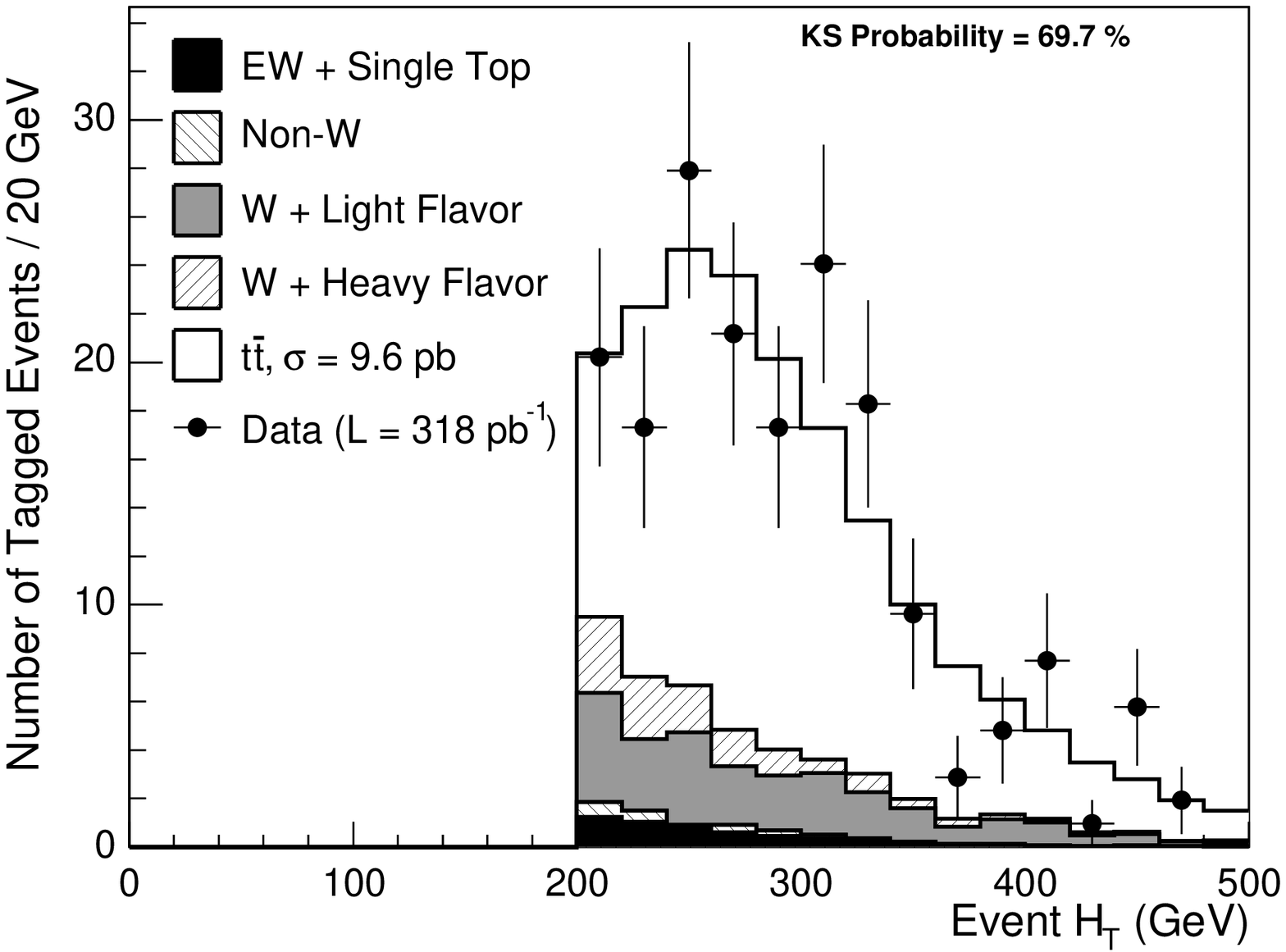}
  \includegraphics[width=8.0cm,clip=]{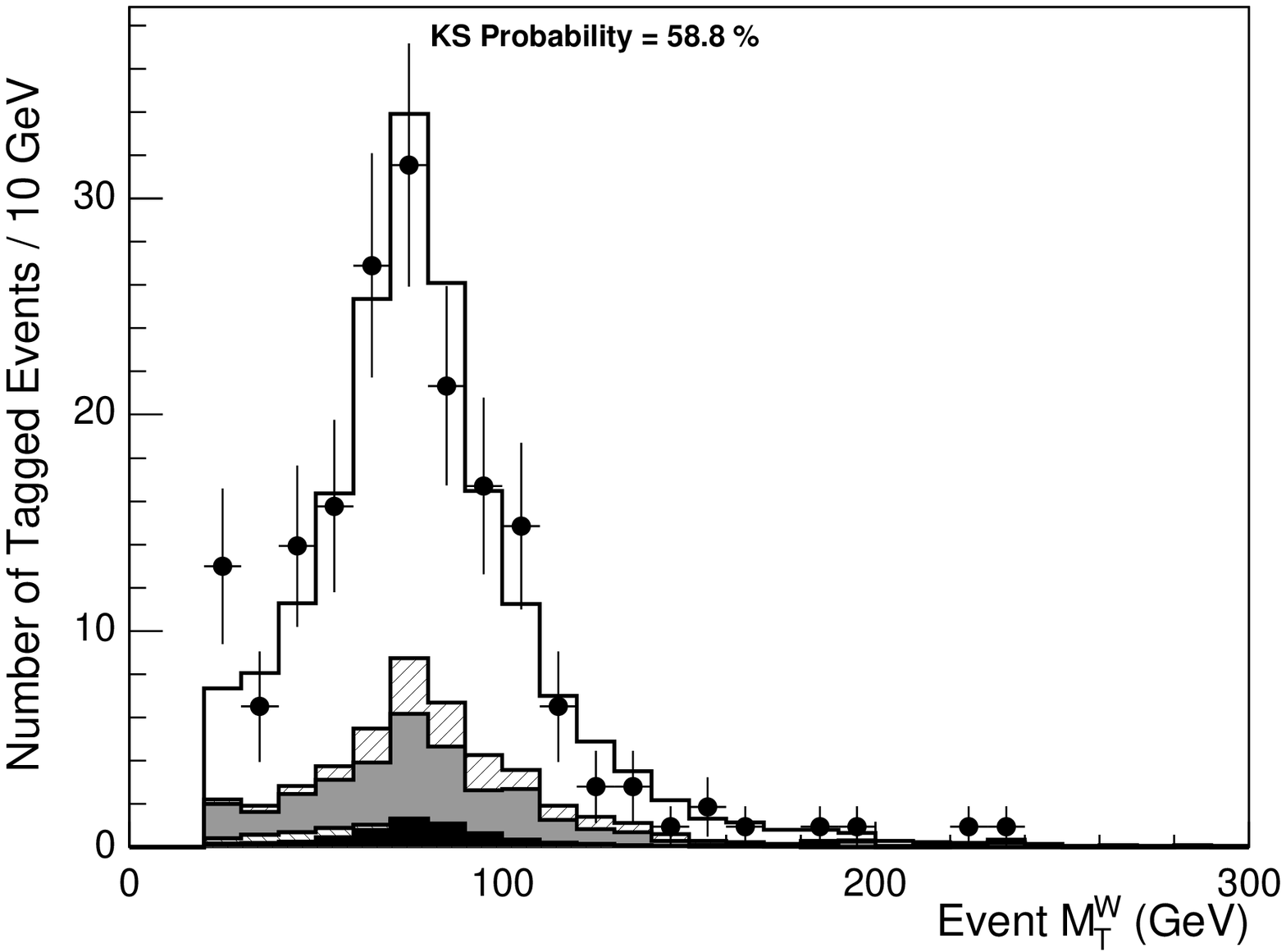}
  \includegraphics[width=8.0cm,clip=]{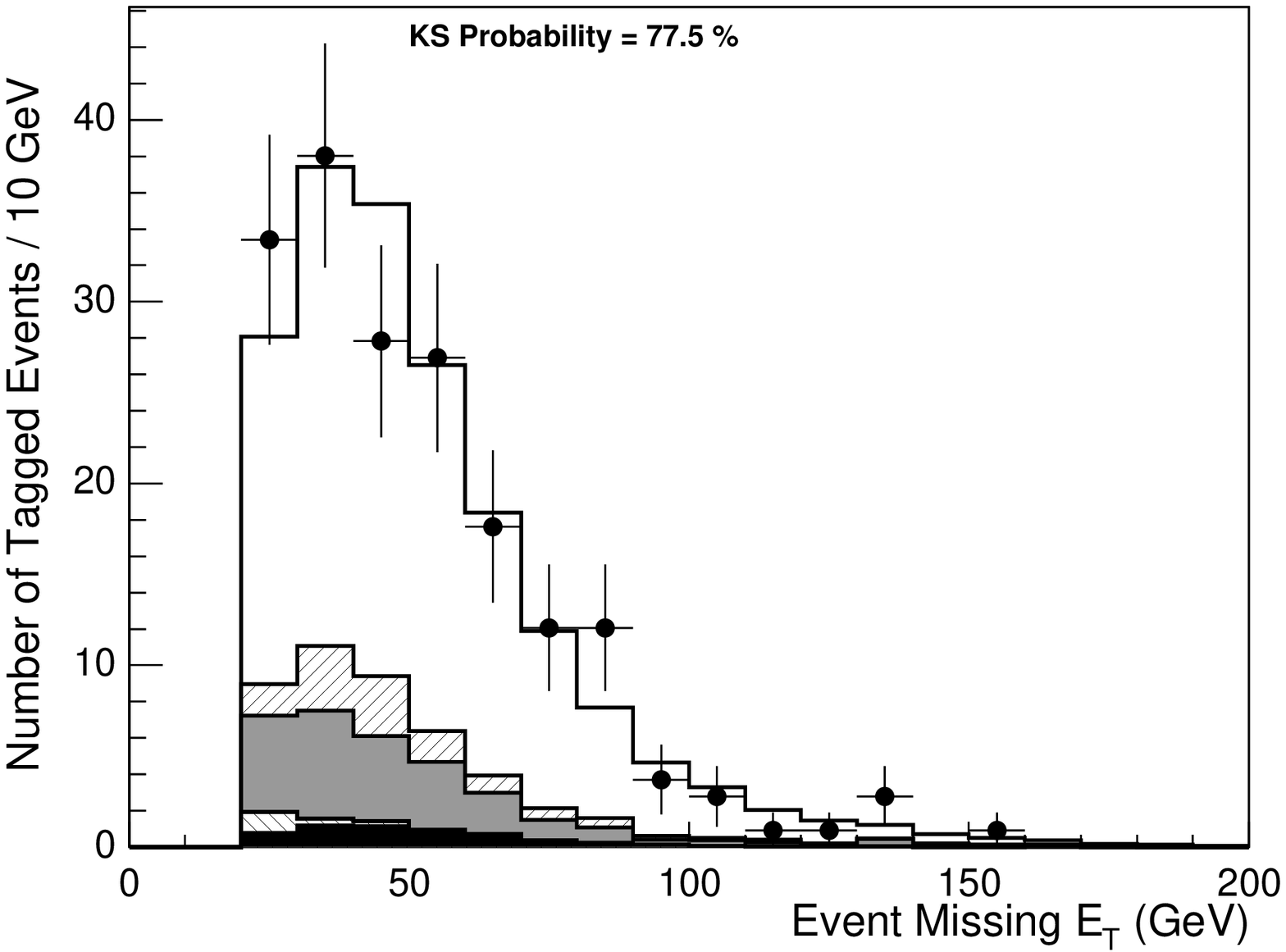}
  \includegraphics[width=8.0cm,clip=]{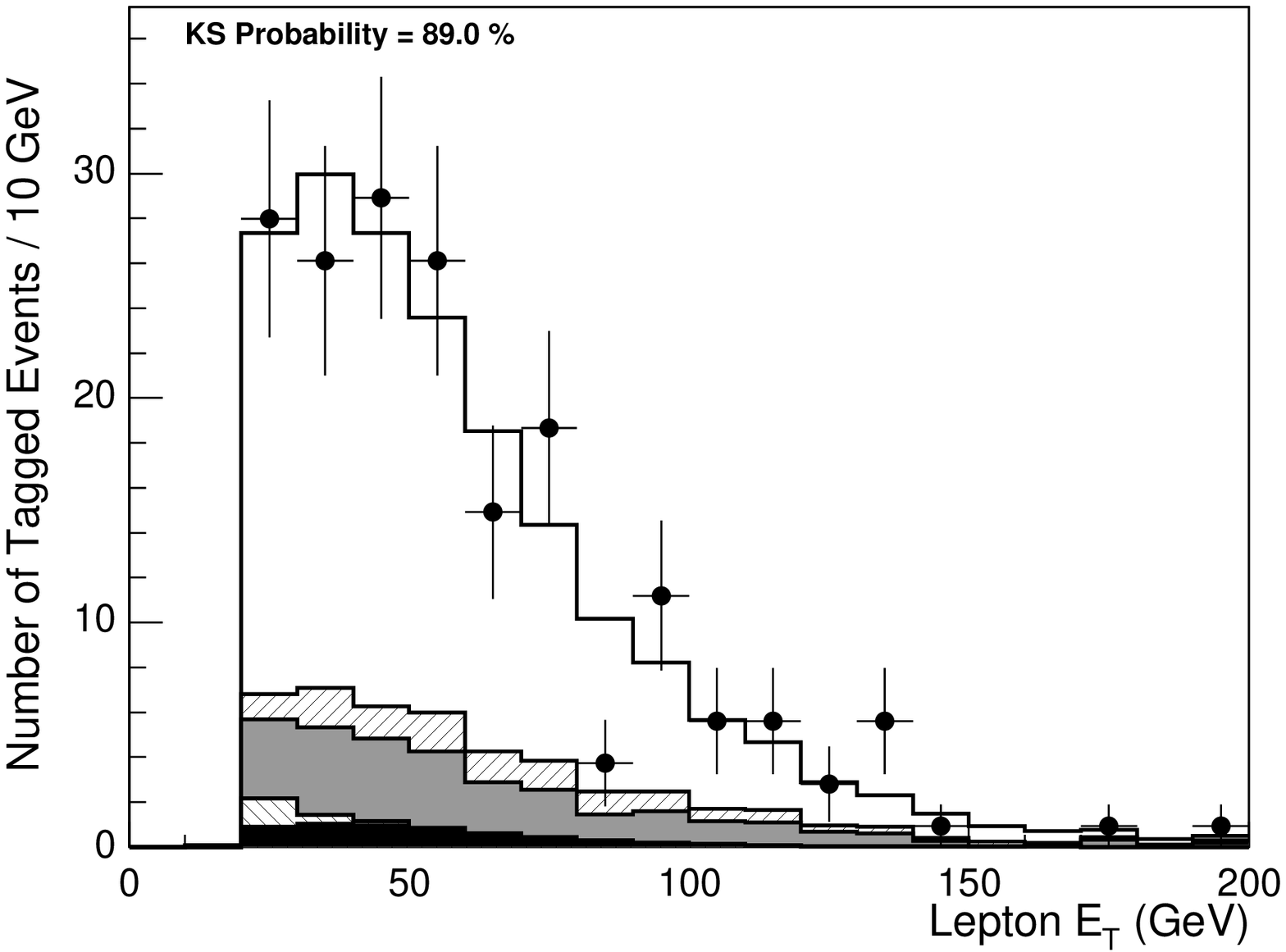}
  \includegraphics[width=8.0cm,clip=]{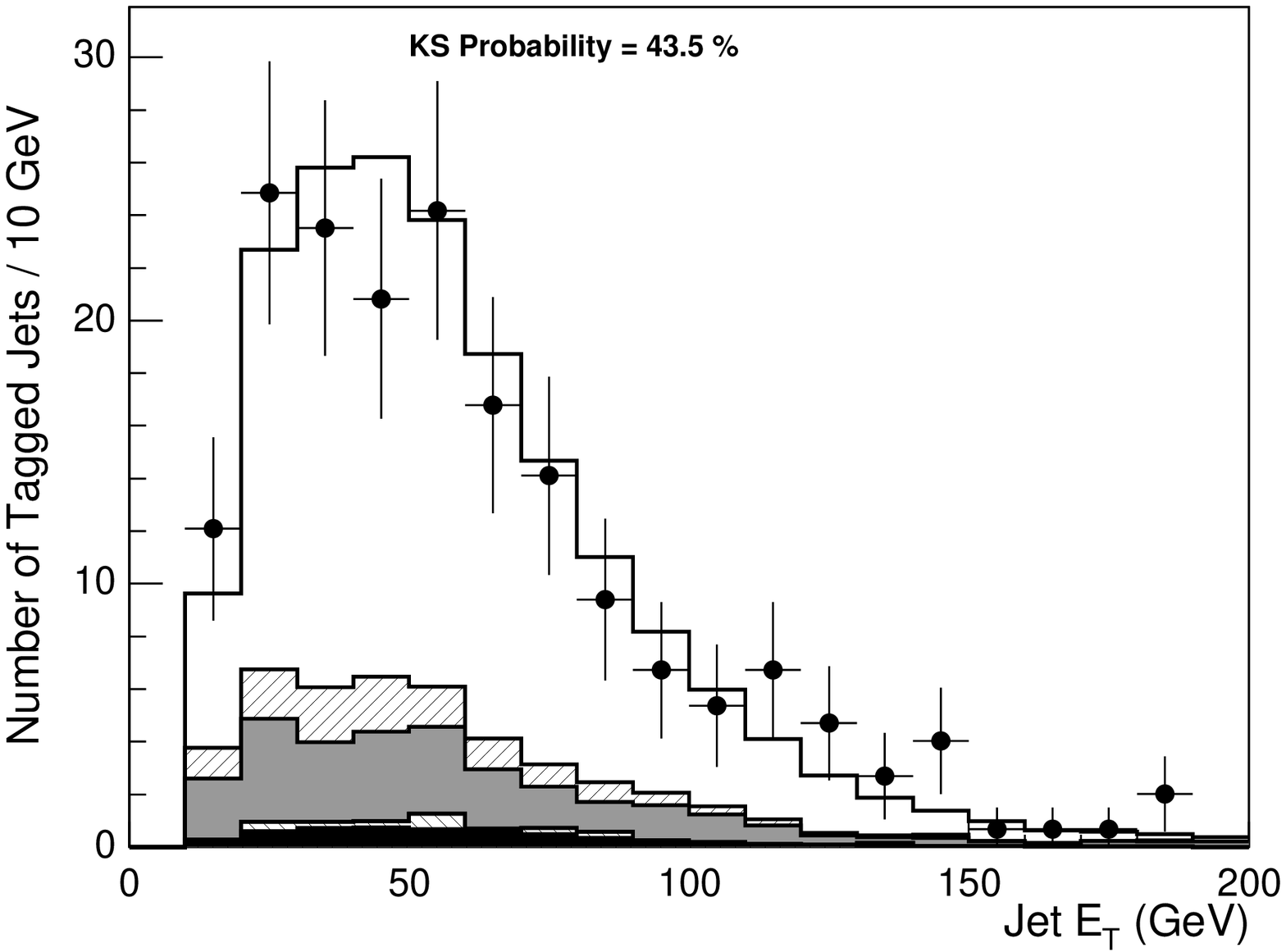}
  \includegraphics[width=8.0cm,clip=]{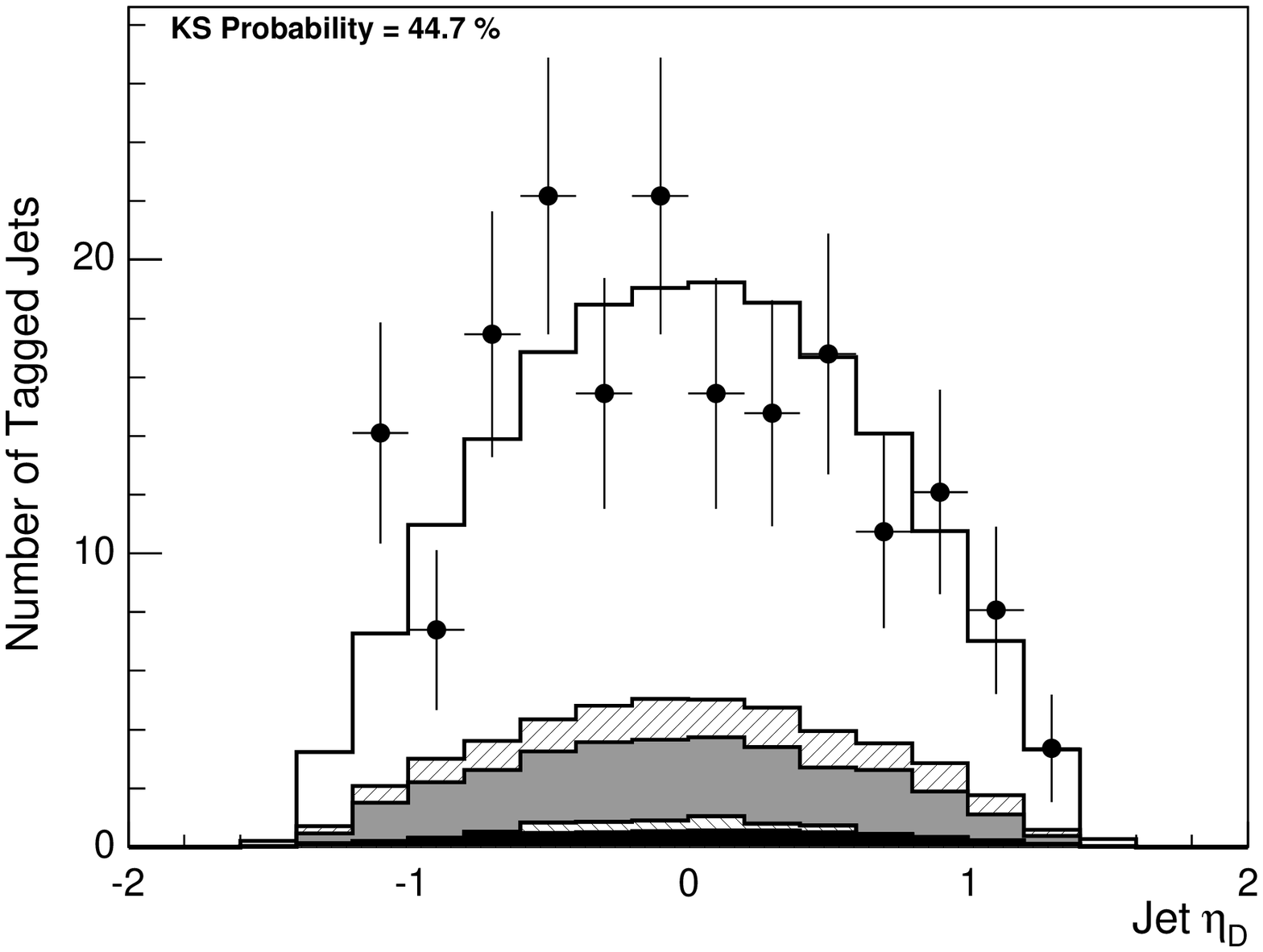}
  \caption{Comparison of kinematic distributions in data to signal and 
background expectations for events in the single tag sample ($P_J<$ 5\%).
}
  \label{fig:jp5_1t}
\end{figure*}
 
\begin{figure*}[!htb]
  \includegraphics[width=8.0cm,clip=]{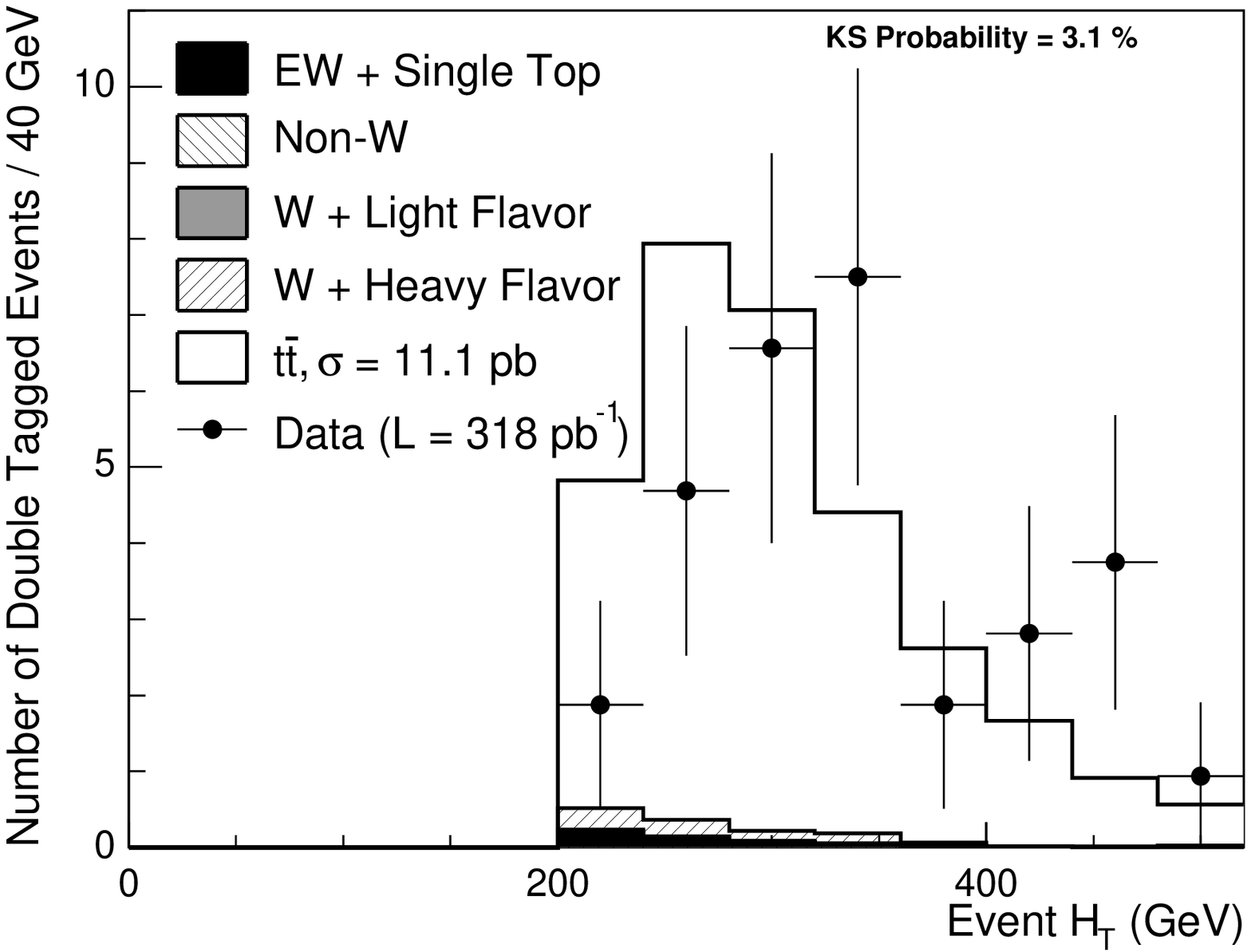}
  \includegraphics[width=8.0cm,clip=]{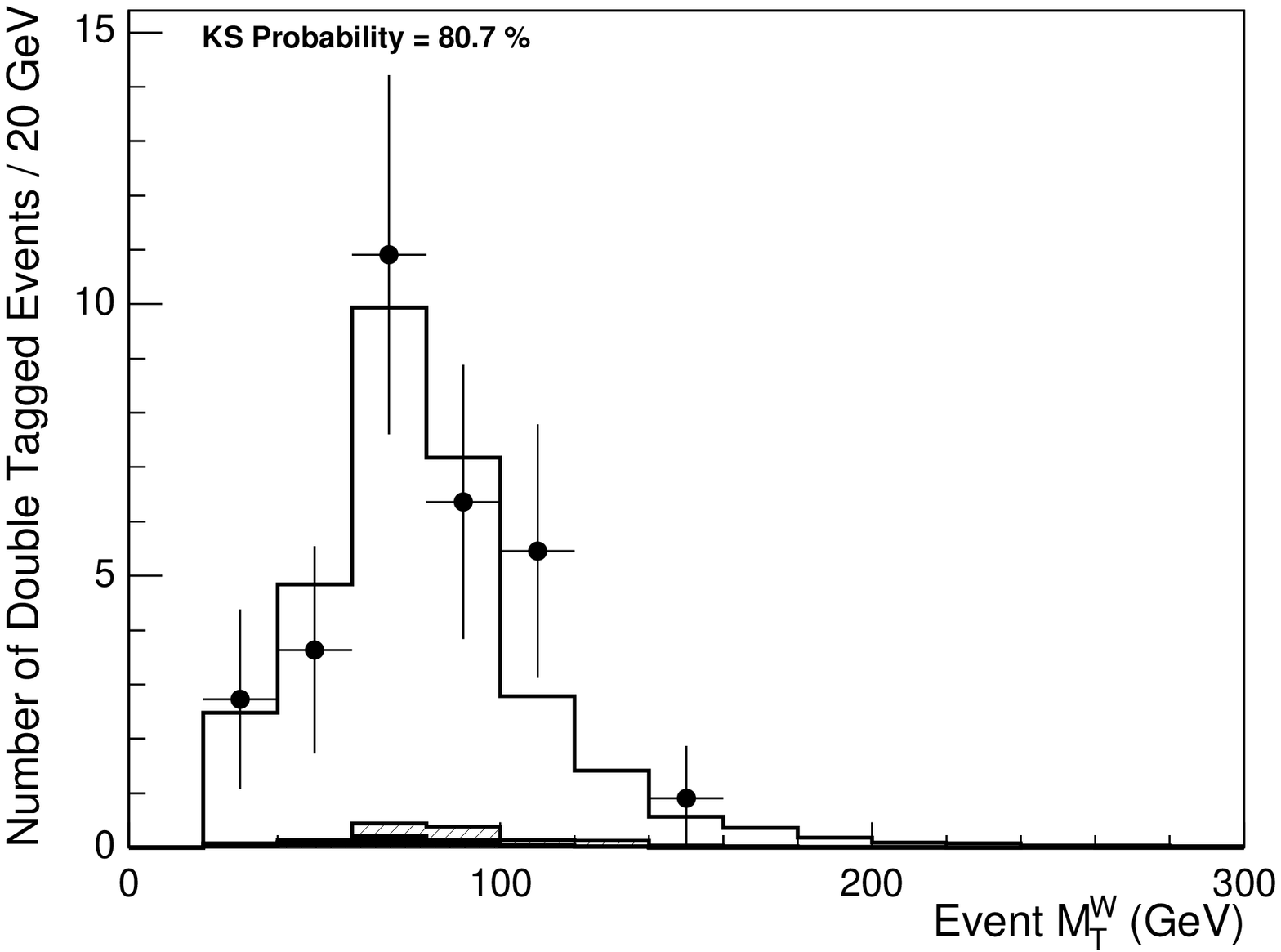}
  \includegraphics[width=8.0cm,clip=]{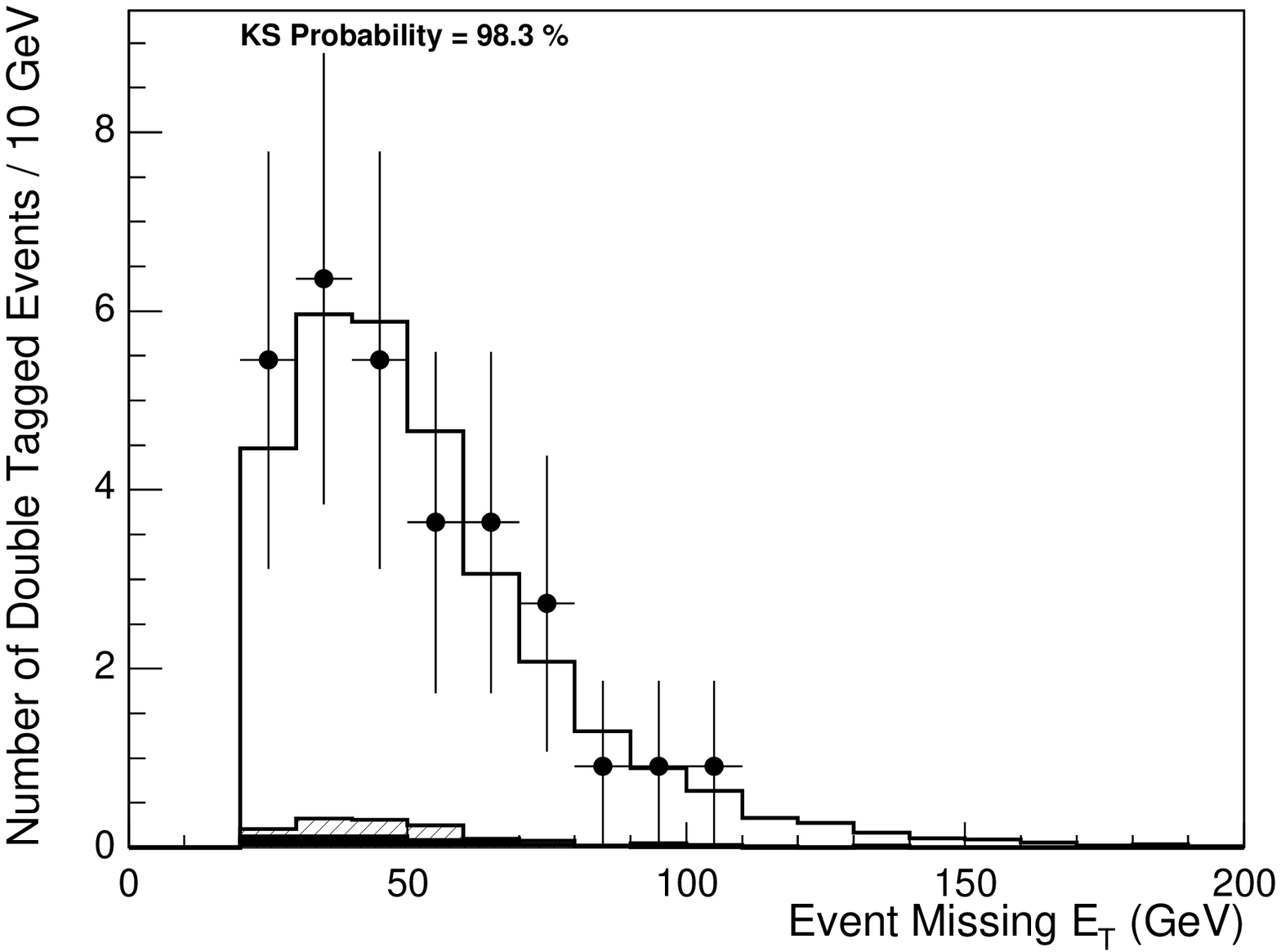}
  \includegraphics[width=8.0cm,clip=]{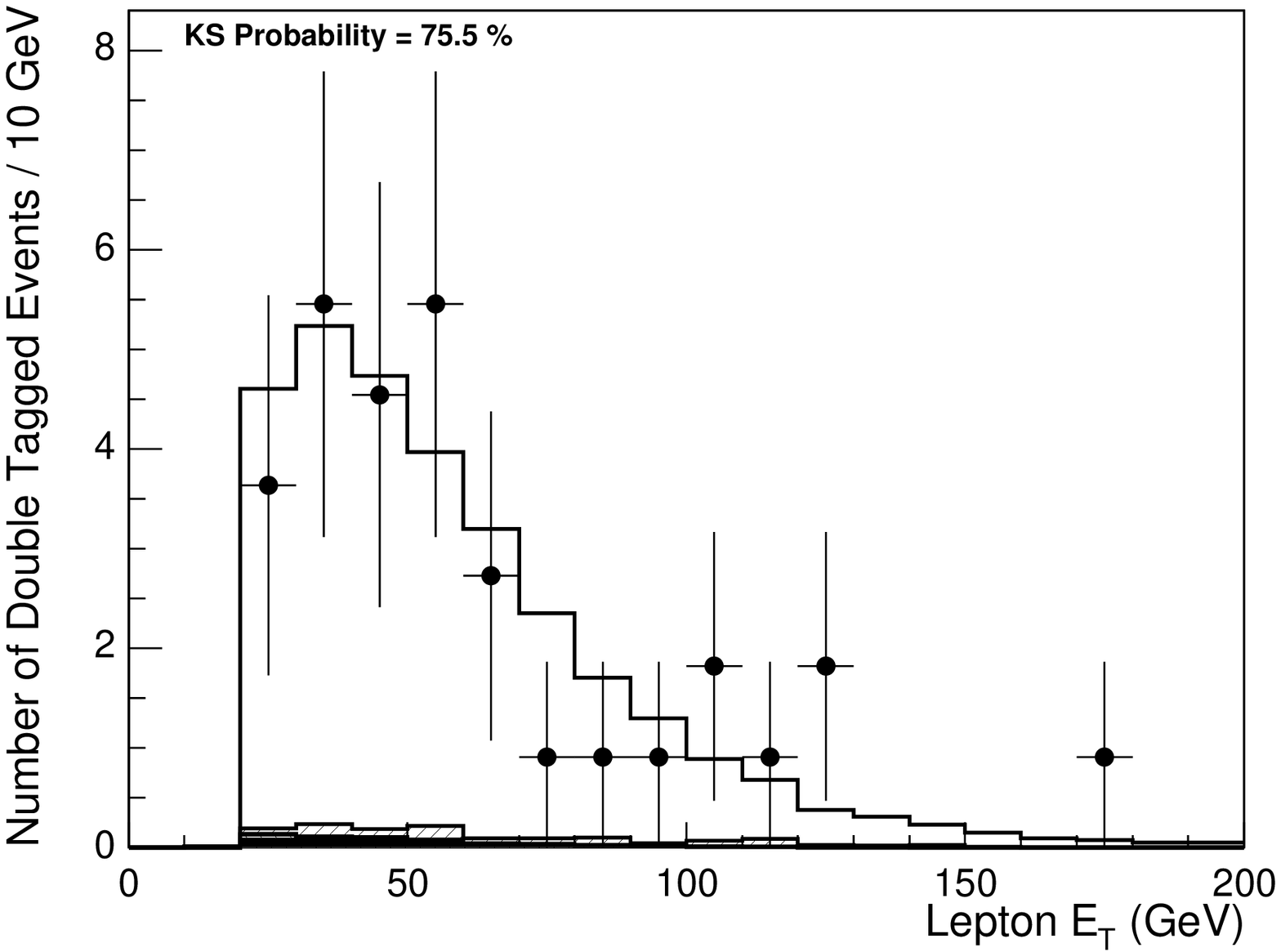}
  \includegraphics[width=8.0cm,clip=]{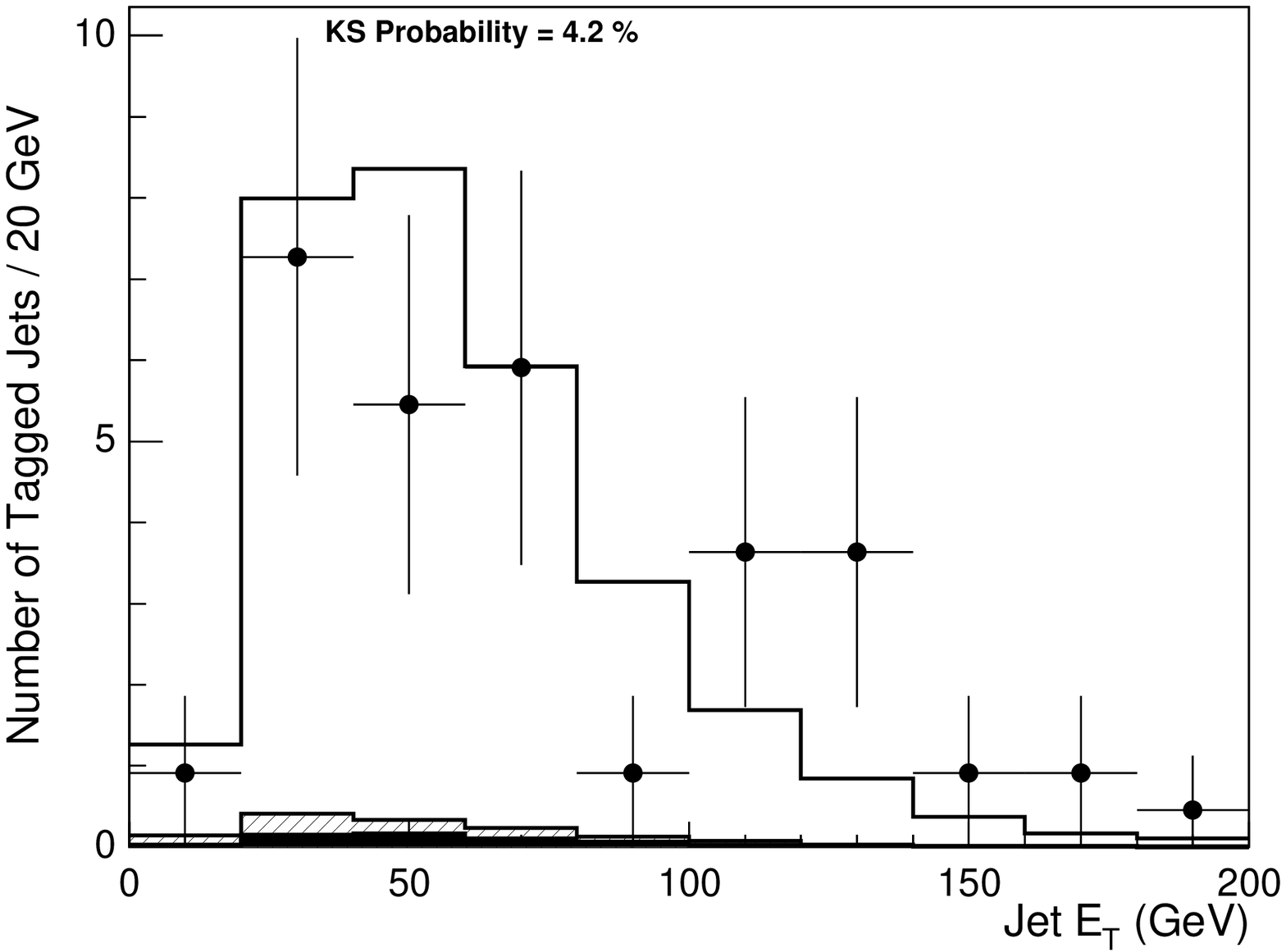}
  \includegraphics[width=8.0cm,clip=]{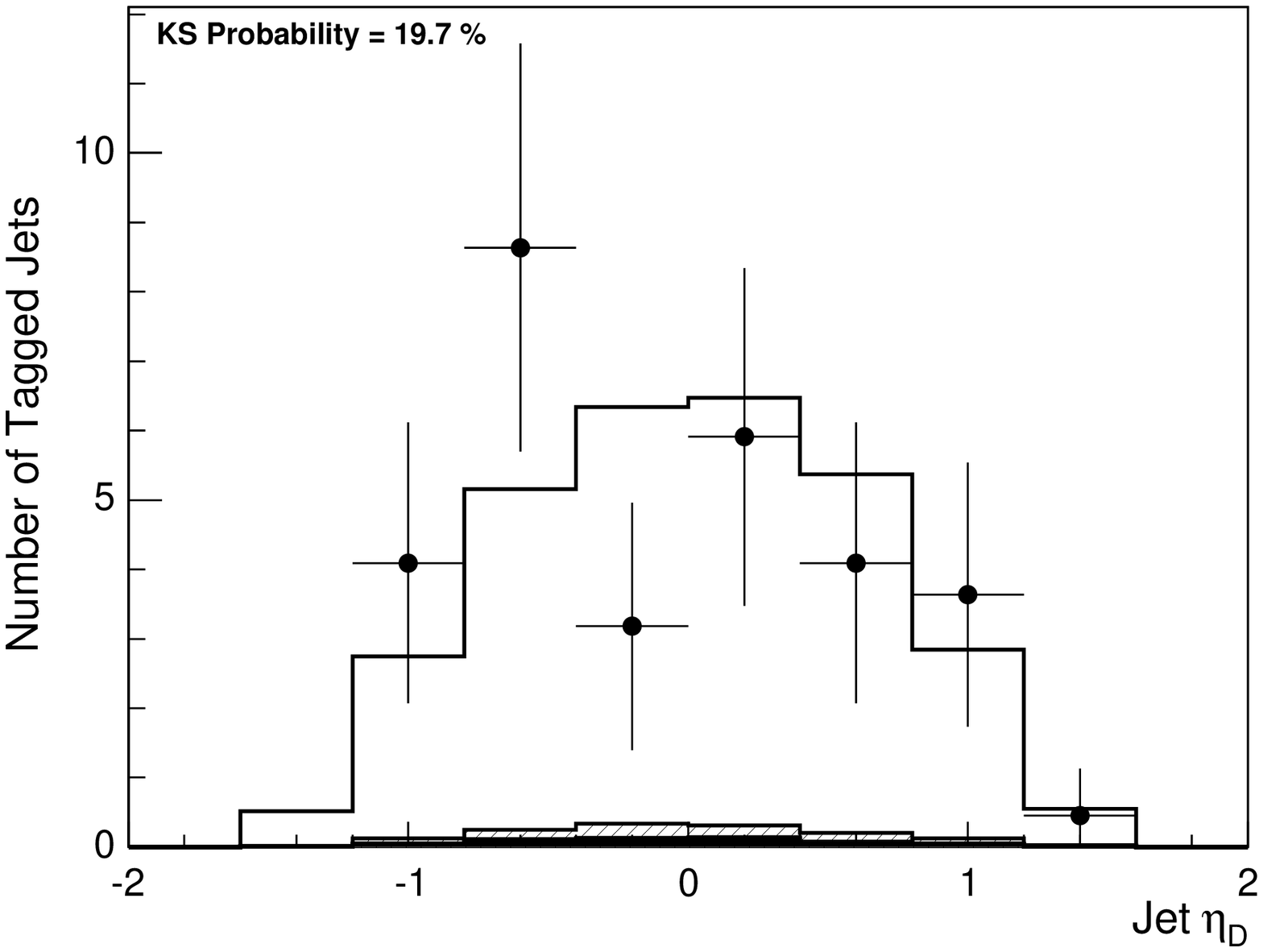}
  \caption{Comparison of kinematic distributions in data to signal and 
background expectations for events in the double tag sample ($P_J<$ 1\%).
}
  \label{fig:jp1_2t}
\end{figure*}

\begin{figure*}[!htb]
  \includegraphics[width=8.0cm,clip=]{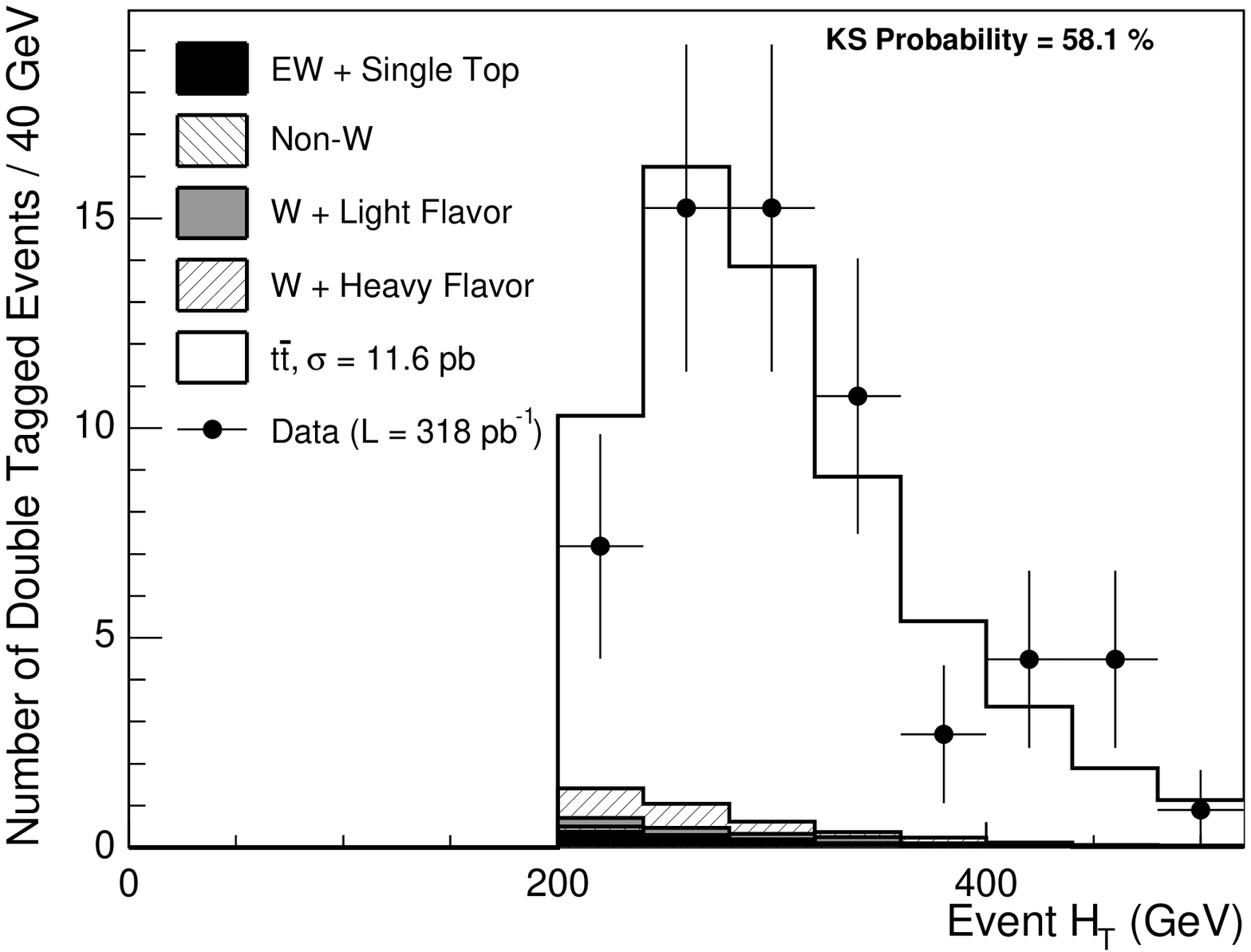}
  \includegraphics[width=8.0cm,clip=]{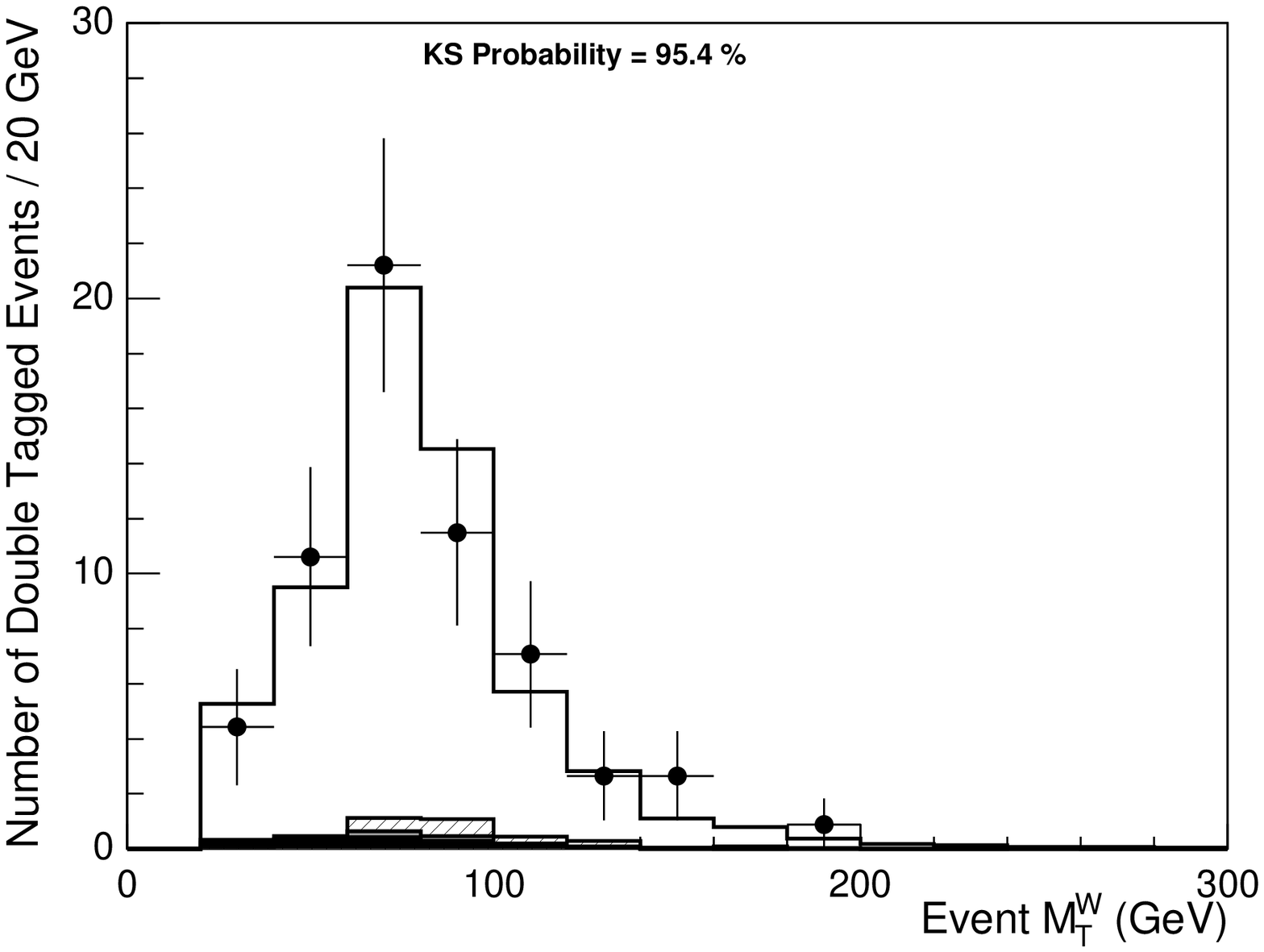}
  \includegraphics[width=8.0cm,clip=]{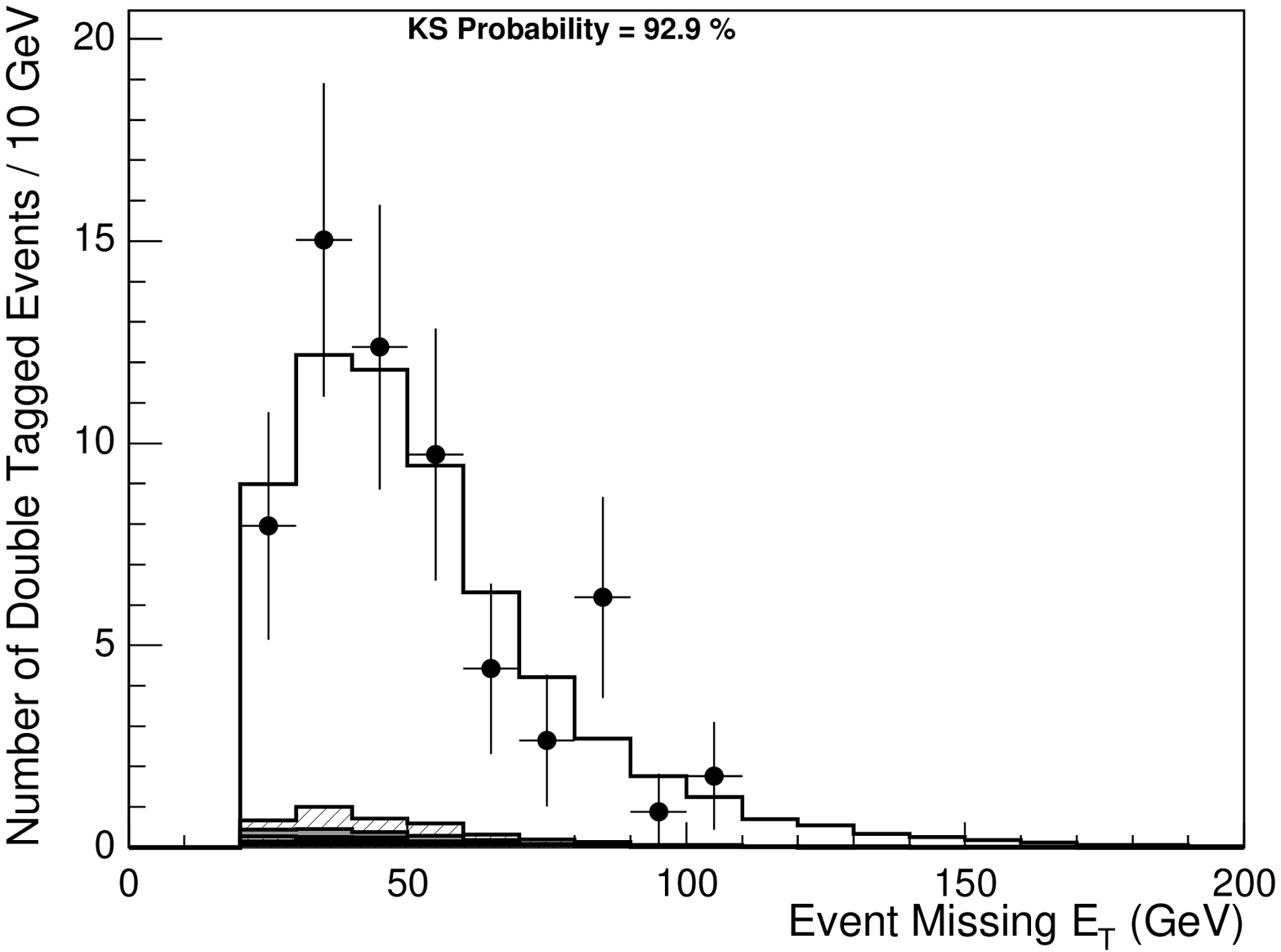}
  \includegraphics[width=8.0cm,clip=]{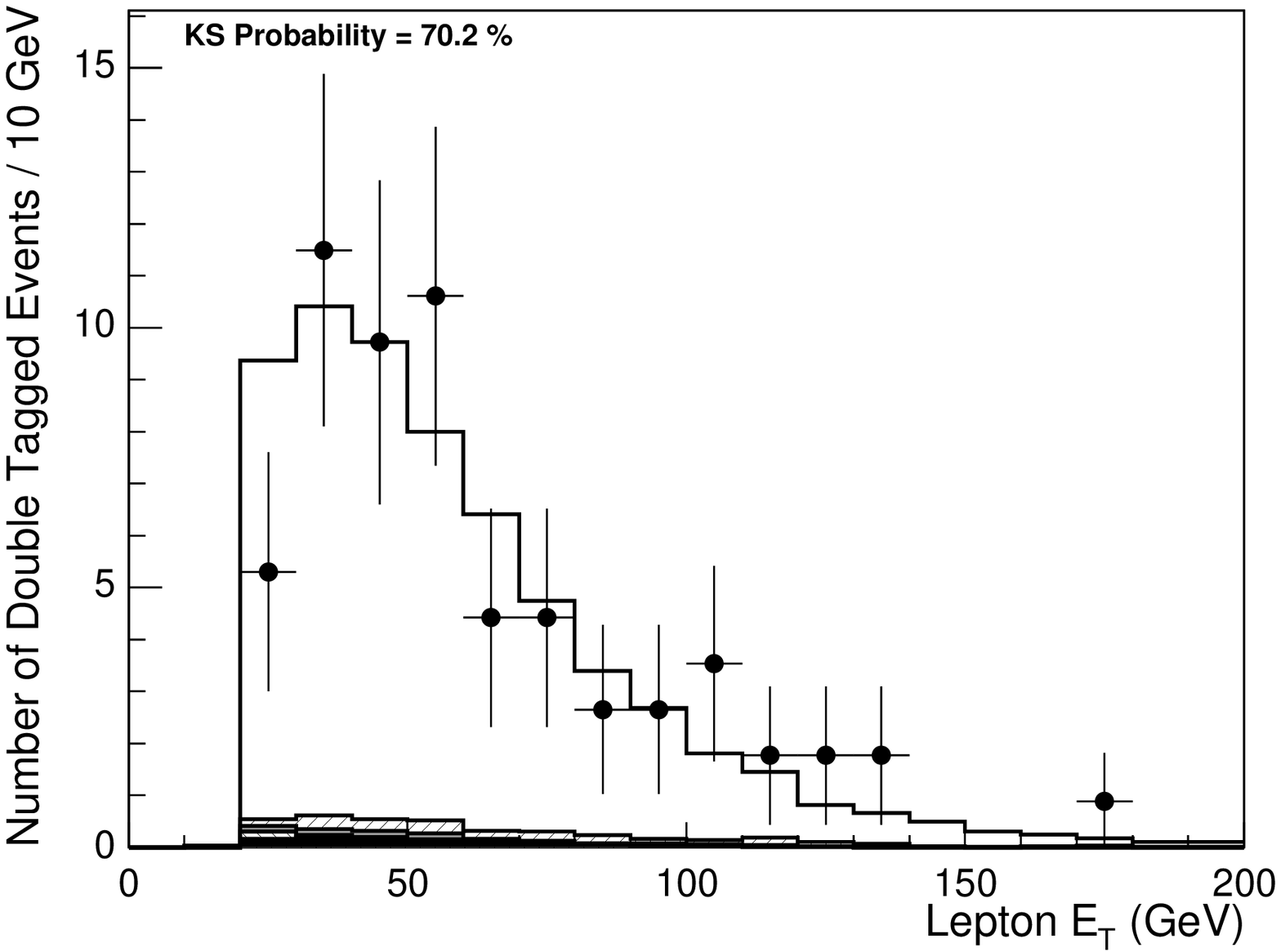}
  \includegraphics[width=8.0cm,clip=]{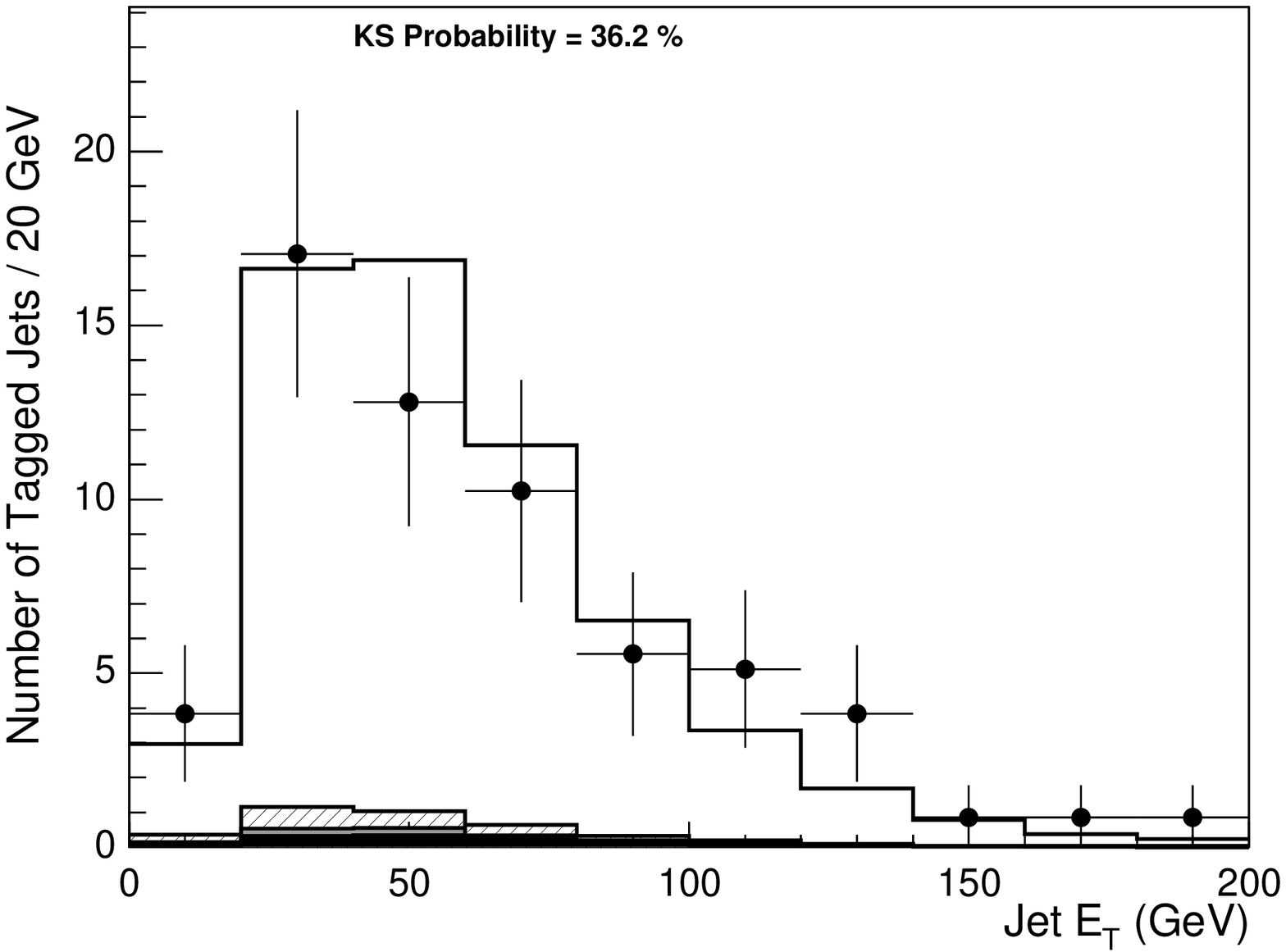}
  \includegraphics[width=8.0cm,clip=]{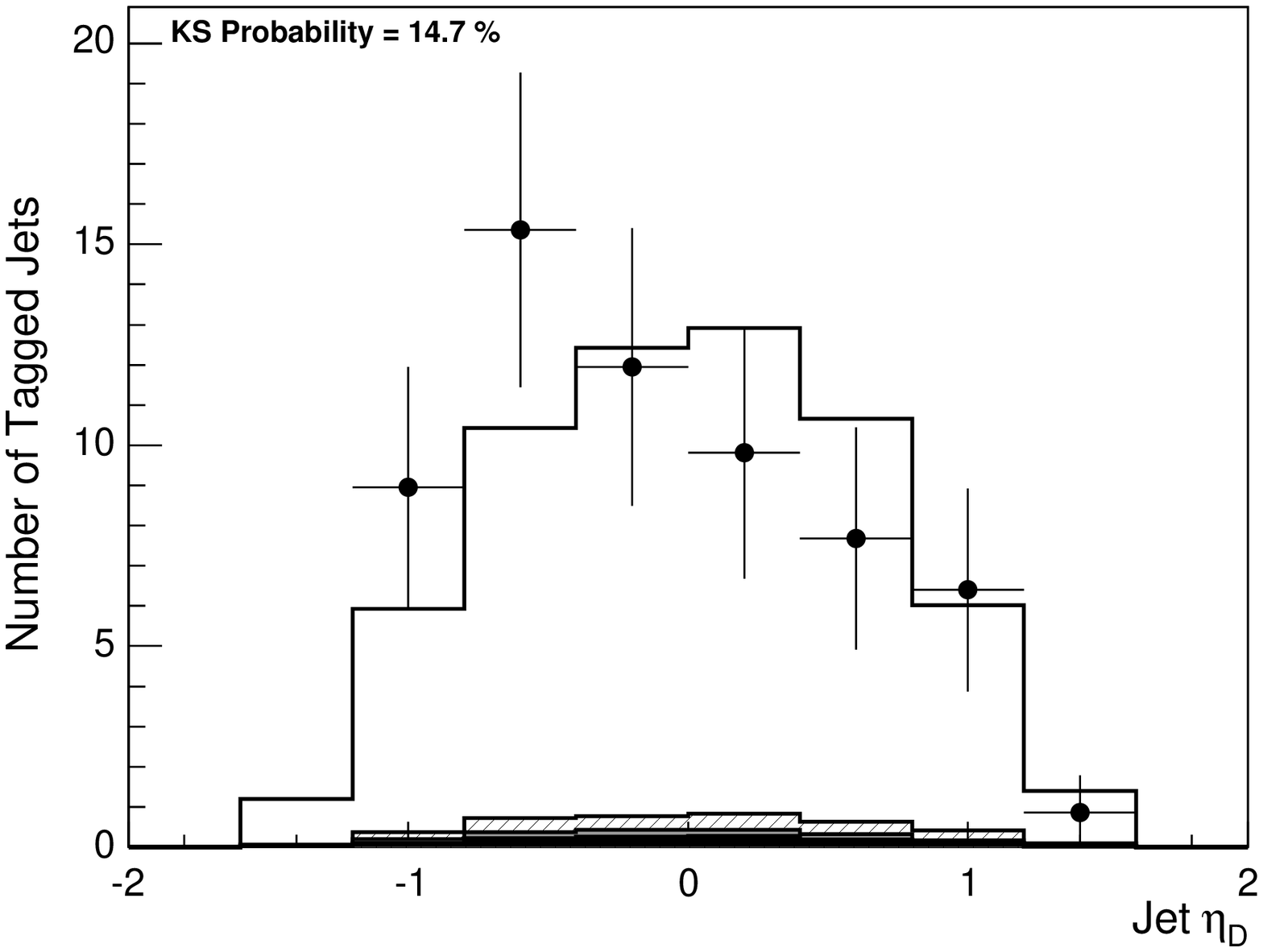}
  \caption{Comparison of kinematic distributions in data to signal and 
background expectations for events in the double tag sample ($P_J<$ 5\%).
}
  \label{fig:jp5_2t}
\end{figure*}

\end{document}